\documentclass[11pt,preprint]{aastex} 

\usepackage{natbib}
\received{}
\accepted{}
\shorttitle{$H_0$ from Refurbished Distance Ladder}
\shortauthors{Riess et al.}

\newcommand{\bq}{\begin{equation}} 
\newcommand{\eq}{\end{equation}}   
 
\newcommand{\ho}{$74.2 \pm 3.6$ km s$^{-1}$ Mpc$^{-1}$}
\newcommand{\statho}{$74.2 \pm 3.4$ km s$^{-1}$ Mpc$^{-1}$}
\newcommand{\unc}{4.8\%}  
\newcommand{\uncs}{4.6\%}  

\newcommand{\beq}{\begin{equation}}
\newcommand{\eeq}{\end{equation}}
\newcommand{\beqa}{\begin{eqnarray}}
\newcommand{\eeqa}{\end{eqnarray}}

\newcommand{\PL}{$P\hbox{-}L \ $}

\begin{document} 

\title{A Redetermination of the Hubble Constant with the Hubble Space
Telescope from a Differential Distance Ladder\altaffilmark{1}}

\author{Adam G. Riess\altaffilmark{2,3}, Lucas Macri\altaffilmark{4},
Stefano Casertano\altaffilmark{3}, Megan Sosey\altaffilmark{3}, Hubert Lampeitl\altaffilmark{3,9},
Henry C. Ferguson\altaffilmark{3}, Alexei V. Filippenko\altaffilmark{5},
Saurabh W. Jha\altaffilmark{6}, Weidong Li\altaffilmark{5}, 
Ryan Chornock\altaffilmark{5}, and Devdeep Sarkar\altaffilmark{8}}

\altaffiltext{1}{Based on observations with the NASA/ESA {\it Hubble Space
Telescope}, obtained at the Space Telescope Science Institute, which is
operated by AURA, Inc., under NASA contract NAS 5-26555.}
\altaffiltext{2}{Department of Physics and Astronomy, Johns Hopkins
University, Baltimore, MD 21218.}
\altaffiltext{3}{Space Telescope Science Institute, 3700 San Martin
Drive, Baltimore, MD 21218; ariess@stsci.edu .}
\altaffiltext{4}{George P. and Cynthia W. Mitchell Institute for Fundamental Physics and Astronomy, Department of Physics,
Department of Physics, Texas A\&M University, 4242
TAMU, College Station, TX 77843-4242.}
\altaffiltext{5}{Department of Astronomy, University of California,
Berkeley, CA 94720-3411.}
\altaffiltext{6}{Department of Physics and Astronomy, Rutgers University,
136 Frelinghuysen Road, Piscataway, NJ 08854.}
\altaffiltext{7}{Smithsonian Astrophysical Observatory, Cambridge, MA}
\altaffiltext{8}{UC Irvine}
\altaffiltext{9}{ Institute of Cosmology and Gravitation, University of Portsmouth, Portsmouth, PO1 3FX, UK}










\begin{abstract} 

This is the second of two papers reporting results from a program to
determine the Hubble constant to $\sim$ 5\% precision from a
refurbished distance ladder based on extensive use of differential measurements.  Here we report observations of 240
Cepheid variables obtained with the Near Infrared Camera and
Multi-Object Spectrometer (NICMOS) Camera 2 through the $F160W$ filter
on the {\it Hubble Space Telescope (HST)}.  The Cepheids are
distributed across six recent hosts of Type Ia supernovae (SNe~Ia) and
the ``maser galaxy'' NGC 4258, allowing us to {\it directly} calibrate
the peak luminosities of the SNe~Ia from the precise, geometric
distance measurements provided by the masers.  
New features of our measurement include the use of the same instrument for all Cepheid measurements across the distance ladder and homogeneity of the Cepheid periods and metallicities thus necessitating only a {\it differential} measurement of Cepheid fluxes and reducing the largest systematic uncertainties in the determination of
the fiducial SN Ia luminosity.  In addition, the NICMOS measurements
reduce the effects of differential extinction in the host galaxies by
a factor of $\sim$5 over past optical data. Combined with a greatly expanded
of 240 SNe~Ia at $z<0.1$ which define their magnitude-redshift relation,
we find $H_0=$\ho, a 4.8\%
uncertainty including both statistical and systematic errors.  To
independently test the maser calibration, we use the ten individual
parallax measurements of Galactic Cepheids obtained with the {\it HST} Fine Guidance Sensor and
find similar results.  We show that the factor of 2.2 improvement in
the precision of $H_0$ is a significant aid to the determination of
the equation-of-state parameter of dark energy, $w = P/(\rho
c^2)$. Combined with the Wilkinson Microwave Anisotropy Probe
5-year measurement of $\Omega_M h^2$, we find $w= -1.12 \pm 0.12$ independent of any information from high-redshift SNe~Ia or baryon acoustic
oscillations (BAO).  This result is also consistent with analyses based on the combination of  high-redshift SNe~Ia and BAO.  The
constraints on $w(z)$ now including high-redshift SNe~Ia and BAO are
consistent with a cosmological constant and are improved by a factor
of 3 due to the refinement in $H_0$ alone.  We show that future
improvements in the measurement of $H_0$ are likely and should
further contribute to multi-technique studies of dark energy.

\end{abstract} 

\keywords{galaxies: distances and redshifts --- cosmology:
observations --- cosmology: distance scale --- supernovae: general}

\section{Introduction} 

The {\it Hubble Space Telescope (HST)} established a cornerstone in
the foundations of cosmology by observing Cepheid variables beyond the
Local Group, leading to a measurement of the Hubble constant ($H_0$)
with 10\%--15\% precision \citep{freedman01,sandage06}.  This
measurement resolved decades of extreme uncertainty about the scale
and age of the Universe.  The discovery of cosmic acceleration
and the dark energy that drives it (Riess et al. 1998; Perlmutter et
al.  1999; see Friemann, Huterer, \& Turner 2008 and Filippenko 2005 for reviews) has intensified the need for ever-higher-precision measurements of $H_0$ to constrain
and test the new cosmological models.  Observations are essential to
determine, empirically, aspects of the new model including its geometry,
age, mass density, and the dark energy equation-of-state parameter, $w
= P/(\rho c^2)$, where $P$ is its pressure and $\rho$ its energy
density.  Perhaps the most fundamental question is whether dark energy
is a static, cosmological constant or a dynamical, inflation-like
scalar field --- or whether it can be accommodated at all within the
framework of General Relativity.

While measurements of the high-redshift Universe from the cosmic
microwave background (CMB), baryon acoustic oscillations (BAO), and
Type Ia supernovae (SNe~Ia) in concert with a fully parameterized
cosmological model can be used to {\it predict} the Hubble constant
\citep[e.g.,][]{spergel07,komatsu08}, they are not a substitute for
its {\it measurement} in the local Universe.  Using all of these
measures and the {\it assumptions} that $w=-1$ and that space is flat,
a predicted precision of 2\% in the Hubble constant may be inferred (see Table 1 for $\Lambda$CDM)
\citep{komatsu08}. However, significant tension (at the 3$\sigma$
level) exists in the value of $H_0$ predicted from CMB+BAO and CMB+high-$z$ SNe Ia when the other cosmological parameters such as curvature and $w$ are constrained only by data
(see Table 1; OWCDM).  This suggests that something interesting about the model
or the measurements would be learned from an independent determination
of $H_0$ of comparable precision.



\begin{table}[th]
\begin{center}
\def\d#1 #2 #3 {{$ #1 ^{+#2} _{-#3} $}}
\def\s#1 #2 {{$ #1 \pm #2 $}}
\begin{tabular}{lcc}
\multicolumn{3}{c}{Table 1: $H_0$ Inferred from 5-year WMAP Combined with the
Most Constraining Data} \\
\hline
\hline
Data set               & $\Lambda$CDM  & OWCDM  \\
                       & $(i.e., \Omega_K \equiv 0, w \equiv -1)$ & $(i.e., \Omega_K = {\rm free}, w = {\rm free})$ \\
\hline
WMAP5                  & \d 71.9 2.6 2.7 & \d 47 14 12     \\
WMAP5 + BAO            & \s 70.9 1.5     & \d 81.7 6.5 6.4 \\
WMAP5 + high-$z$ SNe       & \s 69.6 1.7     & \s 57.5 4.8     \\
WMAP5 + BAO + high-$z$ SNe & \s 70.1 1.3     & \d 68.7 1.9 2.0 \\
\hline
\end{tabular}
\end{center}
\vspace {-5pt}
\noindent 
Note: The constraints on $H_0$ are based on
the WMAP team's analysis of the 5-year WMAP data combined with other
data sets \citep{komatsu08}, as listed; see {\tt
http://lambda.gsfc. nasa.gov/product/map/dr3/parameters.cfm}  High-$z$ SNe refers to measurements of the magnitude$-z$ relation of SNe without reference to their distance scale.
\end{table}

Increasing the precision of the measurement of the Hubble constant
requires reducing systematic uncertainties which dominate the error budget along the conventional
distance ladder \citep{freedman01,leonard03}.  As the Hubble diagram of SNe Ia establishes
the {\it relative} expansion rate to an unprecedented uncertainty of
$<$1\% (e.g., Hicken et al. 2009) the calibration of the luminosity of SNe Ia affords the greatest potential for precision in measuring $H_0$.  As we show in \S 4, the largest sources
of systematic error along this route come from the use of uncertain transformations to meld
heterogeneous samples of Cepheids observed with different photometric systems in the anchor
galaxy and SN~Ia hosts.

\subsection{The SHOES Program}

The goal of the SHOES Program (Supernovae and $H_0$ for the Equation
of State) ({\it HST} Cycle 15, GO-10802) is to measure $H_0$
to $<$5\% precision by mitigating the dominant systematic
errors.\footnote{The {\it HST} observations were also designed to find
SNe~Ia at $z>1$ with coordinated ACS parallel observations. Two high-$z$
SNe were found before the failure of ACS on 2007 Feb. 1.}
To obviate the limited accuracy of photographic SN data, 
we have been calibrating recent SNe~Ia recorded with modern detectors
and acquiring uniform samples of Cepheids observed in the SN~Ia hosts
and in the anchor galaxy.  Progress in the former was presented by
\citet{riess05,riess09}, more than doubling the sample of high-quality calibrators by providing
reliable calibration for four modern SNe~Ia. Here we address the
latter, reporting the results of infrared (IR) observations of
Cepheids which are homogeneous in their periods, metallicities, and
measurements in both the anchor (NGC 4258) and the SN hosts.

NGC 4258 offers attractive benefits over the use of the Large
Magellanic Cloud (LMC) or the Milky Way Galaxy as an anchor of the
distance ladder: (1) all of its Cepheids can be treated as being at a
{\it single distance} determined {\it geometrically} from the Keplerian motion of its masers as 7.2 $\pm0.5$ Mpc (Herrnstein et al. 1999); (2) more than a decade of tracking its masers has resulted in little change to its distance while steadily increasing its {\it precision} from 7\% (Herrnstein et al. 1999) to 5.5\% (Humphreys et al. 2005), to
3\% (Humphreys et al. 2008, 2009, Greenhill et al. 2009) (3) the geometric distance measurement can be internally  cross-checked via proper motion and centripetal acceleration and the method can be externally tested by measurements of other maser systems (Braatz et al. 2008, Greenhill et al. 2009); (4) its Cepheids have a
metallicity similar to those found in the hosts of SNe~Ia (Riess et al. 2009); (5) {\it
HST} observations of NGC 4258 from Cycles 12, 13, and 15 provide the
largest sample of long-period ($P > 10$~d) Cepheids
\citep{macri06,macri09}; and (6) its Cepheids can be observed with
{\it HST} in exactly the same manner as those in SN~Ia hosts.  In \S
4, we independently test the use of the distance to NGC 4258 by
adopting the individual parallax measurements of Galactic Cepheids
from \citet{benedict07}.

The IR observations of Cepheids presented in \S 3 provide additional
advantages over those in the optical: (1) reducing the differential
extinction by a factor of five over visual data, and (2) reducing the
dependence of Cepheid magnitudes on chemical composition
\citep{marconi05}. The resulting, refurbished distance ladder builds
on past work while removing four of the largest systematic sources of
uncertainty in $H_0$. In \S 3 and \S 4 we show that the total uncertainty in
the measurement of $H_0$ has been reduced from 11\% \citep{freedman01} to \unc .

\section{NICMOS Cepheid Observations of the SHOES Program}

In \citet{riess09}, we used {\it HST}/ACS+WFPC2 observations to
discover Cepheids in two new SN~Ia hosts and to expand previous
samples in four other SN~Ia hosts with newly discovered, longer-period
($P > 60$~d) variables. In \citet{macri09} we used {\it HST}/ACS+WFPC2
observations to augment the Cepheid sample in NGC 4258. These new
observations, together with those from \citet{saha96,saha97,saha01},
\citet{gibson00}, \citet{stetson01}, and \citet{macri06} provide the
position, period, and phase of 450 Cepheids in 6 hosts with reliable SN Ia data and NGC 4258, each with typically 14 epochs of
{\it HST} imaging with $F555W$ and 1--5 epochs with $F814W$ (except
for NGC 4258, which has 12 epochs of $F814W$ data).

The Near Infrared Camera and Multi-Object
Spectrometer (NICMOS) on {\it HST} provides the means to obtain near-IR measurements
of optically identified Cepheids.  
\citet{macri01} used short exposures ($\sim$ 1 ksec) with NICMOS to measure 70 extragalactic Cepheids in 14 galaxies at an average distance of 5 Mpc (including 2
Cepheids in NGC 4536) to verify the Galactic extinction
law.  

In {\it HST} Cycle 15 the SHOES program obtained deep (10 to 35 ksec),
near-IR observations of the Cepheids in these SN~Ia hosts. The
SN~Ia in each host was chosen for meeting the following criteria: (1) has modern data (i.e.,
photoelectric or CCD), (2) was observed before maximum brightness, (3)
has low reddening, (4) is spectroscopically typical, and (5) has
optical {\it HST}-based observations of Cepheids in its host. The
resulting sample consists of 6 SN Ia hosts given in Table 2. The six
members and their SNe are shown in Figure 1.  In Cycle 15 we also obtained 2 ksec NICMOS imaging of Cepheids in NGC 4258 to augment that obtained
in Cycle 13 by GO 10399 (P.I. Greenhill).

\begin{table}[h]
\tablenum{2}
\begin{small}
\begin{center}
\vspace{0.4cm}
\begin{tabular}{cccccc}
\multicolumn{5}{c}{Table 2: Cepheid Hosts Observed by SHOES} \\
\hline
\hline
Host & SN Ia &  Initial optical & HST Cycle & Reobservation, Cycle 15  & Observation near-IR, Cycle 15 \\
\hline
NGC 4536 & SN 1981B     &  WFPC2 & 4 & WFPC2     & NIC2 \\
NGC 4639 & SN 1990N     &  WFPC2 & 5 & ACS       & NIC2 \\
NGC 3982 & SN 1998aq    &  WFPC2 & 8 & ACS       & NIC2 \\
NGC 3370 & SN 1994ae    &  ACS  & 11 & ACS       & NIC2 \\
NGC 3021 & SN 1995al    &  ACS   & 14 & ACS       & NIC2 \\
NGC 1309 & SN 2002fk    &  ACS   & 14 & ACS       & NIC2 \\
NGC 4258 & ------------ &  ACS  & 12 & ACS/WFPC2 & NIC2$^a$ \\
\hline
\hline
 \multicolumn{6}{l}{$^a$Some NIC2 data obtained in Cycle 13} \\
\end{tabular}
\end{center}
\end{small} 
\end{table}

\subsection{NICMOS Data Reduction}

Groupings of optically-characterized Cepheids were observed using
the NICMOS Camera 2 and the $F160W$ filter.  This camera offers the
best compromise of area and sampling of the point-spread function (PSF)
of the three NICMOS cameras.  For each SN host galaxy we selected four to
five 0.1 sq arcmin pointings of 3--12 orbit depth (10--35 ks) to
contain multiple previously identified long-period Cepheids.  The
pointing centers and total integration times are given in Table 3. The
imaging configurations are shown in Figures 2 to 7.  The observations
were obtained in single-orbit visits spread over $\sim$ 2 months.

\begin{table}[h]
\tablenum{3}
\begin{small}
\begin{center}
\vspace{0.4cm}
\begin{tabular}{llll}
\multicolumn{4}{c}{Table 3: NIC2 $F160W$ Cepheid Observations} \\
\hline
\hline
Target & $\alpha$(J2000) & $\delta$(J2000) & Exp time (s) \\
\hline
NGC1309-BLUE &        3$^h$   22$^m$   3.499$^s$ &  $-15^\circ$  $24'$  $4.242''$ &       10,239.5 \\
NGC1309-YELLOW &        3     22      7.040 &      $-15$      24      26.800 &       23,031.0 \\
NGC1309-CYAN2 &        3      22      8.790 &      $-15$      24      46.148 &       34,174.3 \\
NGC1309-GREEN &        3      22      9.862 &      $-15$      23      48.548 &       10,239.5 \\
NGC3021-BLUE &        9      50      55.493 &       33      32      48.598 &       10,239.5 \\
NGC3021-CYAN &        9      50      54.079 &       33      33      28.190 &       10,239.5 \\
NGC3021-GREEN &        9      50      59.984 &       33      33      5.202 &       20,478.9 \\
NGC3021-RED &        9     50     57.330 &       33      33      36.883 &       20,158.9 \\
NGC3370-BLUE &       10      47      4.334 &       17      17      3.726 &       20,478.9 \\
NGC3370-CYAN &       10      47      6.432 &       17      15      26.429 &       10,239.5 \\
NGC3370-GREEN &       10      47      1.673 &       17      16      56.221 &       26,238.7 \\
NGC3370-RED &       10     47      8.354 &       17      15      48.264 &       10,239.5 \\
NGC3982-BLUE &       11      56      24.458 &       55       7      13.644 &       17,919.0 \\
NGC3982-CYAN-COPY &       11      56      32.368 &       55    7      30.329 &       7,679.6 \\
NGC3982-GREEN &       11      56      21.375 &       55      7      23.696 &       7,679.6 \\
NGC3982-RED-COPY &       11      56      23.150 &       55     6      38.776 &       17,919.0 \\
NGC3982-YELLOW &       11      56      30.253 &       55      7      54.595 &       10,239.5 \\
NGC4536-BLUE &       12     34      18.360 &        2      11      35.570 &       17,919.0 \\
NGC4536-CYAN &       12      34      21.395 &        2      13      10.230 &       12,799.3 \\
NGC4536-GREEN &       12      34      17.313 &        2      13      6.429 &       17,919.0 \\
NGC4536-RED &       12      34     21.371 &        2      12      6.289 &       17,919.0 \\
NGC4639-BLUE &       12      42      53.114 &       13      14      54.342 &       10,239.5 \\
NGC4639-BLUE-LATE$^a$ &       12      42      52.946 &       13      14      55.207 &       10,239.5 \\
NGC4639-CYAN &       12      42      49.300 &       13      16      9.272 &       19,966.9 \\
NGC4639-GREEN &       12      42      52.163 &       13      16      30.699 &       20,478.9 \\
NGC4639-RED &       12      42     56.438 &       13      15      20.369 &       10,239.5 \\
NGC4258-NIC-POS4	&12 18 50.76	&47 19 24.2	&2559.8 \\
NGC4258-NIC-POS5	&12 18 57.92	&47 20 35.5	&2559.8 \\
NGC4258-NIC-POS3	&12 18 47.52	&47 20 05.1	&2559.8 \\
NGC4258-NIC-POS2	&12 18 50.09	&47 20 43.1	&2559.8 \\
NGC4258-NIC-POS1	&12 18 50.77	 &47 21 09.3	&2559.8 \\
NGC4258-NIC-POS13	&12 18 54.90	&47 21 47.9	&2559.8 \\
NGC4258-NIC-POS11	&12 19 20.28	&47 14 54.0	&2559.8 \\
NGC4258-NIC-POS10&	12 19 22.72	&47 14 44.6	&2559.8 \\
NGC4258-NIC-POS9	&12 19 08.93	&47 12 25.2	&2559.8 \\
NGC4258-NIC-POS8	&12 19 12.02	&47 12 21.1	&2559.8 \\
NGC4258-NIC-POS12&	12 19 25.32	&47 13 44.2	&2559.8 \\
NGC4258-NIC-POS6	&12 19 21.03	&47 10 21.5	&2559.8 \\
NGC4258-NIC-POS7	&12 19 25.39	&47 09 41.2	&2559.8 \\
NGC4258-INNER-NIC-09	&12 18 53.21	&47 18 43.5	&2559.8 \\
NGC4258-INNER-NIC-10	&12 18 54.50	&47 19 00.9	&2559.8 \\
NGC4258-INNER-NIC-08	&12 18 51.38	&47 18 42.1	&2559.8 \\
NGC4258-INNER-NIC-12	&12 18 54.72	&47 19 16.5	&2559.8 \\
NGC4258-INNER-NIC-05	&12 18 48.98	&47 19 13.2	&2559.8 \\
NGC4258-INNER-NIC-13	&12 18 55.88	&47 20 17.8	&2559.8 \\
NGC4258-INNER-NIC-06	&12 18 48.99	&47 19 47.4	&2559.8 \\
NGC4258-INNER-NIC-04	&12 18 48.21	&47 20 10.1	&2559.8 \\
NGC4258-INNER-NIC-11	&12 18 54.55	&47 20 41.7	&2559.8 \\
NGC4258-INNER-NIC-01	&12 18 45.28	&47 20 02.2	&2559.8 \\
NGC4258-INNER-NIC-07	&12 18 50.33	&47 21 06.3	&2559.8 \\
NGC4258-INNER-NIC-02	&12 18 47.64	&47 20 58.2	&2559.8 \\
NGC4258-INNER-NIC-03	&12 18 48.07	&47 21 19.7	&2559.8 \\
NGC4258-OUTER-NIC-04&	12 19 20.62	&47 13 12.4	&2559.8 \\
NGC4258-OUTER-NIC-02&	12 19 15.93	&47 12 00.0	&2559.8 \\
NGC4258-OUTER-NIC-07&	12 19 21.22	&47 11 41.4	&2559.8 \\
NGC4258-OUTER-NIC-05&	12 19 20.18	&47 11 25.6	&2559.8 \\
NGC4258-OUTER-NIC-10&	12 19 26.03	&47 12 02.2	&2559.8 \\
NGC4258-OUTER-NIC-08&	12 19 24.35	&47 11 34.2	&2559.8 \\
NGC4258-OUTER-NIC-13&	12 19 20.73	&47 10 54.6	&2559.8 \\
NGC4258-OUTER-NIC-06&	12 19 20.95	&47 10 20.1	&2559.8 \\
NGC4258-OUTER-NIC-09&	12 19 24.82	&47 10 14.0	&2559.8 \\
\hline
 \hline
 \multicolumn{4}{l}{$^a$Same region as NGC4639-BLUE with different orientation.} \\
\end{tabular}
\end{center}
\end{small}
\end{table}

We developed an automated pipeline to calibrate the raw NICMOS frames.
The first step subtracted one of two ``superdarks'' produced from archival data
obtained after the installation of the NICMOS Cooling System, corresponding to the closest temperature state (of two
primary temperature regimes) at which the data were obtained.  Next,
the data were processed through the STScI-supported CALNICA pipeline
with the following additions. The STSDAS routine BIASEQ
\citep{bushouse00} was used after the corrections for bias, dark
counts, and linearity to account for stochastic changes in quadrant
bias level.  After flat-fielding, cosmic-ray rejection, and count-rate
conversion, the images were corrected for the count-rate nonlinearity
as calibrated by \citet{dejong06}.  The remaining quadrant-dependent
linear DC bias was fit and removed using the PEDSUB task.  Any data
obtained soon after a passage by {\it HST} through the South Atlantic
Anomaly (SAA) were corrected for the persistence of cosmic rays using
a post-SAA dark frame and the routine SAACLEAN \citep{bergeron03}.

Approximately 10\% of our images were contaminated by charge
persistence after the detector was exposed to the bright limb of the
Earth in the preceding orbit.  The structure of the persistence image
is time independent and is a map of the density of charge
traps saturated by the Earth light.  A persistence image was produced
from the data which was then scaled and subtracted from the affected
data as described by \citet{riess08}.

Residual amplifier glow and its persistence were removed by
subtracting a model of the sky image from the combination of all
exposures in a visit.  The model was smoothed with a ring filter
(larger in diameter than the PSF) to ensure that stellar sources in
the data were not present in this sky model.

Next we combined the exposures from each visit to produce a full image
combination for each pointing listed in Table 3. We first registered
the exposures within a visit using the dither positions indicated in
the image headers.  To register images between visits we used between
30 and 100 bright sources to empirically measure the shifts and
rotations between visits (we also verified that scale variations
between orbits were negligible).  The typical root-mean square (rms)
deviation of sources between our visit-to-visit registration solutions
was 0.2-0.3 pixels, yielding an error in the mean of less than 0.05
pixels.  The final image combination was resampled on a pixel scale of
$0.038''$ using the drizzle algorithm \citep{fruchter02}.

Because the PSF of NIC2 with $F160W$ is well sampled, our image
combination should cause little broadening of the PSF.  To test this,
we measured the difference in the photometry of non-variable
supergiants in single epochs and in the full image
combinations. The median difference was $\sim$0.003 mag (in the sense of the combination being brighter, opposite the expected direction if the effect were real) and
consistent with zero to within the statistical uncertainty.   Thus we concluded there was no loss in accuracy of the photometry obtained by combining images 
from individual visits.

To identify the precise positions of Cepheids in the NICMOS image
combinations, we derived the geometric transformation from the {\it
HST} $F814W$ images to the $F160W$ images, iteratively matching bright
to faint sources to find sources in common.  This registration
empirically determined the difference in plate scale between ACS,
WFPC2, and NIC2.  Typically we identified more than 100 sources in
common, resulting in an uncertainty in the mean Cepheid position of
$<$0.03 pixels (1 milliarcsec).

\subsection{NICMOS Cepheid Photometry}

We developed software to measure Cepheid photometry in crowded NICMOS
images based on the procedures established for {\it HST} optical
photometry \citep{stetson94,saha96}. Since we know {\it a priori} the
precise position of the Cepheids in our NICMOS data, we can fix the
positions in the NICMOS images to mitigate the measurement bias which
can arise naturally for flux measurements of sources made from the same data used for their discovery
\citep{hogg98}.

We derived a model of the PSF in our NICMOS images using observations
of the bright solar analogue, P330E, averaged over several visits and
processed in the same way our host images.  P330E provides a
fundamental standard for the NICMOS Vega magnitude zeropoint
($F160W=11.45$ mag, Vega system) and our natural system magnitudes are
measured relative to this zeropoint.  However,  the difference between the photometry of Cepheids in NGC 4258 and the SN hosts, employed to measure $H_0$ in \S 3, are {\it independent}  of the adopted zeropoint.

For each known Cepheid, we produced a list of neighboring stars in the NICMOS images within
its ``critical" radius (4 $\times$ FWHM, where FWHM is the full width at
half-maximum intensity) or within that of one of its neighbors.  Together,
these stars and the Cepheid define a ``crowded group'' whose members
must be modeled together.  Initially, we subtracted a PSF model at the
location of the Cepheid (as determined from the optical data)  and
then used the algorithm {\it DAOFIND} to identify neighboring stars
within the critical radius but at least 0.75 $\times$ FWHM beyond each
Cepheid.  Stellar sources within and beyond the group were modeled and
subtracted, and the background level for the group was determined from
the mode of the pixels in an annulus around the Cepheid with an inner
radius of 15 pixels and an outer radius of 20 pixels.

We then used a Levenberg-Maquardt-based algorithm to find the most
likely values and uncertainties of the group parameters by minimizing
the $\chi^2$ statistic between the image and model pixels within the
critical radii of the modeled sources.  For all non-Cepheid sources,
their positions were allowed to vary within 0.5 pixels of their
original detected position and the amplitudes were allowed to vary.
For the Cepheids, only the amplitude of the PSF was varied.  The
Cepheid position determined from the optical images was fixed, as was
the group sky level. Our typical group had 5 to 15 unresolved, modeled
sources besides the Cepheids producing 3 times this number of free
parameters plus one additional parameter for the Cepheid brightness.
The individual pixel noise was relatively uniform, resulting from a
combination of sky, dark current and read noise.

After identifying the optimal solution, we subtracted the model from
the data and inspected the residuals to determine the best set of
global photometry parameters for all images.  In Figure 8 we show
as an example the image, model, and residuals of the groups for one of
the richest NICMOS pointings, NGC3370-GREEN, with 14 Cepheids over a wide range
of periods.

\subsection{Sky Determination and Bias Correction}

The surroundings of the Cepheids are mottled, with
unresolved sources and surface brightness fluctuations whose fluxes
are generally fainter but occasionally brighter than our target
Cepheids (see Fig. 10, NGC 4258 and NGC 3370 GREEN field).  These
scenes pose a challenge to estimating the correct background level for
the Cepheids.  Simply measuring the mean flux in an annulus centered on the
Cepheid would provide an unbiased but very noisy estimate of the
background.

Instead, we follow the conventional approach of determining the sky
level from the sky annulus after first subtracting models of the stellar sources within it.  Because we would expect a similar number of background sources coincident with (yet inseparable from) the Cepheid, we would naturally
{\it underestimate} the sky level for the Cepheid. Though this bias is
ameliorated somewhat by the use of the mode statistic from the
residual image as discussed by Stetson (1987), a bias still
remains. 

In previous work it has been shown that this photometric bias in
optically-selected Cepheid samples is reduced by the act of selecting
Cepheids with strong amplitudes and statistically significant
variations in flux \citep{ferrarese00}.  The addition of significant,
blended flux would reduce the amplitude of the Cepheid, increase the
model uncertainty and reduce the significance of true variations.
However, this mechanism does not apply to the NICMOS images as they
were not used to select Cepheids.  Indeed, Cepheids are bluer than a
common source of blending, red giants, so the blending bias in the
NICMOS data can be significant. \citet{macri01} found this photometric
bias to vary from negligible to 0.1 mag for the Cepheids discovered
with WFPC2 and reobserved with NICMOS, and measured the impact to artificial stars 
injected in the vicinity of each Cepheid to correct for this
effect. We adopt the same approach here.

In addition, we can improve our estimate of the blending bias.   On average, the displacement of a Cepheid's centroid in the NICMOS data 
relative to its optically 
determined position correlates with the degree of blending in the NICMOS data.  For randomly located sources of blending,
brighter blended sources cause larger Cepheid displacements 
and bias.  For artificial stars rediscovered within $\sim$ 0.1 pixels of
their injected position we find $<0.1$ mag of blending bias.  The
photometric bias grows linearly with the displacement of the centroid,
rising to $\sim$ 0.3 mag for a full pixel (0.038$\arcsec$) displacement.
Beyond a pixel, the
recovered star is often not the same as the one injected (as occurs
when the injected star is too faint to be found), and any relation between the
displacement and bias dissipates.  Rarely, a Cepheid  will be {\it exactly} coincident with a bright 
source causing it to be an outlier in the \PL relation.  Such complete blends are later
eliminated from both our sample (and the artificial star simulations)
with a 2.5$\sigma$ rejection from the mean, a threshold based on
Chauvenet's criterion (i.e., less than half a Cepheid would be
expected to exceed the outlier limit for a Gaussian distribution of
residuals).

To determine the individual photometric correction for each Cepheid we
added 1000 artificial stars at random positions within a radius
of $0.05''$ to $0.75''$ from each Cepheid.  The magnitudes of the
artificial stars were given by the Cepheid magnitude {\it predicted}
by its period using an initial fit (i.e., uncorrected) to the
period-magnitude relations.  After correcting the Cepheid magnitudes
for the measured bias, these relations were refit and the process was
repeated until convergence.  The dispersion of the artificial stars
was used to estimate the uncertainty in the magnitude of the Cepheid
by adding this term in quadrature to the Cepheid measurement
uncertainty. An example of this artificial-star analysis is shown in
Figure 9 for a Cepheid in our second-most-distant galaxy,
NGC 3021. 

For NGC 4258, our anchor galaxy, the median bias correction was 0.14
mag ($\pm 0.014$) and for the SN hosts the median was 0.16 ($\pm 0.021$) mag.  Thus we conclude
that {\it the photometric corrections for the Cepheids in NGC 4258 and
the SN hosts are extremely similar}.  This is not surprising as the apparent stellar density of the fields is also quite similar as seen in Figure 10.
Although NGC 4258 is closer than the SN hosts, reducing
its relative crowding, the NGC 4258 inner fields \citep{macri06} are
closer to the nucleus where the true stellar density is greater. 

Because the luminosity calibration of SNe~Ia
depends on only {\it the difference} in the magnitudes of the Cepheids in the
anchor galaxy and the SN hosts, the net effect of blending even without correction is quite small: $\sim$0.02 mag, or about 1\% in the
distance scale.  However, our corrections account for this small difference as
further addressed in \S 4.2.

Artificial-star simulations cannot account for blending which is local to the Cepheids
(i.e., binarity or cluster companions).  However, we expect little net
effect from such blending (after the removal of outliers) as such blending is likely to occur with similar frequency in the anchor and SN hosts and thus would largely cancel in their difference.

Because the amplitudes of IR light curves are $< 0.3$ mag, 
even magnitudes measured at random phases provide comparable precision to the mean flux 
for determining the period-luminosity (\PL) relation \citep{madore91}.
As our exposures were obtained over $\sim 2$ months the magnitude measured from the mean image will have
a dispersion of $< 0.08$ mag around the mean flux.  
To account for this error, we correct the measured
magnitude to the mean-phase
magnitude using the Cepheid phase, period, and amplitude from the
optical data, the dates of the NICMOS observations, and the Fourier
components of \citet{soszynski05} which quantify the relations between Cepheid light curves in the optical and near-IR. These phase corrections were found to be insignificant in the subsequent analysis.

Table 4 contains the aforementioned parameters for each Cepheid.  The Cepheid's NIC2 field, position, identification number (from Riess et al. 2009 and Macri et al. 2006), period, mean $V-I$ color, $F160W$ mag and its uncertainty are given in the first 8 columns.  Column 9 contains the displacement of the Cepheid in the NICMOS data from optical position in pixels of 0.038$\arcsec$ size.  Column 10 gives the photometric bias determined from the artificial star tests for the Cepheid's environment and displacement and are already added to determine column 7.  Column 11 contains the correction from the sampled phase to the mean and has already been subtracted to determine column 7.  Column 12 contains the metallicity parameter, 12+log[O/H], inferred at the position of each Cepheid.  Column 13 and 14 contains the rejection flag employed and the source of the Cepheid detection as noted in Table 4, respectively.

\tabletypesize{\scriptsize}
\tablewidth{0pt}
\begin{deluxetable}{llllllllllllll}
\tablenum{4}
\tablecaption{NICMOS Cepheids}
\tablehead{\colhead{Field}&\colhead{$\alpha$}&\colhead{$\delta$}&\colhead{Id}&\colhead{P}&\colhead{$V-I$}&\colhead{$F160W$}&\colhead{$\sigma$}&\colhead{Offset} &\colhead{Bias}& \colhead{Phase}&\colhead{[O/H]}&\colhead{Flag$^a$}&\colhead{Src$^*$} \\ 
\colhead{ }&\colhead{(J2000)}&\colhead{(J2000)}&\colhead{ }&\colhead{(days)}&\colhead{(mag)}&\colhead{(mag)}&\colhead{(mag)}&\colhead{(pix)} &\colhead{(mag)}& \colhead{(mag)}&\colhead{ }&\colhead{ }&\colhead{ }}
\startdata
N4536-B &        188.57747 &        2.193860 &       9491 &  19.79  &  1.10  &  24.20  &  0.24  &  0.52  &  0.06  &  0.06  &  8.69  &   & psl \nl
N4536-B &        188.57760 &        2.192130 &       9935 &  42.81  &  1.05  &  23.59  &  0.19  &  0.35  &  0.04  &  -0.09  &  8.68  &   & psl \nl
N4536-B &        188.57383 &        2.194050 &       12075 &  49.70  &  1.17  &  23.26  &  0.22  &  0.48  &  0.15  &  0.02  &  8.64  &   & psl \nl
N4536-C &        188.58908 &        2.221220 &       3633 &  23.37  &  0.82  &  23.91  &  0.20  &  0.57  &  0.06  &  0.14  &  8.75  &   & psl \nl
N4536-G &        188.57352 &        2.219870 &       6827 &  24.54  &  0.74  &  23.86  &  0.31  &  0.70  &  0.28  &  0.19  &  8.71  &   & psl \nl
N4536-G &        188.57237 &        2.216330 &       5630 &  55.24  &  1.03  &  22.55  &  0.19  &  0.10  &  0.05  &  -0.09  &  8.72  &   & psl \nl
N4536-R &        188.58741 &        2.202540 &       3571 &  20.98  &  0.78  &  24.02  &  0.39  &  0.95  &  0.29  &  0.07  &  8.89  &   & psl \nl
N4536-R &        188.58842 &        2.204110 &       2692 &  31.59  &  1.09  &  23.71  &  0.20  &  0.57  &  0.20  &  -0.04  &  8.90  &   & psl \nl
N4536-R &        188.59047 &        2.200050 &       6914 &  34.63  &  1.22  &  22.72  &  0.22  &  0.37  &  0.13  &  -0.03  &  8.92  & rej & psl \nl
N4536-R &        188.59103 &        2.202740 &       4786 &  37.45  &  0.93  &  23.67  &  0.22  &  0.46  &  0.13  &  -0.03  &  8.93  &   & psl \nl
N4536-B &        188.57908 &        2.191510 &       9023 &  19.90  &  0.51  &  24.40  &  0.30  &  0.72  &  0.22  &  0.04  &  8.69  &   & lm \nl
N4536-C &        188.58946 &        2.220100 &       3607 &  80.17  &  0.69  &  22.79  &  0.25  &  0.34  &  0.03  &  0.01  &  8.76  &   & lm \nl
N4536-G &        188.57149 &        2.218370 &       6582 &  38.84  &  1.43  &  23.57  &  0.20  &  0.24  &  0.05  &  0.00  &  8.70  &   & lm \nl
N4536-R &        188.58811 &        2.199800 &       6233 &  25.30  &  0.64  &  23.31  &  0.23  &  0.65  &  0.25  &  0.09  &  8.89  &   & lm \nl
N4536-B &       188.57932 &       2.199800 &       13 &  18.54  &  0.72  &  24.56  &  0.38  &  0.60  &  0.16  &  0.00  &  8.71  &   & ps \nl
N4536-R &       188.58854 &       2.192987 &       5 &  30.25  &  1.14  &  23.64  &  0.22  &  0.52  &  0.15  &  0.00  &  8.89  &   & ps \nl
N4639-B &        190.72103 &        13.24735 &       11893 &  26.59  &  1.09  &  25.83  &  0.47  &  1.15  &  0.29  &  0.06  &  8.98  & rej & psl \nl
N4639-B &        190.72412 &        13.24921 &       8651 &  27.54  &  0.91  &  24.58  &  0.54  &  0.74  &  0.44  &  0.08  &  9.05  &   & psl \nl
N4639-B &        190.71891 &        13.25026 &       16601 &  42.20  &  1.18  &  23.54  &  0.56  &  0.54  &  0.18  &  0.01  &  9.09  &   & psl \nl
N4639-B &        190.72047 &        13.24661 &       12394 &  54.82  &  1.06  &  24.10  &  0.47  &  2.69  &  0.53  &  0.05  &  8.95  &   & psl \nl
N4639-B-L &        190.72412 &        13.24921 &       8651 &  27.54  &  0.91  &  24.68  &  0.47  &  0.36  &  0.59  &  0.11  &  9.05  &   & psl \nl
N4639-B-L &        190.71891 &        13.25026 &       16601 &  42.20  &  1.18  &  23.84  &  0.34  &  0.41  &  0.29  &  -0.02  &  9.09  &   & psl \nl
N4639-B-L &        190.72047 &        13.24661 &       12394 &  54.82  &  1.06  &  23.80  &  0.33  &  8.76  &  0.06  &  -0.03  &  8.95  &   & psl \nl
N4639-C &        190.70392 &        13.26827 &       40321 &  37.27  &  1.24  &  23.89  &  0.25  &  0.20  &  0.11  &  0.02  &  8.83  &   & psl \nl
N4639-C &        190.70509 &        13.26793 &       39829 &  39.41  &  0.91  &  23.79  &  0.22  &  0.24  &  0.07  &  -0.07  &  8.86  &   & psl \nl
N4639-C &        190.70451 &        13.26849 &       40158 &  52.16  &  1.09  &  24.26  &  0.27  &  0.32  &  0.16  &  -0.09  &  8.84  &   & psl \nl
N4639-C &        190.70539 &        13.26932 &       61786 &  56.31  &  1.18  &  24.04  &  0.24  &  0.87  &  0.11  &  0.13  &  8.84  &   & psl \nl
N4639-G &        190.71845 &        13.27400 &       30160 &  51.11  &  1.11  &  24.21  &  0.19  &  0.16  &  0.04  &  -0.00  &  8.69  &   & psl \nl
N4639-R &        190.73302 &        13.25647 &       4481 &  34.24  &  0.97  &  24.08  &  0.19  &  0.08  &  0.16  &  0.00  &  8.88  &   & psl \nl
N4639-B &        190.72165 &        13.25058 &       12430 &  39.53  &  0.93  &  25.23  &  0.67  &  2.33  &  0.70  &  0.01  &  9.11  & rej & lm \nl
N4639-B &        190.72020 &        13.24847 &       13602 &  47.27  &  1.44  &  23.92  &  0.31  &  0.56  &  0.16  &  0.04  &  9.02  &   & lm \nl
N4639-B-L &        190.72020 &        13.24847 &       13602 &  47.27  &  1.44  &  24.06  &  0.30  &  0.50  &  0.03  &  -0.02  &  9.02  &   & lm \nl
N4639-R &        190.73314 &        13.25609 &       4383 &  42.43  &  1.09  &  23.98  &  0.22  &  0.38  &  0.09  &  -0.03  &  8.88  &   & lm \nl
N3370-B &        161.76928 &        17.28204 &       24497 &  16.78  &  0.87  &  25.89  &  0.68  &  2.63  &  0.39  &  -0.05  &  8.86  & low P & lm \nl
N3370-B &        161.76766 &        17.28206 &       21444 &  19.64  &  0.92  &  26.21  &  0.42  &  3.98  &  0.03  &  -0.06  &  8.91  & low P & lm \nl
N3370-B &        161.76554 &        17.28584 &       15081 &  32.56  &  1.22  &  25.56  &  0.39  &  0.98  &  0.34  &  0.00  &  8.80  &   & lm \nl
N3370-B &        161.76923 &        17.28640 &       21445 &  37.10  &  1.07  &  24.67  &  0.20  &  0.47  &  0.14  &  -0.14  &  8.68  &   & lm \nl
N3370-B &        161.76844 &        17.28412 &       21506 &  38.54  &  0.75  &  24.63  &  0.30  &  0.60  &  0.25  &  0.13  &  8.80  &   & lm \nl
N3370-B &        161.76774 &        17.28324 &       20732 &  41.55  &  1.15  &  24.48  &  0.27  &  4.33  &  0.17  &  -0.06  &  8.86  &   & lm \nl
N3370-B &        161.76869 &        17.28313 &       22612 &  69.35  &  1.04  &  23.83  &  0.22  &  0.37  &  0.14  &  0.01  &  8.83  &   & lm \nl
N3370-C &        161.77931 &        17.25660 &       50670 &  20.52  &  0.84  &  25.50  &  0.41  &  2.96  &  0.13  &  0.00  &  8.64  & low P & lm \nl
N3370-C &        161.77441 &        17.25595 &       47059 &  24.49  &  0.93  &  24.95  &  0.34  &  1.10  &  0.31  &  -0.01  &  8.68  &   & lm \nl
N3370-C &        161.77627 &        17.25957 &       47494 &  24.43  &  1.22  &  24.57  &  0.40  &  0.07  &  0.28  &  -0.01  &  8.77  & rej & lm \nl
N3370-C &        161.78015 &        17.25611 &       51334 &  28.79  &  0.98  &  24.88  &  0.32  &  1.31  &  0.25  &  -0.06  &  8.61  &   & lm \nl
N3370-C &        161.77485 &        17.25741 &       46992 &  29.60  &  0.95  &  24.94  &  0.27  &  2.27  &  -0.02  &  -0.03  &  8.72  &   & lm \nl
N3370-C &        161.77587 &        17.25844 &       47492 &  39.41  &  1.15  &  24.84  &  0.22  &  0.48  &  0.08  &  -0.05  &  8.75  &   & lm \nl
N3370-C &        161.77827 &        17.26024 &       48903 &  51.68  &  1.09  &  24.76  &  0.28  &  0.99  &  0.09  &  -0.00  &  8.74  &   & lm \nl
N3370-C &        161.77799 &        17.26004 &       48741 &  96.49  &  0.96  &  23.90  &  0.25  &  0.44  &  0.11  &  -0.01  &  8.75  &   & lm \nl
N3370-R &        161.78414 &        17.26088 &       52428 &  33.48  &  1.04  &  24.95  &  0.32  &  0.74  &  0.15  &  -0.06  &  8.54  &   & lm \nl
N3370-R &        161.78547 &        17.26580 &       52279 &  33.69  &  1.03  &  24.93  &  0.26  &  0.63  &  0.13  &  -0.03  &  8.48  &   & lm \nl
N3370-G &        161.75400 &        17.28417 &       2638 &  17.46  &  0.76  &  26.08  &  0.63  &  6.05  &  0.47  &  0.00  &  8.71  & low P & lm \nl
N3370-G &        161.75882 &        17.28007 &       8807 &  23.72  &  1.04  &  26.46  &  0.41  &  0.87  &  0.61  &  -0.08  &  8.92  & rej & lm \nl
N3370-G &        161.75875 &        17.28389 &       61720 &  25.43  &  0.90  &  26.21  &  0.47  &  1.70  &  0.61  &  -0.02  &  8.86  & rej & lm \nl
N3370-G &        161.75791 &        17.28025 &       62219 &  29.43  &  1.10  &  25.25  &  0.49  &  1.29  &  0.61  &  -0.07  &  8.89  &   & lm \nl
N3370-G &        161.75647 &        17.28052 &       5744 &  27.74  &  1.06  &  25.35  &  0.28  &  5.16  &  0.18  &  -0.09  &  8.83  &   & lm \nl
N3370-G &        161.75620 &        17.28353 &       4345 &  34.07  &  1.21  &  25.85  &  0.31  &  1.05  &  0.37  &  -0.12  &  8.79  & rej & lm \nl
N3370-G &        161.75647 &        17.28320 &       4710 &  32.62  &  1.07  &  25.30  &  0.39  &  0.63  &  0.39  &  0.07  &  8.80  &   & lm \nl
N3370-G &        161.75710 &        17.28387 &       59919 &  36.99  &  0.88  &  25.22  &  0.25  &  3.02  &  0.21  &  -0.04  &  8.81  &   & lm \nl
N3370-G &        161.76072 &        17.28069 &       10677 &  35.24  &  1.11  &  23.49  &  0.36  &  10.5  &  0.29  &  -0.00  &  8.97  & rej & lm \nl
N3370-G &        161.75713 &        17.28309 &       5439 &  45.82  &  0.98  &  23.66  &  0.28  &  0.95  &  0.31  &  0.05  &  8.83  & rej & lm \nl
N3370-G &        161.75761 &        17.28213 &       6440 &  43.94  &  1.17  &  24.37  &  0.28  &  0.32  &  0.31  &  0.08  &  8.86  &   & lm \nl
N3370-G &        161.75980 &        17.28234 &       9014 &  45.10  &  1.12  &  24.60  &  0.27  &  1.47  &  0.28  &  -0.06  &  8.91  &   & lm \nl
N3370-G &        161.75525 &        17.28084 &       4367 &  52.72  &  1.22  &  24.94  &  0.28  &  0.09  &  0.05  &  0.04  &  8.78  &   & lm \nl
N3370-G &        161.75695 &        17.28274 &       5361 &  50.60  &  0.90  &  23.99  &  0.27  &  0.28  &  0.05  &  -0.06  &  8.83  &   & lm \nl
N3370-G &        161.75874 &        17.28466 &       6706 &  64.79  &  0.96  &  25.00  &  0.40  &  1.65  &  0.31  &  0.09  &  8.84  & rej & lm \nl
N3370-G &        161.75677 &        17.28193 &       5501 &  62.71  &  1.26  &  23.58  &  0.25  &  0.24  &  0.04  &  0.02  &  8.83  &   & lm \nl
N3982-B &        179.09710 &        55.12080 &       9531 &  25.83  &  0.87  &  25.45  &  0.35  &  0.76  &  0.00  &  -0.11  &  8.70  &   & psl \nl
N3982-B &        179.09800 &        55.11996 &       32398 &  45.43  &  0.87  &  24.33  &  0.19  &  0.36  &  0.04  &  -0.10  &  8.72  &   & psl \nl
N3982-B &        179.10593 &        55.12011 &       9075 &  37.04  &  0.80  &  24.32  &  0.33  &  8.02  &  0.23  &  -0.00  &  8.97  &   & psl \nl
N3982-B &        179.10563 &        55.11935 &       9114 &  53.86  &  1.07  &  23.52  &  0.34  &  0.73  &  0.13  &  -0.09  &  8.93  &   & psl \nl
N3982-C &        179.13277 &        55.12384 &       12134 &  19.08  &  0.86  &  24.92  &  0.67  &  0.80  &  0.64  &  -0.03  &  8.90  &   & psl \nl
N3982-C &        179.13148 &        55.12278 &       627 &  38.86  &  1.15  &  24.28  &  0.39  &  0.47  &  0.21  &  0.00  &  8.93  &   & psl \nl
N3982-C &        179.13803 &        55.12526 &       281 &  40.88  &  1.31  &  24.36  &  0.25  &  0.52  &  0.23  &  0.03  &  8.72  &   & psl \nl
N3982-C &        179.13666 &        55.12775 &       43380 &  44.22  &  0.99  &  23.97  &  0.33  &  0.21  &  0.21  &  0.10  &  8.77  &   & psl \nl
N3982-G &        179.09010 &        55.12513 &       32813 &  30.04  &  1.16  &  25.22  &  0.41  &  4.81  &  -0.09  &  -0.01  &  8.48  &   & psl \nl
N3982-G &        179.08546 &        55.12250 &       33304 &  32.02  &  0.97  &  24.56  &  0.21  &  0.38  &  0.11  &  -0.05  &  8.32  &   & psl \nl
N3982-R &        179.09379 &        55.10950 &       9610 &  38.11  &  1.23  &  24.40  &  0.20  &  0.24  &  0.06  &  -0.13  &  8.28  &   & psl \nl
N3982-R &        179.09561 &        55.11319 &       32634 &  41.00  &  0.89  &  24.85  &  0.34  &  0.69  &  0.03  &  -0.11  &  8.46  &   & psl \nl
N3982-C &        179.13316 &        55.12718 &       11942 &  25.57  &  1.49  &  25.28  &  0.37  &  1.59  &  0.56  &  -0.02  &  8.89  &   & lm \nl
N3982-R &        179.09459 &        55.11199 &       9584 &  61.79  &  1.16  &  24.59  &  0.24  &  0.48  &  0.15  &  -0.07  &  8.39  & rej & lm \nl
N3982-Y &        179.12839 &        55.13052 &       12782 &  24.11  &  1.11  &  25.47  &  0.46  &  1.53  &  0.85  &  -0.14  &  8.99  &   & lm \nl
N3982-Y &        179.12240 &        55.13018 &       2002 &  37.57  &  1.35  &  22.81  &  0.37  &  1.07  &  0.42  &  -0.08  &  9.14  & rej & lm \nl
N3982-Y &        179.12400 &        55.12933 &       1434 &  75.40  &  1.86  &  22.33  &  0.25  &  0.69  &  0.21  &  -0.06  &  9.14  & rej & lm \nl
N3982-Y &       179.13114 &       55.13153 &       32 &  29.53  &  0.89  &  23.90  &  0.50  &  10.4  &  0.13  &  0.00  &  8.88  & rej & ps \nl
N3021-B &        147.72778 &        33.54702 &       30672 &  13.92  &  0.48  &  25.88  &  0.37  &  0.87  &  0.24  &  -0.04  &  8.29  & low P & lm \nl
N3021-B &        147.73211 &        33.54878 &       26946 &  26.84  &  0.85  &  25.04  &  0.32  &  0.95  &  0.22  &  0.04  &  8.76  &   & lm \nl
N3021-B &        147.72812 &        33.54750 &       30428 &  32.60  &  0.78  &  25.03  &  0.30  &  1.08  &  0.17  &  -0.01  &  8.37  &   & lm \nl
N3021-B &        147.73249 &        33.54885 &       26545 &  39.57  &  0.87  &  24.79  &  0.34  &  4.08  &  0.08  &  0.05  &  8.79  &   & lm \nl
N3021-C &        147.72678 &        33.55614 &       32088 &  25.77  &  0.98  &  26.35  &  0.59  &  1.79  &  0.59  &  -0.03  &  8.89  & rej & lm \nl
N3021-C &        147.72586 &        33.55581 &       32375 &  24.01  &  0.82  &  24.46  &  0.45  &  1.37  &  0.11  &  -0.01  &  8.82  & rej & lm \nl
N3021-C &        147.72645 &        33.56000 &       32380 &  25.18  &  0.72  &  25.27  &  0.31  &  0.91  &  0.08  &  -0.09  &  8.69  &   & lm \nl
N3021-C &        147.72789 &        33.55893 &       31803 &  37.27  &  0.89  &  25.30  &  0.28  &  1.04  &  -0.16  &  -0.05  &  8.83  &   & lm \nl
N3021-G &        147.74838 &        33.55002 &       8621 &  15.37  &  0.75  &  25.94  &  0.64  &  3.49  &  0.70  &  -0.06  &  8.94  & low P & lm \nl
N3021-G &        147.74935 &        33.55170 &       8102 &  18.71  &  0.62  &  26.10  &  0.47  &  2.77  &  0.48  &  0.08  &  8.90  & low P & lm \nl
N3021-G &        147.74871 &        33.55237 &       8636 &  24.36  &  0.72  &  24.91  &  0.60  &  4.36  &  0.10  &  -0.02  &  8.93  &   & lm \nl
N3021-G &        147.74791 &        33.55032 &       9028 &  31.89  &  0.72  &  24.70  &  0.32  &  1.53  &  0.60  &  0.13  &  8.98  &   & lm \nl
N3021-G &        147.74757 &        33.55109 &       9402 &  39.77  &  1.15  &  23.98  &  0.33  &  5.43  &  0.13  &  0.12  &  9.02  & rej & lm \nl
N3021-G &        147.74740 &        33.55142 &       9611 &  40.49  &  0.63  &  25.30  &  0.49  &  3.20  &  0.05  &  0.05  &  9.04  &   & lm \nl
N3021-G &        147.74683 &        33.55170 &       10203 &  95.91  &  0.85  &  24.05  &  0.25  &  5.58  &  0.13  &  0.00  &  9.08  &   & lm \nl
N3021-G &        147.75116 &        33.55414 &       7098 &  82.66  &  0.72  &  24.18  &  0.25  &  0.55  &  0.12  &  0.00  &  8.67  &   & lm \nl
N3021-G &        147.74734 &        33.55075 &       9558 &  88.18  &  1.43  &  23.96  &  0.25  &  1.28  &  0.33  &  -0.05  &  9.03  &   & lm \nl
N3021-R &        147.73688 &        33.55930 &       23149 &  32.52  &  0.92  &  25.59  &  0.40  &  2.34  &  0.37  &  0.06  &  8.92  &   & lm \nl
N3021-R &        147.73982 &        33.56093 &       19817 &  68.61  &  1.06  &  23.47  &  0.24  &  0.66  &  0.18  &  -0.02  &  8.61  &   & lm \nl
N1309-B &        50.513220 &       -15.40390 &       52566 &  47.41  &  0.53  &  24.79  &  0.43  &  0.64  &  0.12  &  -0.01  &  8.70  &   & lm \nl
N1309-B &        50.513500 &       -15.39881 &       52170 &  47.99  &  0.97  &  24.63  &  0.24  &  0.39  &  0.07  &  -0.01  &  8.73  &   & lm \nl
N1309-B &        50.512020 &       -15.39909 &       53187 &  59.75  &  0.57  &  24.91  &  0.35  &  0.81  &  0.14  &  -0.00  &  8.68  &   & lm \nl
N1309-B &        50.516480 &       -15.40236 &       49485 &  74.19  &  0.39  &  23.72  &  0.37  &  0.24  &  -0.05  &  0.01  &  8.83  &   & lm \nl
N1309-C &        50.535980 &       -15.41296 &       6737 &  25.45  &  0.71  &  26.03  &  0.33  &  3.53  &  0.03  &  0.03  &  8.77  & low P & lm \nl
N1309-C &        50.535850 &       -15.41538 &       7224 &  30.90  &  0.82  &  25.23  &  0.31  &  0.71  &  0.06  &  0.09  &  8.70  & low P & lm \nl
N1309-C &        50.535240 &       -15.41099 &       7989 &  39.41  &  0.90  &  24.70  &  0.24  &  1.33  &  0.17  &  -0.06  &  8.85  &   & lm \nl
N1309-C &        50.535740 &       -15.41413 &       59151 &  32.61  &  0.59  &  24.85  &  0.22  &  4.19  &  0.01  &  -0.02  &  8.74  & low P & lm \nl
N1309-C &        50.537010 &       -15.41209 &       4882 &  48.91  &  0.76  &  25.25  &  0.19  &  0.11  &  0.12  &  -0.05  &  8.78  &   & lm \nl
N1309-C &        50.536060 &       -15.41233 &       6542 &  59.12  &  0.92  &  24.57  &  0.22  &  0.45  &  0.12  &  -0.03  &  8.79  &   & lm \nl
N1309-C &        50.535980 &       -15.41154 &       6581 &  58.98  &  0.79  &  24.82  &  0.22  &  0.36  &  0.11  &  0.12  &  8.82  &   & lm \nl
N1309-C &        50.535540 &       -15.41410 &       7702 &  73.76  &  0.86  &  24.36  &  0.25  &  0.06  &  0.18  &  -0.04  &  8.74  &   & lm \nl
N1309-G &        50.540170 &       -15.39411 &       2032 &  42.53  &  0.72  &  24.91  &  0.26  &  0.74  &  0.11  &  -0.03  &  8.85  &   & lm \nl
N1309-G &        50.541640 &       -15.39645 &       1166 &  41.11  &  1.10  &  24.84  &  0.44  &  4.43  &  -0.23  &  0.05  &  8.83  &   & lm \nl
N1309-Y &        50.528160 &       -15.40843 &       23076 &  30.66  &  0.82  &  25.99  &  0.49  &  2.33  &  0.52  &  0.00  &  9.00  & low P & lm \nl
N1309-Y &        50.525250 &       -15.40856 &       30349 &  33.51  &  0.61  &  26.04  &  0.68  &  3.78  &  -0.22  &  0.00  &  8.96  & rej,low P & lm \nl
N1309-Y &        50.531480 &       -15.40689 &       15346 &  46.85  &  0.81  &  25.53  &  0.51  &  1.63  &  0.13  &  0.01  &  9.04  &   & lm \nl
N1309-Y &        50.528240 &       -15.40865 &       22918 &  42.03  &  0.81  &  24.97  &  0.38  &  0.76  &  0.54  &  0.02  &  8.99  &   & lm \nl
N1309-Y &        50.528300 &       -15.40526 &       68817 &  49.93  &  0.53  &  24.63  &  0.62  &  10.6  &  0.40  &  -0.01  &  9.11  &   & lm \nl
N1309-Y &        50.526610 &       -15.40578 &       71911 &  51.99  &  0.75  &  24.07  &  0.44  &  4.48  &  -0.18  &  -0.02  &  9.07  &   & lm \nl
N1309-Y &        50.526040 &       -15.40768 &       28132 &  52.24  &  1.00  &  24.89  &  0.35  &  1.29  &  0.24  &  0.01  &  9.00  &   & lm \nl
N1309-Y &        50.528080 &       -15.40923 &       69494 &  60.17  &  0.93  &  24.97  &  0.46  &  6.65  &  -0.05  &  -0.00  &  8.97  &   & lm \nl
N1309-Y &        50.529580 &       -15.40892 &       19918 &  64.94  &  0.80  &  24.48  &  0.33  &  0.26  &  0.08  &  0.02  &  8.98  &   & lm \nl
N1309-Y &        50.531070 &       -15.40794 &       64757 &  65.03  &  1.09  &  24.27  &  0.29  &  1.12  &  0.24  &  0.02  &  9.01  &   & lm \nl
IN-NIC-01 &        184.68938 &        47.33554 &       118961 &  12.82  &  1.26  &  22.58  &  0.28  &  0.76  &  0.16  &  -0.06  &  8.90  & rej & lm \nl
IN-NIC-02 &        184.70089 &        47.34927 &       110213 &  11.60  &  1.18  &  23.20  &  0.32  &  0.71  &  0.20  &  -0.01  &  8.91  &   & lm \nl
IN-NIC-02 &        184.69963 &        47.35114 &       113982 &  11.64  &  0.84  &  24.81  &  0.48  &  3.12  &  -0.14  &  0.04  &  8.90  & rej & lm \nl
IN-NIC-05 &        184.70311 &        47.31951 &       83857 &  28.13  &  0.76  &  22.05  &  0.25  &  4.61  &  0.32  &  -0.04  &  8.93  &   & lm \nl
IN-NIC-05 &        184.70506 &        47.32093 &       80885 &  65.23  &  1.13  &  20.91  &  0.20  &  0.32  &  0.14  &  -0.07  &  8.94  &   & lm \nl
IN-NIC-06 &        184.70488 &        47.33184 &       91209 &  10.80  &  0.88  &  23.59  &  0.37  &  1.14  &  0.35  &  -0.09  &  8.94  &   & lm \nl
IN-NIC-06 &        184.70320 &        47.32795 &       91129 &  11.16  &  1.00  &  23.25  &  0.39  &  1.40  &  0.40  &  -0.03  &  8.93  &   & lm \nl
IN-NIC-07 &        184.71071 &        47.34988 &       95403 &  11.58  &  1.03  &  24.41  &  0.56  &  0.99  &  0.38  &  -0.00  &  8.91  & rej & lm \nl
IN-NIC-07 &        184.70793 &        47.34997 &       100093 &  12.02  &  0.71  &  24.78  &  0.59  &  7.06  &  0.82  &  0.12  &  8.91  & rej & lm \nl
IN-NIC-08 &        184.71066 &        47.31184 &       57246 &  10.90  &  0.89  &  22.91  &  0.63  &  0.33  &  0.66  &  -0.05  &  8.94  & rej & lm \nl
IN-NIC-08 &        184.71169 &        47.30965 &       51416 &  11.09  &  0.77  &  22.81  &  0.44  &  5.43  &  0.40  &  0.06  &  8.94  & rej & lm \nl
IN-NIC-08 &        184.71509 &        47.31114 &       43119 &  23.81  &  0.85  &  22.56  &  0.32  &  5.17  &  0.65  &  0.15  &  8.96  &   & lm \nl
IN-NIC-08 &        184.71112 &        47.31241 &       56661 &  23.83  &  0.68  &  23.04  &  0.36  &  0.22  &  0.05  &  -0.09  &  8.94  &   & lm \nl
IN-NIC-08 &        184.71746 &        47.31136 &       36357 &  29.48  &  0.86  &  22.33  &  0.43  &  1.78  &  0.11  &  -0.10  &  8.96  &   & lm \nl
IN-NIC-08 &        184.71239 &        47.30964 &       49279 &  36.12  &  0.94  &  21.44  &  0.19  &  1.74  &  0.03  &  0.27  &  8.94  &   & lm \nl
IN-NIC-08 &        184.71440 &        47.31272 &       47358 &  50.89  &  0.85  &  21.63  &  0.20  &  0.30  &  0.16  &  -0.04  &  8.96  &   & lm \nl
IN-NIC-08 &        184.71707 &        47.31160 &       37841 &  66.89  &  1.13  &  21.61  &  0.25  &  0.89  &  0.21  &  0.02  &  8.96  &   & lm \nl
IN-NIC-08 &        184.71324 &        47.31224 &       50193 &  93.23  &  0.83  &  21.17  &  0.25  &  0.34  &  0.06  &  -0.00  &  8.95  &   & lm \nl
IN-NIC-08 &        184.71528 &        47.31138 &       42837 &  95.92  &  0.93  &  20.93  &  0.25  &  0.34  &  0.14  &  0.03  &  8.96  &   & lm \nl
IN-NIC-09 &        184.71932 &        47.31409 &       34408 &  22.42  &  0.85  &  22.18  &  0.43  &  0.63  &  0.49  &  -0.08  &  8.97  &   & lm \nl
IN-NIC-09 &        184.71746 &        47.31136 &       36357 &  29.48  &  0.86  &  22.16  &  0.29  &  0.58  &  0.42  &  -0.04  &  8.96  &   & lm \nl
IN-NIC-09 &        184.71962 &        47.31407 &       33434 &  34.48  &  0.78  &  20.99  &  0.38  &  0.98  &  0.20  &  -0.01  &  8.98  & rej & lm \nl
IN-NIC-09 &        184.72344 &        47.31211 &       19435 &  39.53  &  0.82  &  20.70  &  0.33  &  0.39  &  0.18  &  -0.00  &  8.99  & rej & lm \nl
IN-NIC-09 &        184.71707 &        47.31160 &       37841 &  66.89  &  1.13  &  21.48  &  0.25  &  6.30  &  0.79  &  -0.10  &  8.96  &   & lm \nl
IN-NIC-10 &        184.72821 &        47.31332 &       6616 &  16.99  &  0.93  &  22.83  &  0.37  &  5.96  &  0.64  &  0.02  &  9.00  &   & lm \nl
IN-NIC-10 &        184.72937 &        47.31691 &       8052 &  20.76  &  0.90  &  22.17  &  0.49  &  6.21  &  0.23  &  0.02  &  9.00  &   & lm \nl
IN-NIC-10 &        184.72616 &        47.31461 &       14643 &  22.04  &  1.92  &  22.05  &  0.63  &  2.99  &  -0.30  &  -0.04  &  8.99  &   & lm \nl
IN-NIC-10 &        184.72389 &        47.31625 &       23741 &  22.68  &  1.43  &  22.77  &  0.32  &  1.78  &  0.03  &  -0.16  &  8.99  &   & lm \nl
IN-NIC-10 &        184.72948 &        47.31746 &       8361 &  23.79  &  1.22  &  21.97  &  0.33  &  0.89  &  0.35  &  0.11  &  8.99  &   & lm \nl
IN-NIC-10 &        184.72728 &        47.31776 &       15470 &  50.70  &  1.48  &  22.19  &  0.41  &  1.83  &  0.24  &  -0.04  &  8.99  &   & lm \nl
IN-NIC-10 &        184.72834 &        47.31586 &       9633 &  69.46  &  1.19  &  20.75  &  0.26  &  0.48  &  0.12  &  0.13  &  9.00  &   & lm \nl
IN-NIC-11 &        184.72657 &        47.34519 &       54398 &  15.73  &  1.06  &  22.48  &  0.50  &  1.82  &  0.47  &  0.10  &  8.91  & rej & lm \nl
IN-NIC-12 &        184.72567 &        47.32182 &       25760 &  9.979  &  0.92  &  24.06  &  0.44  &  5.12  &  0.69  &  -0.07  &  8.98  &   & lm \nl
IN-NIC-12 &        184.72505 &        47.32063 &       25811 &  10.30  &  0.90  &  23.26  &  0.68  &  2.82  &  0.37  &  -0.07  &  8.98  &   & lm \nl
IN-NIC-12 &        184.72545 &        47.32166 &       26176 &  18.28  &  1.67  &  22.49  &  0.59  &  0.46  &  0.15  &  0.21  &  8.98  &   & lm \nl
IN-NIC-12 &        184.72759 &        47.31971 &       17151 &  22.35  &  1.65  &  22.46  &  0.39  &  0.08  &  0.47  &  -0.10  &  8.99  &   & lm \nl
IN-NIC-12 &        184.72948 &        47.31746 &       8361 &  23.79  &  1.22  &  22.41  &  0.40  &  0.98  &  0.35  &  -0.01  &  8.99  &   & lm \nl
IN-NIC-12 &        184.73086 &        47.32120 &       9241 &  27.25  &  0.72  &  22.96  &  0.47  &  3.40  &  0.57  &  -0.01  &  8.98  &   & lm \nl
IN-NIC-12 &        184.73086 &        47.32065 &       8480 &  37.63  &  0.76  &  21.29  &  0.26  &  0.36  &  0.04  &  -0.08  &  8.98  & rej & lm \nl
IN-NIC-12 &        184.72728 &        47.31776 &       15470 &  50.70  &  1.48  &  22.49  &  0.28  &  2.30  &  0.42  &  -0.06  &  8.99  & rej & lm \nl
IN-NIC-01 &        184.68739 &        47.33468 &       121078 &  18.42  &  0.97  &  22.61  &  0.21  &  0.55  &  0.00  &  0.18  &  8.89  &   & lm \nl
IN-NIC-01 &        184.69005 &        47.33265 &       116159 &  21.87  &  0.89  &  23.09  &  0.21  &  0.17  &  0.11  &  -0.13  &  8.90  &   & lm \nl
IN-NIC-02 &        184.70103 &        47.35073 &       111064 &  14.59  &  0.98  &  22.99  &  0.34  &  2.11  &  0.30  &  0.16  &  8.90  &   & lm \nl
IN-NIC-05 &        184.70263 &        47.31955 &       84934 &  15.64  &  0.97  &  23.35  &  0.45  &  1.28  &  0.42  &  -0.12  &  8.93  &   & lm \nl
IN-NIC-05 &        184.70646 &        47.32085 &       77610 &  42.82  &  1.00  &  22.14  &  0.25  &  0.71  &  0.17  &  -0.15  &  8.94  &   & lm \nl
IN-NIC-06 &        184.70453 &        47.32912 &       89618 &  12.47  &  1.01  &  23.06  &  0.31  &  0.77  &  0.39  &  -0.01  &  8.94  &   & lm \nl
IN-NIC-06 &        184.70347 &        47.33143 &       93585 &  18.19  &  0.97  &  22.90  &  0.36  &  1.15  &  0.48  &  -0.00  &  8.93  &   & lm \nl
IN-NIC-07 &        184.70981 &        47.35380 &       99783 &  14.31  &  0.80  &  23.68  &  0.35  &  3.80  &  0.13  &  0.14  &  8.90  &   & lm \nl
IN-NIC-07 &        184.71165 &        47.35150 &       95003 &  20.57  &  1.21  &  23.00  &  0.26  &  0.78  &  0.14  &  0.23  &  8.90  &   & lm \nl
IN-NIC-07 &        184.70864 &        47.35115 &       99756 &  29.05  &  1.04  &  22.70  &  0.23  &  0.59  &  0.10  &  -0.00  &  8.90  &   & lm \nl
IN-NIC-08 &        184.71348 &        47.31023 &       46762 &  12.65  &  0.71  &  23.65  &  0.56  &  6.64  &  -0.38  &  -0.02  &  8.95  &   & lm \nl
IN-NIC-08 &        184.71404 &        47.31339 &       49332 &  16.00  &  0.71  &  23.63  &  0.68  &  5.62  &  0.45  &  0.10  &  8.96  &   & lm \nl
IN-NIC-08 &        184.71249 &        47.31037 &       49942 &  16.34  &  1.03  &  23.68  &  0.39  &  0.63  &  0.62  &  -0.03  &  8.95  &   & lm \nl
IN-NIC-08 &        184.71407 &        47.31168 &       46945 &  25.12  &  0.90  &  22.83  &  0.31  &  0.61  &  0.35  &  -0.00  &  8.95  &   & lm \nl
IN-NIC-08 &        184.71470 &        47.30831 &       189390 &  34.41  &  0.95  &  21.96  &  0.22  &  0.27  &  0.02  &  -0.02  &  8.95  &   & lm \nl
IN-NIC-09 &        184.71988 &        47.31322 &       31615 &  23.98  &  1.32  &  22.07  &  0.45  &  2.34  &  0.40  &  0.22  &  8.98  &   & lm \nl
IN-NIC-09 &        184.72036 &        47.31321 &       30136 &  26.07  &  1.22  &  22.36  &  0.40  &  1.02  &  0.35  &  -0.13  &  8.98  &   & lm \nl
IN-NIC-09 &        184.72467 &        47.31105 &       14316 &  29.63  &  1.13  &  22.26  &  0.50  &  2.13  &  0.08  &  0.11  &  8.99  &   & lm \nl
IN-NIC-09 &        184.72402 &        47.31193 &       17423 &  34.57  &  1.28  &  22.79  &  0.33  &  1.54  &  0.19  &  0.24  &  8.99  &   & lm \nl
IN-NIC-09 &        184.72228 &        47.31203 &       22927 &  33.99  &  1.04  &  22.70  &  0.49  &  2.72  &  -0.09  &  0.19  &  8.98  &   & lm \nl
IN-NIC-09 &        184.71972 &        47.31095 &       29058 &  40.54  &  0.95  &  22.56  &  0.32  &  4.97  &  -0.28  &  -0.19  &  8.97  &   & lm \nl
IN-NIC-09 &        184.71854 &        47.31213 &       34159 &  41.57  &  1.01  &  21.91  &  0.24  &  0.38  &  0.13  &  -0.17  &  8.97  &   & lm \nl
IN-NIC-10 &        184.72694 &        47.31762 &       16299 &  12.28  &  1.28  &  22.86  &  0.53  &  6.31  &  0.54  &  0.17  &  8.99  &   & lm \nl
IN-NIC-10 &        184.72795 &        47.31605 &       11066 &  15.91  &  1.21  &  23.80  &  0.65  &  13.7  &  0.60  &  0.16  &  9.00  &   & lm \nl
IN-NIC-10 &        184.72657 &        47.31756 &       17357 &  22.45  &  1.21  &  22.39  &  0.58  &  4.39  &  -0.11  &  -0.06  &  8.99  &   & lm \nl
IN-NIC-10 &        184.72375 &        47.31687 &       24960 &  25.49  &  1.06  &  22.19  &  0.50  &  5.02  &  0.14  &  -0.07  &  8.99  &   & lm \nl
IN-NIC-12 &        184.72694 &        47.31762 &       16299 &  12.28  &  1.28  &  23.33  &  0.60  &  2.37  &  0.83  &  0.03  &  8.99  &   & lm \nl
IN-NIC-12 &        184.73048 &        47.31999 &       8723 &  35.57  &  0.96  &  21.46  &  0.29  &  0.33  &  0.10  &  0.03  &  8.99  &   & lm \nl
IN-NIC-13 &        184.73382 &        47.33895 &       24365 &  22.38  &  0.93  &  23.12  &  0.39  &  0.59  &  0.44  &  -0.10  &  8.91  &   & lm \nl
IN-NIC-13 &        184.73069 &        47.33826 &       32759 &  42.31  &  0.95  &  22.14  &  0.26  &  0.99  &  0.19  &  -0.10  &  8.92  &   & lm \nl
IN-NIC-03 &        184.69895 &        47.35497 &       117710 &  14.30  &  0.84  &  23.56  &  0.26  &  0.51  &  0.21  &  -0.06  &  8.89  &   & lm \nl
IN-NIC-03 &        184.69870 &        47.35607 &       118782 &  25.56  &  0.89  &  22.34  &  0.20  &  0.41  &  0.09  &  -0.03  &  8.89  &   & lm \nl
IN-NIC-04 &        184.69961 &        47.33411 &       102255 &  12.25  &  1.04  &  24.23  &  0.42  &  1.68  &  0.28  &  -0.10  &  8.92  & rej & lm \nl
IN-NIC-04 &        184.69929 &        47.33694 &       105183 &  23.00  &  1.11  &  22.85  &  0.29  &  1.18  &  0.26  &  -0.14  &  8.92  &   & lm \nl
IN-NIC-04 &        184.69827 &        47.33338 &       104131 &  22.89  &  0.87  &  22.21  &  0.23  &  3.02  &  0.14  &  0.24  &  8.92  &   & lm \nl
IN-NIC-04 &        184.70092 &        47.33803 &       103070 &  24.86  &  1.12  &  22.88  &  0.25  &  0.47  &  0.24  &  -0.05  &  8.92  &   & lm \nl
NIC-POS10 &        184.84596 &        47.24667 &       310420 &  45.40  &  0.86  &  21.34  &  0.18  &  0.17  &  0.03  &  0.19  &  8.74  &   & lm \nl
NIC-POS11 &        184.83328 &        47.24912 &       312665 &  39.09  &  0.80  &  23.20  &  0.19  &  0.25  &  0.05  &  0.11  &  8.78  & rej & lm \nl
NIC-POS12 &        184.85726 &        47.22978 &       307758 &  32.40  &  1.02  &  22.23  &  0.18  &  0.20  &  0.00  &  0.22  &  8.71  &   & lm \nl
NIC-POS13 &        184.72838 &        47.36342 &       999999 &  92.00  &  1.00  &  20.26  &  0.25  &  0.23  &  0.02  &  0.20  &  8.84  &   & lm \nl
NIC-POS1 &        184.70981 &        47.35380 &       99783 &  14.31  &  0.80  &  23.41  &  0.36  &  5.59  &  0.16  &  -0.01  &  8.90  &   & lm \nl
NIC-POS1 &        184.71165 &        47.35150 &       95003 &  20.57  &  1.21  &  23.24  &  0.26  &  0.92  &  0.13  &  -0.00  &  8.90  &   & lm \nl
NIC-POS1 &        184.70864 &        47.35115 &       99756 &  29.05  &  1.04  &  22.38  &  0.21  &  0.48  &  0.06  &  0.06  &  8.90  &   & lm \nl
NIC-POS1 &        184.71263 &        47.35496 &       95995 &  36.79  &  1.05  &  22.25  &  0.21  &  0.21  &  -0.00  &  -0.15  &  8.89  &   & lm \nl
NIC-POS2 &        184.71006 &        47.34746 &       94632 &  16.03  &  1.14  &  23.17  &  0.38  &  2.89  &  0.02  &  -0.12  &  8.91  &   & lm \nl
NIC-POS2 &        184.70643 &        47.34573 &       99411 &  17.02  &  0.98  &  22.97  &  0.28  &  0.59  &  0.16  &  -0.12  &  8.92  &   & lm \nl
NIC-POS2 &        184.70905 &        47.34317 &       92950 &  16.70  &  1.01  &  23.39  &  0.35  &  0.71  &  0.09  &  0.00  &  8.92  &   & lm \nl
NIC-POS2 &        184.70773 &        47.34559 &       97135 &  31.78  &  1.20  &  22.43  &  0.28  &  0.57  &  0.15  &  -0.11  &  8.92  &   & lm \nl
NIC-POS3 &        184.69961 &        47.33411 &       102255 &  12.25  &  1.04  &  23.66  &  0.30  &  0.46  &  0.52  &  -0.13  &  8.92  &   & lm \nl
NIC-POS3 &        184.69929 &        47.33694 &       105183 &  23.00  &  1.11  &  22.85  &  0.28  &  1.06  &  0.15  &  -0.12  &  8.92  &   & lm \nl
NIC-POS3 &        184.69827 &        47.33338 &       104131 &  22.89  &  0.87  &  22.03  &  0.25  &  0.98  &  0.09  &  0.21  &  8.92  &   & lm \nl
NIC-POS3 &        184.69648 &        47.33310 &       106960 &  28.26  &  0.97  &  22.51  &  0.19  &  0.40  &  0.08  &  0.18  &  8.92  &   & lm \nl
NIC-POS4 &        184.70931 &        47.32443 &       74725 &  11.99  &  0.81  &  23.18  &  0.68  &  0.14  &  0.15  &  -0.08  &  8.95  &   & lm \nl
NIC-POS4 &        184.71440 &        47.32211 &       59576 &  12.65  &  1.10  &  23.43  &  0.36  &  7.48  &  0.70  &  0.15  &  8.96  &   & lm \nl
NIC-POS4 &        184.70898 &        47.32425 &       75254 &  16.52  &  0.98  &  23.02  &  0.26  &  0.34  &  0.42  &  -0.04  &  8.95  &   & lm \nl
NIC-POS4 &        184.71341 &        47.32261 &       62769 &  33.02  &  0.77  &  21.93  &  0.25  &  0.66  &  0.09  &  -0.05  &  8.96  &   & lm \nl
NIC-POS5 &        184.74252 &        47.34238 &       2686 &  44.05  &  1.06  &  21.57  &  0.19  &  0.28  &  0.05  &  0.10  &  8.88  &   & lm \nl
NIC-POS6 &        184.83891 &        47.17151 &       107069 &  83.40  &  1.58  &  20.56  &  0.25  &  0.06  &  0.01  &  -0.10  &  8.67  &   & lm \nl
NIC-POS7 &        184.85593 &        47.16107 &       104251 &  33.29  &  1.04  &  22.56  &  0.18  &  0.03  &  0.01  &  -0.13  &  8.64  &   & lm \nl
NIC-POS8 &        184.79990 &        47.20616 &       220789 &  21.29  &  1.04  &  23.31  &  0.19  &  0.35  &  0.04  &  0.16  &  8.77  &   & lm \nl
NIC-POS8 &        184.79992 &        47.20734 &       220887 &  31.29  &  1.26  &  22.70  &  0.18  &  0.15  &  0.00  &  -0.14  &  8.77  &   & lm \nl
NIC-POS8 &        184.80002 &        47.20401 &       220576 &  101.9  &  1.92  &  20.22  &  0.25  &  0.11  &  0.01  &  -0.02  &  8.76  &   & lm \nl
OUT-NIC-02 &        184.81866 &        47.19866 &       34729 &  14.92  &  1.12  &  23.44  &  0.23  &  0.22  &  0.07  &  -0.01  &  8.75  &   & lm \nl
OUT-NIC-04 &        184.83594 &        47.22012 &       28606 &  53.88  &  1.03  &  21.76  &  0.18  &  0.25  &  0.03  &  0.09  &  8.76  &   & lm \nl
OUT-NIC-05 &        184.83415 &        47.19030 &       21109 &  8.503  &  0.86  &  23.98  &  0.27  &  8.81  &  0.34  &  0.04  &  8.72  &   & lm \nl
OUT-NIC-06 &        184.83914 &        47.17078 &       12705 &  9.942  &  0.84  &  24.23  &  0.25  &  0.31  &  0.07  &  -0.08  &  8.67  &   & lm \nl
OUT-NIC-06 &        184.83644 &        47.17427 &       14709 &  10.97  &  0.84  &  23.68  &  0.22  &  0.42  &  -0.00  &  0.10  &  8.68  &   & lm \nl
OUT-NIC-06 &        184.83530 &        47.17378 &       15276 &  16.43  &  0.87  &  23.62  &  0.21  &  0.62  &  0.03  &  -0.07  &  8.68  &   & lm \nl
OUT-NIC-07 &        184.83867 &        47.19483 &       19312 &  13.55  &  0.82  &  23.32  &  0.21  &  0.52  &  0.07  &  -0.08  &  8.72  &   & lm \nl
OUT-NIC-08 &        184.85030 &        47.19245 &       11990 &  8.024  &  0.65  &  23.78  &  0.22  &  0.34  &  0.09  &  0.10  &  8.71  &   & lm \nl
OUT-NIC-09 &        184.85468 &        47.16904 &       5713 &  31.74  &  0.96  &  22.41  &  0.18  &  0.14  &  0.02  &  -0.15  &  8.66  &   & lm \nl
OUT-NIC-10 &        184.85857 &        47.20057 &       9786 &  8.920  &  0.72  &  23.85  &  0.24  &  0.28  &  0.14  &  -0.05  &  8.70  &   & lm \nl
OUT-NIC-13 &        184.83969 &        47.18170 &       14656 &  8.779  &  0.80  &  24.23  &  0.35  &  1.52  &  0.06  &  0.03  &  8.70  &   & lm \nl
\tableline
\multicolumn{14}{l}{$^*$ Source of Optical Cepheid parameters.  lm=L.M. from Riess et al. 2009 and Macri et al. 2009, psl=P.B.S. from Gibson \& Stetson 2000 } \\
\multicolumn{14}{l}{$^a$ Cepheid Rejection Flag.  $<$P indicates that the period is shorter than the optical completeness from Riess et al. 2009, rej is $\sigma$ clipped } \\
\enddata
\end{deluxetable}


\subsection{Near-Infrared Cepheid Relations}

In the {\it i}th galaxy, for a set of Cepheids with periods {\bf P}, 
with mean magnitudes {\bf m}$_X$, the pulsation equation
leads to a period-luminosity (\PL) relation of the form

\bq {\bf m_X}=zp_{X,i}+b_X \ {\rm log} {\bf \ P}, \eq where $zp_{X,i}$ is the
intercept of the \PL relation, and $b_X$ is its slope for passband $X$.
We will make use of multi-linear regressions to {\it simultaneously} fit
the Cepheid data (and in the next section the SN data) and to propagate
the covariance of the data and model to the fitted parameters.

It is convenient to express the \PL relation for the {\it j}th Cepheid $j$ in the {\it i}th host as

\bq m_{X,i,j}=(zp_{X,i}-zp_{X,4258})+zp_{X,4258}+b_X \ {\rm log} \
P_{i,j}. \eq
\noindent

Although the $H$-band \PL relation is expected to be relatively
insensitive to metallicity  as compared to the
visible, where metal-line blanketing influences opacity \citep{marconi05}, we will 
{\it not assume the LMC slope applies to our
more metal-rich Cepheid sample}.\footnote{The slope in the $H$ band, $b_H$,
has been measured by \citet{persson04} to be $-3.234 \pm 0.042$ based
on 88 Cepheids in the LMC. Limiting the sample to 75 variables with
$P > 10$~d yields the same result.}  Instead, we will determine the slope for
the narrow range of solar-like metallicity of our sample.

We rewrite equation (2) in matrix form to allow
a single, unknown value of $b_H$,

\bq \pmatrix{m_{H_1}\cr m_{H_2} \cr . \cr m_{H_n}\cr
m_{H_{4258}}}=\pmatrix{1&0&0&0&0&0&1 & {\rm log} P_1 \cr 0&1&0&0&0&0&1
& {\rm log} P_2 \cr .&.&.&.&.&.&. & . \cr 0&0&0&0&0&1&1 & {\rm log}
P_n \cr 0&0&0&0&0&0&1 & {\rm log} P_{4258}} \ \ \ \ * \ \ \ \
\pmatrix{zp_{H,1}-zp_{H,4258} \cr zp_{H,2}-zp_{H,4258} \cr .  \cr
zp_{H,n}-zp_{H,4258} \cr zp_{H,4258} \cr b_H } \ + n(t)\eq
\noindent
Referring to this matrix equation symbolically as ${\bf y=Lq}$, we define:
{\bf y} is the column of measured magnitudes, {\bf L} is the
2-dimensional ``design matrix'' with entries that arrange the
operations,  and {\bf q} is the set of
free parameters.  With these definitions and {\bf C} as the matrix of measurement errors, we write the $\chi^2$ statistic as

\bq \chi^2=({\bf y}-{\bf Lq})^T{\bf C}^{-1}({\bf y}-{\bf Lq}).\eq
\noindent
The minimization of $\chi^2$ with respect to {\bf q} gives the following expression
for the maximum likelihood estimator of {\bf q}:

\bq {\bf \hat{q}}= ({\bf L}^T{\bf C}^{-1}{\bf L})^{-1}{\bf L}^T{\bf
C}^{-1}[{\bf y}] \eq
\noindent
The standard errors for the parameters in ${\bf \hat{q}}$ are given by
 the covariance matrix,\\ $({\bf L}^T{\bf C}^{-1}{\bf L})^{-1}$ (Rybick \& Press 1992).

The seven individual \PL relations fitted with a common slope are shown in
Figure 11.  While 240 Cepheids previously identified in the optical (Riess et al. 2009) could be
measured in the NICMOS data, it is apparent from Figure 11 that
$\sim$10\% appear as outliers in the relations.  This is not
surprising as we expect outliers to occur from (1) a complete blend
with a bright, red source such as a red giant or  (2) objects
misidentified as Cepheids in the optical or with the wrong period.  To
reject these outliers we performed an iterative rejection of objects
$>$ 0.75 mag from the \PL relations, resulting in a reduction
of the sample to 209.  In the next section we consider the effect of
this rejection on the the determination of $H_0$.

For the sample we find $b_H = -3.09 \pm 0.11$, in good agreement with
the value of $-3.23 \pm 0.04$ from the LMC \citep{persson04}.

To determine the difference in distances between the anchor galaxy and the SN hosts we now account for interstellar extinction of the Cepheids.
Although such extinction is a factor of $\sim$5  smaller in the $H$ band  than in the optical and might be ignored (an option we consider in \S 4), the difference in what remains directly impacts the determination of $H_0$ at the few percent level. 

The use of two or more passbands allows for the
measurement of reddening and the associated correction for extinction.
For each Cepheid we use the measurement of its mean $V-I$ color from WFPC2 or ACS.  Following
\citet{madore82} we define a ``Wesenheit reddening-free'' mean
magnitude,

\bq {\bf m_W} ={\bf m_H} - R({\bf m_V - m_I}), \eq where $R \equiv A_H /
(A_V - A_I)$.
\noindent
For a \citet{cardelli89} reddening law and a Galactic-like value of 
$R_V = 3.1$, $R=0.479$.  In the next section we consider the sensitivity of $H_0$ to the value of $R_V$. 

To account for possible differences in the Cepheid photometry measured with
ACS and WFPC2, we compared the photometry of 8711 non-variable stellar
sources in the field of NGC 3982 observed with both cameras through 
$F555W$ and $F814W$.  Using the master catalog of WFPC2
photometry used by \citet{gibson00} and \citet{stetson01} we find the mean
difference between our WFPC2 and ACS $V-I$ colors of these
sources to be $0.054 \pm 0.005$ mag (WFPC2 is bluer), with no dependence
on source color or magnitude.  The origin of this difference largely resides in the specific zeropoints adopted by \citet{stetson01} and those from Riess et al. (2009) and Macri et al. (2006,2009).
Because our goal is limited to placing the WFPC2 Cepheid colors on the
same photometric scale as the ACS data to measure distances
relative to NGC 4258, we corrected the WFPC2 Cepheid data of
\citet{gibson00} for NGC 4639, NGC 4536, and NGC 3982 to 
the ACS color scale.  The mean $V-I$ colors of the Cepheids are given in
Table 4.  We propagate a {\it systematic} 0.02 mag error in the
difference between $V-I$ colors measured with WFPC2 and ACS in the
next section, though the net effect on $m_W$ amounts to only
$0.02R \approx 0.01$
mag.\footnote{For analysis using the optical relation $m_W = m_V -
2.45(m_V - m_I)$, differences in color measurements between different
photometric systems are $\sim$ 5 times larger with additional uncertainties due to the
difficulty in cross-calibrating ground-based and space-based systems.
The resulting systematic uncertainty is typically 0.10 mag, one of the
leading systematic errors in the determination of $H_0$.}

Substituting the values of $m_W$ for $m_H$ in equation (3), we find
$b_W = -3.23 \pm 0.11$.  

Differences in $zp_{W}$ between galaxies are equivalent
to differences in distances, which follows from equation (1) and    $ \mu_0=m_W-M_W $.
Therefore we can now substitute ($zp_{W,i}-zp_{W,4258}$)=($\mu_{0,i}-\mu_{0,4258}$) to
derive reddening-free distances, $\mu_{0,i}$, for the
SN hosts relative to NGC 4258 from the Cepheids,
$\mu_{0,i}-\mu_{0,4258}$.
The results are
given in Table 5, column 6.

To account for the possible dependence of Cepheid magnitude on
metallicity even over the narrow range of metallicity in our Cepheid sample,
we express $m_W$ as

\bq m_{W,i,j}=(\mu_{0,i}-\mu_{0,4258})+zp_{W,4258}+b_W \ {\rm log} \
P_{i,j}+Z_W \ \Delta {\rm log[O/H]}_{i,j}, \eq
\noindent
where the individual values of $\Delta {\rm log[O/H]}_{i,j}$ were derived
from the metallicity values and gradients for the Cepheid hosts given
by \citet{riess09}. These values are listed in Table 4. 

For the parameter $Z_W$ we find $-0.27 \pm 0.18$ in the sense that
metal-rich Cepheids have a brighter value of $m_W$, though this
relation is not significant.  Indeed, the benefit of using Cepheids across the distance ladder
with similar metallicities is that, as shown in the next section, their relative distance measures are insensitive to the uncertainty in their metallicity relation.

We now move to the joint use of the Cepheid and SN Ia data for
deriving the Hubble constant.

\section{Measuring the Hubble Constant}

\subsection{Type Ia Supernova Magnitudes}

Distance estimates from SN~Ia light curves are derived
 from the luminosity distance,

\bq d_{L} = \left(\frac{L}{4 \pi {\cal
F}}\right)^{\frac{1}{2}}, \eq
\noindent
where $L$ and ${\cal F}$ are the intrinsic luminosity and the
absorption-free flux within a given passband, respectively.
Equivalently, logarithmic measures of the flux in passband (e.g., $V$)
(apparent magnitude, $m_V$) and luminosity (absolute magnitude, $M_V$)
are used to derive extinction-corrected distance moduli,

\bq \mu_0=m_V^0-M_V^0=5\log d_L +25 \eq
\noindent
($d_L$ in units of Mpc), where $m_V^0$ derives from
$m_V$ corrected for selective absorption through the use of colors and a reddening law.

We may relate the observables of SN Ia distance and redshift, $z$, to the scale factor of the Universe, $a$,
by expanding $a(t)$ using the definitions

\bq
H(t) = +{\dot a} /a , q(t) = -({\ddot a}/a)({\dot a}/a)^{-2}, j(t) = +({\dot {\ddot a}}/a)({\dot a}/a)^{-3}
\eq
(cf. Visser 2004).

For $z \approx 0$

\bq d_l={cz \over H_0},\eq
\noindent

where $H_0$ is the present expansion rate ($z=0$)
of the Universe.  

Allowing for changes in the expansion rate at $z>0$ 

\bq
d_L(z) =  {c\; z\over H_0}
\left\{ 1 + {1\over2}\left[1-q_0\right] {z} 
-{1\over6}\left[1-q_0-3q_0^2+j_0 \right] z^2
+ O(z^3) \right\}
\eq

or

\bq \mu_0=m_V^0-M_V^0=5 \ {\rm log} {cz \over H_0}
\left\{ 1 + {1\over2}\left[1-q_0\right] {z} 
-{1\over6}\left[1-q_0-3q_0^2+j_0 \right] z^2
+ O(z^3) \right\} + 25, \eq

Using empirical relations between SN Ia light curve shape and luminosity
allows for a modest correction of individual SN Ia magnitudes to relate them to a fiducial luminosity, $M_V^0$, at a fiducial epoch (by convention, $B$-band peak). 
For the
multi-color light-curve shape \citep[MLCS;][]{riess96} method of
fitting SN~Ia light curves, $M_V^0$ is the $V$-band peak absolute
magnitude for a SN Ia matching the template light curve shape (i.e., the light curve parameter $\Delta=0$).  The value $m_V^0$ is the maximum light apparent $V$-band brightness of the fiducial SN~Ia at the
time of $B$-band peak if it had $A_V=0$ and $\Delta=0$.  This
quantity is determined from a full light-curve fit, so that it is a
weighted average, not a measurement at a single epoch.

We can rewrite equation (13) to move the intercept of the
magnitude-redshift relation to the left,

\bq {\rm log} \ cz
\left\{ 1 + {1\over2}\left[1-q_0\right] {z} 
-{1\over6}\left[1-q_0-3q_0^2+j_0 \right] z^2
+ O(z^3) \right\} - 0.2m_V^0 = {\rm log} \ H_0-0.2 \ M_V^0 - 5, \eq
\noindent

and define the intercept of the $log cz$-$0.2m_V^0$ relation, $a_v$, 

\bq a_v = {\rm log} \ cz
\left\{ 1 + {1\over2}\left[1-q_0\right] {z} 
-{1\over6}\left[1-q_0-3q_0^2+j_0 \right] z^2
+ O(z^3) \right\} - 0.2m_V^0. \eq
\noindent
The intercept, $a_v$, is an apparent quantity which is measured from the set of ($z,m_V^0$) independent of any absolute
(i.e., luminosity or distance) scale. We use the kinematic expansion of $a_v$ to include terms of order $z^2$ and $z^3$ rather than the Friedmann relation (i.e., $\Omega_M$, $\Omega_\Lambda$ or $w=P/(\rho
c^2)$) to retain its conventional definition (and measurement) as an apparent (not inferred) quantity. In practice the difference between the kinematic and Friedmann relations is negligible in the range $z<0.1$ where we determine $a_v$.  \footnote{It is worth noting that terms of order $z^2$ were not included in the use of 
of $a_v$ and SNe Ia by Freedman et al. (2001) from Suntzeff et al. (1999) and Phillips et al. (1999), tantamount to setting $q_0=1$ or $\Omega_M=2$ and reducing $H_0$ by $\sim$ 3\%.}

Figure 12 shows a Hubble diagram for 240 SNe~Ia from Hicken et al. (2009) whose intercept determines
the value of $a_v$.  The magnitude-$z$ relation was determined with the fiducial parameters in MLCS2k2
\citep{jha07}.   Limiting the sample to $0.023<z<0.1$  (to avoid the possibility of a local, coherent flow) leaves 140 SNe Ia  where $z$ is the redshift in the restframe of the CMB, the present acceleration $q_0=-0.55$ and prior deceleration $j_0=1$ (Riess et al. 2007) 
yields $a_v=0.698 \pm 0.00225$.   The sensitivities of $a_v$ to the cosmological model, the minimum redshift and the MLS2k2 parameters are discussed in the next section. 

For the $i$th member of a set of nearby SNe Ia whose luminosities
are calibrated by independent estimates of the distances to their hosts,  the Hubble constant is given from equation (14) and (15) as

\bq {\rm \log} \ H_0^i={(m_{v,i}^0-\mu_{0,i})+5a_v+25 \over 5}. \eq
\noindent
The terms $\mu_{0,i}$, determined from Cepheid data, were discussed in
the previous section (e.g., equation (7)).

Because the selection of the fiducial SN~Ia along the luminosity vs.
light-curve shape relation is arbitrary, the value of $a_v$ is also
arbitrary.  However, the inferred value of $H_0$ is independent of
this choice because the luminosity of the fiducial cancels in the sum
$m_{v,i}^0 +5a_v$ in equation (16).  For each SN Ia, the sum
$m_{v,i}^0 +5a_v$ in equation (16) is a {\it fundamental} measure of
its distance (in magnitudes) in the sense that it is independent, in
principle, of the various approaches used to relate SN Ia light curves and their luminosity.  It is also
independent of bandpass.  This sum makes it clear that the measurement
of $H_0$ depends only on the apparent {\it differences} between SN Ia
distances in the calibration set and the Hubble-flow
set.\footnote{This SN difference measurement is similar to the way SNe
are used at high redshift to measure dark energy, but without the
complexity of significant SN evolution, reddening-law evolution, $K$-corrections, time dilation
changes in demographics, or gravitational lensing.}  Systematic errors
may arise from a combination of inaccuracies in the light-curve fitter
and differences in the mean properties of the calibration and
Hubble-flow samples.  We will explore the size of these errors in the
next subsection by varying the assumptions of the light-curve fitter
and by using a different one, SALT II (Guy et al. 2005).

In Table 5 we give the quantities $m_{v,i}^0 +5a_v$ for each of the
SHOES SNe~Ia. 

\begin{table}[h]
\begin{small}
\begin{center}
\vspace{0.4cm}
\begin{tabular}{llllll}
\multicolumn{5}{c}{Table 5: Distance Parameters} \\
\hline
\hline
Host & SN Ia &  filters   &  $m_{v,i}^0 +5a_v$  & $\sigma^a$ & $\mu_{0,i}-\mu_{0,4258}$ \\
\hline
NGC 4536 & SN 1981B  &  UBVR   &  15.156 & 0.145 & 1.145 (0.0845) \\
NGC 4639 & SN 1990N  &  UBVRI  &  16.059 & 0.111 & 2.185 (0.0963) \\
NGC 3982 & SN 1998aq &  UBVRI  &  15.976 & 0.091 & 2.473 (0.101)  \\
NGC 3370 & SN 1994ae &  UBVRI  &  16.578 & 0.102 & 2.831 (0.0771) \\
NGC 3021 & SN 1995al &  UBVRI  &  16.726 & 0.113 & 2.914 (0.101)  \\
NGC 1309 & SN 2002fk &   BVRI  &  16.806 & 0.103 & 3.261 (0.0861) \\
Weighted Mean     &  -----    &  -----  &  -----  & 0.0448 & ------ (0.0367) \\
\hline
\hline
\multicolumn{5}{l}{$^a$ For MLCS2k2, 0.08 mag added in quadrature 
to fitting error.} \\
\end{tabular}
\end{center}
\end{small} 
\end{table}

In Figure 13 we compare the relative distances determined strictly from Cepheids, $\mu_{0,i}-\mu_{0,4258}$, and from SNe Ia, $m_{v,i}^0 +5a_v$.  
These quantities are relative in the sense that they both involve purely differential measurements of like quantities and benefit from the cancellation
of systematic errors associated with the determination of absolute quantities.
The dispersion between these relative distances is 0.08 mag, somewhat smaller than the mean SN distance error of 0.11 mag. 

\subsection{Global Fit for $H_0$}

For convenience we define a parameter ($m^0_{v,4258}$) which is the
expected reddening-free, fiducial, peak magnitude of a SN~Ia appearing
in NGC 4258.   We then express $m^0_v$ for the $ith$ SN Ia as

\bq m_{v,i}^0=( \mu_{0,i}-\mu_{0,4258})+m^0_{v,4258}. \eq
\noindent
Combining the two equations for apparent magnitudes;  for SNe Ia, equation (17), and for Cepheids, equation (7), we write one
matrix equation,

    {\footnotesize      
          \bq \pmatrix{m_{w,1,1}\cr  m_{w,1,2} \cr ... \cr m_{w,1,r_1} \cr m_{w,2,1} \cr ... \cr m_{w,2,r_2} \cr ... \cr m_{w,n,1} \cr ... \cr m_{w,n,r_n} \cr m_{w,4258,1} \cr ... \cr m_{w,4258,r_0} \cr m_{v,1}^0 \cr m_{v,2}^0 \cr ... \cr m_{v,n}^0 }=\pmatrix{
          1&0&...&0&1 & log P_{1,1} & 0 & \Delta log [O/H]_{1,1} \cr 
          1&0&...&0&1 & log P_{1,2} & 0 & \Delta log [O/H]_{1,2} \cr 
          ...&...&...&...&... & ...  & ... & ... \cr 
          1&0&...&0&1 & log P_{1,r_1} & 0 & \Delta log [O/H]_{1,r_1} \cr 
          0&1&...&0&1 & log P_{2,1} & 0 & \Delta log [O/H]_{2,1} \cr 
          ...&...&...&...&... & ...  & ... & ... \cr 
          0&1&...&0&1 & log P_{2,r_2} & 0 & \Delta log [O/H]_{2,r_2} \cr 
          ...&...&...&...&... & ...  & ... & ... \cr 
          0&0&...&1&1 & log P_{n,1} & 0 & \Delta log [O/H]_{n,1} \cr 
          ...&...&...&...&... & ...  & ... & ... \cr 
          0&0&...&1&1 & log P_{n,r_n} & 0 & \Delta log [O/H]_{n,r_n} \cr 
          0&0&...&0&1 & log P_{4258,1} & 0 & \Delta log [O/H]_{4258,1} \cr 
          ...&...&...&...&... & ...  & ... & ... \cr 
          0&0&...&0&1 & log P_{4258,r_0} & 0 & \Delta log [O/H]_{4258,r_0} \cr 
         1&0&...&0&0&0&1&0 \cr  
         0&1&...&0&0&0&1&0  \cr
         ...&...&...&...&...&...&...&... \cr 
         0&0&...&1&0&0&1&0 }   
\ \ \ \ * \ \ \ \ \pmatrix{\mu_{0,1}-\mu_{0,4258} \cr  \mu_{0,2}-\mu_{0,4258} \cr ...  \cr \mu_{0,n}-\mu_{0,4258} \cr zp_{w,4258} \cr b_W \cr m^0_{v,4258} \cr Z_W} \ + {\bf noise}, \eq }
\noindent
for $n$ SN host galaxies ($n=0$ will correspond to NGC 4258), with host $i$ having $r_i$ Cepheids.  Thus we
have $t = n + \displaystyle\sum_{i=0}^n r_i$ equations to solve
simultaneously.  The only term of significance for the determination
of $H_0$ is $m^0_{v,4258}$ and its uncertainty derived from the 
covariance matrix of fitted parameters, which propagates the
uncertainties in Cepheid nuisance parameters such as the slope and metallicity
relations for the Cepheid.\footnote{However, unlike one such real
event, the precision of our estimate of $m^0_{v,4258}$ is equivalent to
measuring $n$ such SN Ia events (requiring a millennium to accomplish!),
though modestly diminished by the noise in the Cepheids measurements.}  The meaning of
$m^0_{v,4258}$ can be readily seen from Figure 13 as it connects the Cepheid and SN Ia relative distance measures, i.e.,  $m^0_{v,4258}=m_{v,i}^0-( \mu_{0,i}-\mu_{0,4258})$.

From equation (16) we derive our best estimate of $H_0$ using

\bq {\rm log} \ H_0={(m_{v,4258}^0-\mu_{0,4258})+5a_V+25 \over 5}. \eq
\noindent
Derived this way, the full statistical error in $H_0$ is the quadrature sum of the uncertainty in the three {\it independent} terms
$(\mu_{0,4258},~m^0_{v,4258},~ {\rm and}~ 5a_V) $ where $\mu_{0,4258}$ is the previously discussed geometric distance estimate to NGC 4258 (Herrnstein et al. 1999, Humphreys et al. 2009).
More than a decade of tracking the Keplerian motion of its water masers supports an uncertainty of 3\% ($\sigma=0.06$ mag; Humphreys et al. 2008, 2009, Greenhill et al. 2009).  

Our result is $H_0=$\statho, a \uncs\ measurement.   The uncertainty from the terms independent of the maser distance to NGC 4258, e.g.,  errors due to the form of the \PL relation, metallicity
dependences, photometry bias, and zeropoint errors as well as the SN Ia $m-z$ relation result in a $\pm
3.4$\% uncertainty in $H_0$.  In past
determinations of the Hubble constant, these sources of uncertainty
have been the leading systematic uncertainties.  In this analysis, these
uncertainties have been reduced by matching the distribution of
Cepheid measurements (i.e., metallicity, periods, and photometric
systems) between NGC 4258 and the SN hosts.  However, given the small
uncertainty in $H_0$, it is important to consider a broader
exploration of systematic uncertainties of the type now under
examination for large-scale high-redshift SN~Ia surveys
\citep[e.g.,][]{astier06,woodvasey07,sullivan07}.

\section{Systematics}

In Table 6 we show 22 variants of the previously described
analysis which we use to estimate the systematic error on our
measurement of $H_0$.  Our primary analysis in row 1 of Table 6 is based on our estimation of the best approach.
Column (1) gives the value of $\chi^2_{\nu}$, column (2) the number of
Cepheids in the fit, column (3) the value and total uncertainty in
$H_0$, and column (4) the uncertainty without including the
uncertainty in the maser distance for NGC 4258.  Column (5) gives the
determination of $M_V^0$, a parameter specific to the light curve fitter employed, column (6) the value and uncertainty in the
metallicity dependence, and column (7) the value and uncertainty of
the slope of the Cepheid \PL or $P-W$ relation.  The next seven parameters are used
to indicate variants in the analysis whose impact we now
consider.

\subsection{ SN Systematics}

Following \citet{woodvasey07}, the leading sources of systematic
uncertainty in the cosmological use of SNe~Ia relevant to our analysis
are addressed here.

\noindent {\it Lower Limit in SN redshift used to measure Hubble flow: } The
minimum redshift beyond which SNe Ia measure the Hubble flow has been
an ongoing source of debate. \citet{zehavi98} and later \citet{jha06}
claimed to see a local ``Hubble bubble" with an increased outflow of
$\sim$5\% within a local void ending at $z=0.023$. \citet{conley07}
demonstrated that the evidence for the bubble rested on a set of SNe
Ia at $0.01 < z < 0.023$ with more than average reddening and that the
reality of the bubble depended on the form of their extinction,
whether $R_V$ is Galactic in nature ($R_V=3.1$) or empirically determined by minimizing the scatter in the Hubble flow ($R_V$=2 to 2.5).  We consider both approaches to estimating the extinction in the range of $0.01 < z < 0.023$ when we consider the value of $R_V$ used for the SNe.

We think the safest choice is to begin the measurement of the Hubble
flow at $z>0.023$ to avoid the uncertainty of the Bubble or other coherent large-scale flows.  A number of authors
\citep{hui06,cooray06} have shown that coherent flows like a Hubble
bubble are likely to induce bias at lower redshifts, and we maintain
our view that it is better to restrict our analysis to $z>0.023$ and
avoid this possible bias.  The penalty is a reduction in the statistical precision of the measurement of the Hubble flow, but this
term remains subdominant in the determination of $H_0$.  
However, we also include a number of analyses with $z_{\rm min} =
0.01$, as indicated by column (8) of Table 6.  These have the effect
of raising the Hubble constant by 1.0 to 1.2 km s$^{-1}$ Mpc$^{-1}$ depending on the
aforementioned treatment of extinction.    An alternate selection of the Hubble flow set would be to consider all SNe Ia at $z>0.01$ but limit the selection to those with $A_V<0.5$, making the Hubble flow sample a good match to the calibrators and avoiding the degeneracy between the Hubble Bubble and the extinction law at $z<0.023$.  This approach results in a value of $H_0$ from SNe Ia at $z>0.01$ which is only 0.7 km s$^{-1}$ Mpc$^{-1}$ greater than the nominal fit at $z>0.023$.

\noindent {\it SN-host $R_V$: } In our primary analysis we account for
the difference in SN Ia extinction between the calibration and
Hubble-flow samples using the $UBVRI$ colors of the SNe and the
MLCS2k2 prescription.  For the extinction due to host-galaxy dust our primary analysis
uses a recent ``consensus" value of $R_V = 2.5$ (fit parameter 37 in column 9 of Table 6) for the lines of sight of SNe Ia 
(Kessler et al. 2009), but we also consider values for $R_V$ of 1.5, 2.0, and
3.1 with fit parameters of 29, 28, and 20,
respectively.  The change in $H_0$ is 0.2 km s$^{-1}$ Mpc$^{-1}$ across the range of $1.5 <
R_V < 3.1$ for the SNe.  The effect is so small because the SN colors
for the calibration sample and those in our nominal Hubble-flow sample
are well matched, so altering 
$R_V$ for the SNe provides little change.

\noindent {\it Distribution of host-galaxy extinction:} 
The observed distribution of SN Ia host galaxy extinction is used
as a prior in the determination of the extinction of individual SNe Ia (Riess, Press, Kirshner 1996) and is particularly important in the absence of precise color measurements (e.g., at high redshifts).  However, the prior has little effect on the present analysis because the SN colors at low redshifts are well measured.  To determine the sensitivity to this prior, we varied its
functional form across two
extremes, using either a simulation of the lines-of-sight through galaxies (Deaton, Branch, Fisher 1998; ``glos") which anticipates less extinction on average than the default or
no extinction prior at all, (fit parameters 27 and 26), respectively.  The difference
in $H_0$ is only 0.8 km s$^{-1}$ Mpc$^{-1}$.  We also changed the
algorithm used to fit and compare the SN~Ia light and color curves
from MLCS2k2 to the SALT II \citep{guy05} approach. These fits (fit
parameter 42) reduce $H_0$ by 0.5 km s$^{-1}$ Mpc$^{-1}$ with other variants held fixed.  Overall we find the determination of $H_0$ is insensitive to assumptions about the relation between SN Ia colors and extinction. 

\noindent {\it SN Ia $U$-band:} We also perform an analysis of the SN
data discarding the $U$ band, fit 61, as it should be most sensitive to the form
of the extinction law, changes in SN Ia metallicity, and errors in calibration.  This decreases $H_0$ by 1.0 km s$^{-1}$ Mpc$^{-1}$.  Because both the nearby and Hubble flow samples make use of the same $U$-band calibration (Jha et al. 2005), our results are
insensitive to the parameters of the $U$-band (Kessler et al. 2009).

Other sources of systematic error listed in Wood-Vasey et al (2007)
arise from a large change in redshift
between two samples of SNe Ia (i.e., cross-filter $K$-corrections
and the possibility of SN~Ia evolution)
and are not significant in our analysis as all SN data are at $z<0.1$.  In
general, changes to the treatment of the SN Ia light curves affect
both the calibration and Hubble-flow sample similarly,  largely canceling in the sum $m_v^0+5a_v$ and their impact on  $H_0$.

\subsection{Cepheid Systematics}

For systematic errors relevant to the analysis of Cepheid data, Table 14 in
\citet{freedman01} lists the dominant terms.  The largest terms
relevant to our analysis are considered here.

\noindent {\it Cepheid metallicity: } Metallicity was addressed in
\S 3.  The critical conclusion is that the range in metallicity for
our Cepheid data is small ($\Delta [O/H] \sim 0.1$), a factor of four times
smaller than if LMC Cepheids are used to calibrate SNe Ia.
In addition, the metallicity sensitivity should be further reduced by a significant factor by observing Cepheids in the near-IR (Marconi et al. 2005).
However, we formally include and marginalize over a first-order
metallicity dependence for the Cepheids using our previous
measurements of the host metallicities.  It may be of
interest to remove the metallicity term in the analyses to determine
its impact on $H_0$.  We include a few such entries in Table 6,
indicated by ``$---$'' in the entry for this term.  The result is that
the nominal uncertainty in $H_0$ decreases by 5\% and its value
is reduced by $\sim$1.2.

\noindent {\it Cepheid reddening: } Reddening of the Cepheids is
largely mitigated over optical-based analyses by the use of $H$-band
photometry, which reduces the net by a factor of 5 over the
$V$-band.  The use of Wesenheit magnitudes should account for 
what extinction remains.  However, our knowledge of the reddening law
is imperfect, perhaps resulting in systematic errors.  Previous work has
shown that a Galactic value of $R_V = 3.1$ is appropriate for
extragalactic Cepheids \citep{macri01} and this is used in our primary analysis. 
We also fit the Cepheids with $R_V=2.0$ and
2.5 as indicated in Table 6.  The result is an increase in $H_0$ by
0.5 to 1.0 km s$^{-1}$ Mpc$^{-1}$ when the value of $R_V$ for Cepheids is decreased from 3.1
to 2.0.  As an alternative, we fit the Cepheids with only their
$H$-band magnitudes, which increases $H_0$ by 1.6 km s$^{-1}$ Mpc$^{-1}$, indicating that the
differential extinction of the Cepheids in the $H$ band between NGC
4258 and the SN hosts is $\sim$0.04 mag; we think it prudent to
account for this difference using the colors of the Cepheids.

\noindent {\it Short-end limit of Cepheid periods: } Because the
Cepheids were selected at bluer wavelengths, the bias of selecting
brighter Cepheids at
shorter periods due to a magnitude limit does not necessarily apply
to the $H$ band magnitudes.  The dispersion in magnitude
at a given period arising from
the width of the instability strip will be significantly reduced in
the near-IR as we view Cepheids on their Rayleigh-Jeans tail. In
addition, the use of Wesenheit magnitudes mitigates the contribution to the 
selection 
bias due to the color variation on the instability strip.  However, 11 of
the Cepheids used in our primary analysis have periods which are 
shorter than the low period limits determined in Riess et al. (2009) for the onset of optical selection bias and these are indicated in 
Figure 11.  Rejecting these (resulting in the entry in Table 6 with
199 Cepheids) results in an increase in $H_0$ of 0.4 km s$^{-1}$ Mpc$^{-1}$ and a 2\% increase in
its uncertainty.

\noindent{\it Other:} Other significant terms in \citet{freedman01}
include bulk flows, crowding, and zeropoints.  We 
addressed bulk flows in \S 4.1.  Errors due to crowding
were discussed in \S 2.3.  The 
key points regarding crowding are  (1) we correct each Cepheid statistically for crowding bias, and
(2) $H_0$ is only sensitive to a {\it difference} in crowding between
NGC 4258 and the SN hosts and (3) artificial-star
tests indicate that this difference is 
only 0.02 mag in the photometry of the Cepheids, even before a statistical correction is
applied.   To test for any remaining dependence on $H_0$ on the degree of crowding, we analyzed subsets of Cepheids with the least apparent crowding.
We found that truncating the Cepheid sample to the objects in the lower 40\% or 60\% of the crowding bias ($< 0.12$ or $<0.20$ mag) 
results in a reduction in the Hubble constant by 2.3\% and 0.8\%, respectively.  The overall uncertainty in $H_0$ naturally increases as the Cepheid sample is reduced, rising by 25\% when retaining only 40\% of the original sample. Thus we find the net effect on $H_0$ due to crowding is contained within the statistical uncertainties.\footnote{Implicit in this
analysis is that local blending of Cepheids with binary companions or cluster
companions would also cancel between NGC 4258 and the SN hosts.}  This
is an advantage of the use of NGC 4258 over the LMC and the Galaxy, as
this and other difficulties in achieving accurate photometry of
Cepheids (such as the determination of photometric zeropoints) largely
cancel in the determination of $H_0$.

We also consider the effect on $H_0$ of the rejection
of outliers on the Cepheid \PL relations discussed in \S 2.4.  Including the rejected objects naturally has a
severe impact on the value of $\chi^2_{\nu}$ (where $\nu$ is the
number of degrees of freedom), increasing it from 0.84 to 1.38, with
each rejected object contributing an average of $\chi^2=5$.  This variant is
indicated in Table 6 by the increase in the sample of Cepheids from
209 to 240.  The change in $H_0$ is an increase of 3 km s$^{-1}$ Mpc$^{-1}$, the largest change, but still within the
1$\sigma$ of the statistical error.  However, as discussed in \S 2.4,
such outliers are expected and we think it is sensible to reject them
as they may pull the global solution well beyond their merit.
They are included in Table 4 for those who want to consider them
further.  We also considered a less stringent outlier cut of $\pm 1.0$ mag
resulting in the retention of 229 out of 240 Cepheids, increasing $H_0$ by 1.0 km s$^{-1}$ Mpc$^{-1}$ and demonstrating that most of the change in $H_0$ results from a handful of the most extreme of outliers.  

Historically, the determination of $H_0$ through the Cepheid and SN Ia 
distance ladder has been significantly altered by choices made in the analysis,
with different authors making different (if all reasonable) choices
leading to different results.  Thanks to the greater
homogeneity of the data we are using and the smaller number of steps
needed to proceed from a direct geometric distance determination to the
final measurement of $H_0$, we could expect a priori that different choices
would not have a major impact on our results.  We have already shown that
no single variant described above causes a significant change in $H_0$.  However,
to propagate the systematic uncertainty from variants in the analysis and to consider {\it combinations} of analysis variants, we developed a number of plausible scenarios in which different choices are made and the full analysis is completed to determine $H_0$.  
The results are presented in Table 6.  

Although systematic errors are notoriously difficult to quantify, 
our approach is to use the variation in $H_0$ in the previous analyses to determine the systematic error.
The variation in
the inferred value of $H_0$ is relatively small, with a median and
dispersion of 74.2 +/- 1.0 km  s$^{-1}$ Mpc$^{-1}$.
The median is the same as our
primary determination (thus the changes scatter fairly equally between
increases and decreases), and all inferred values lie within a range of
about +/- 3 km/s/Mpc. 
We take the formal dispersion of 1.0 km/s/Mpc as an estimate of the
 {\it systematic} uncertainties in our determination,
which we then add in quadrature to the statistical uncertainty of the
value derived with our preferred approach, yielding a final estimate of
$H_0$=\ho.

\subsection{Anchor Systematics}

The use of NGC 4258 in lieu of
the LMC or the Galaxy as an anchor to the distance ladder provides a significant enhancement to the
precision and accuracy in the measurement of $H_0$.  Indeed, a 3\% 
uncertainty in the distance to NGC 4258 does not even dominate the
current total uncertainty.  The natural advantages of NGC 4258, including the
sample size, period range, and typical metallicity of its Cepheids, and
the ability to measure them in the same way as those in SN~Ia hosts,
provide for extensive use of differential measurements of the Cepheids in the distance ladder and the means to measure $H_0$ to $<$ 5\%.
In the next section we
discuss the use of additional maser hosts which can serve to
test and improve the maser distance estimates.  However, at present
there is only one thoroughly measured system, and use of the LMC or the
Galaxy as an anchors can still provide a test of the
distance scale set by NGC 4258.  

A set of 10 parallax measurements to Galactic Cepheids was recently
obtained by \citet{benedict07} using the Fine Guidance Sensor on {\it
HST}.  Parallax measurements remain the ``gold standard" of distance
measurements, and unlike previous HIPPARCOS measurements, the {\it
individual precision} of this set of measurements is high, averaging
$\sigma=8$\% for each. We have not made use of additional distance
measures to Galactic Cepheids based on the Baade-Wesselink method or
stellar associations as they are much more uncertain than
well-measured parallaxes, and the former appear to be under refinement
due to uncertainties in their projection factors, as discussed by
Fouqu\'e et al. (2007) and \citet{vanleeuwen07}.

Considered as a set, the Cepheids in \citet{benedict07} have an
uncertainty in their mean distance measure of only 2.5\%, comparable
to the precision of the measurement of NGC 4258.  These Galactic
Cepheids also have metallicities which are
very similar to that of Cepheids in the
SN hosts as discussed by \citet{sandage06}.  Using the values of
$\mu_0$ (including corrections for interstellar extinction and
Lutz-Kelker-Hanson bias) and $V,I$-band magnitudes given by
\citet{benedict07}, as well as $H$-band magnitudes compiled by
\citet{groenewegen99}\footnote{For $\eta$ Gem and W Sgr we determined
$H = 2.18 \pm 0.05$ and $2.87 \pm 0.05$, respectively, based on $J$
and $K$ data from \citet{berdnikov96}.}, we determined the absolute
Wesenheit magnitudes of this set of 10 variables. Their \PL relation
is shown in Figure 11.  Their inclusion in the global fit is achieved
by altering equation (7) for the NICMOS Cepheids to be

\bq m_{W,i,j}=\mu_{0,i}+M_W+b_W \ {\rm log} \ P_{i,j}+Z_W \ \Delta
{\rm log[O/H]}_{i,j}, \eq
\noindent
and equation (17) for the SNe Ia to be

\bq m_{v,i}^0=\mu_{0,i}-M_V^0. \eq
\noindent 
Moreover, for the Galactic Cepheids,

\bq M_{W,i,j}=M_W+b_W \ {\rm log} \ P_{i,j}+Z_W \ \Delta {\rm
log[O/H]}_{i,j}, \eq
\noindent
where $M_W$ is the absolute Wesenheit magnitude for a Cepheid with
$P = 1$~d.

The key parameters in the
determination of $H_0$ change from $m^0_{v,4258}$ and $\mu_{0,4258}$ in equation (19) to $M^0_V$,

\bq {\rm log} \ H_0={M_V^0+5a_v+25 \over 5}. \eq
\noindent  
As before, the statistical error in $M_V^0$ includes all
Cepheid-related uncertainties such as the nuisance parameters like the
slope and metallicity relations, and the uncertainty in $H_0$ comes
from the two independent terms $(M_V^0, ~5a_V)$.  The Cepheids in NGC
4258 still contribute to the global analysis as they help determine
the slope of the \PL relation, though their distance estimate is
immaterial to the determination of $H_0$.  We now include a $\sigma=0.04$ mag
uncertainty in the photometry (i.e., zeropoints and relative crowding) between the space-based Cepheid data
and the ground-based Cepheid data.  These analyses are indicated in Table 6
with the scale given as ``MW.'' Compared to the primary analysis based on the
independent distance measurement to NGC 4258, use of the
\citet{benedict07} parallaxes reduces $H_0$ by 0.9 km s$^{-1}$ Mpc$^{-1}$ with an increase in
the uncertainty of 15\%.

However, there are some ``risks'' in taking this route over that based
on the distance to NGC 4258. The magnitudes of these Galactic
Cepheids, unlike the distant Cepheids, suffer little crowding, and so
we must fully rely on the statistical crowding corrections of mean 0.16 mag in \S 2.3 rather than
the more modest difference in the correction
between NGC
4258 and the SN hosts of 0.02 mag.
(We assumed a systematic uncertainty of 0.03
mag for use of the the full corrections.)  Errors along the magnitude scale from
Galactic Cepheids of $<H>\, = 2$ mag to those in SN hosts of 25 mag
pose another risk in this route.  We estimate 0.03 mag systematic
uncertainty for the magnitude scale which is included in the values in
Table 6.  In addition, the mean period of the \citet{benedict07}
Cepheids, $<P> = 10$~d is significantly lower than the $<P> \approx
35$~d in the SN hosts.  The use of the Cepheids in NGC 4258, even
without the use of its distance, provides an empirical bridge across
this period range.  Still, the assumption of the linearity of the
\PL relation, even in the $H$ band and even for a Wesenheit relation,
is another weakness along this route.  We estimate an 0.04 mag
systematic uncertainty from this mismatch in mean periods.  Including
systematics, the total uncertainty in $H_0$ is 5.8\%, only moderately worse than the NGC 4258 route but the result carries more caveats.  Future measurements from GAIA of precise parallaxes
for $\sim 10^3$ Cepheids over a wide range of periods will provide
increased precision while removing the reliance on the form of the
\PL relation to yield a great improvement to the pursuit of $H_0$ {\it if accompanied with more precise calibration of the near-IR magnitude scale}.

The use of the LMC as our anchor for the distance scale carries
similar risks as those discussed for Milky Way Cepheids with two
significant additions: the metallicity of the LMC differs
substantially from the SN hosts and the distance to the LMC is
uncertain at the $>5$\% level.  Nevertheless, the LMC has a long
history of use as an anchor, and for comparison to previous work it is
valuable to again cast the LMC in that role.  We use the set of
$H$-band Cepheid measurements from \citet{persson04} and the optical
measurements of \citet{sebo02} to extract the 53 Cepheids with
measurements of their mean magnitudes in $VIH$. Due to the significant
difference in metallicity between the LMC and the SN hosts of ${\Delta \rm
[O/H]} \approx 0.4$~dex and our lack of constraint on or detection of
a metallicity parameter, we made no metallicity correction.  This
approach is supported by theory, in which the zeropoints of near-IR
\PL relations are found to vary with chemical composition by a
factor of $\sim 3$ less than those of optical zeropoints (Marconi et al. 2005).  Assuming
that for the LMC, $\mu_0=18.42$  mag based on a set of 4 detached eclipsing 
binaries (Fitzpatrick et al. 2003) and with a generous $\pm 0.10$~mag uncertainty to allow for the wide range of estimates for the LMC distance,
we find $H_0=73.3 \pm 4.6$
km s$^{-1}$ Mpc$^{-1}$ as shown in Table 6, in good accord with the previous two 
anchors.

\tabletypesize{\scriptsize}
\begin{deluxetable}{ccccccccccccccc}
\tablenum{6}
\tablecaption{Fits for $H_0$}
\tablehead{\colhead{$\chi^2_{dof}$}&\colhead{\#}&\colhead{$H_0$}&\colhead{$\sigma_{fit}$}&\colhead{$a_v$}&\colhead{$M_V^0$}&\colhead{$\delta M/ \delta{\rm [O/H]}$} &\colhead{b}& \colhead{$z_{min}$}&\colhead{Fit}&\colhead{Scale}&\colhead{PLW}&\colhead{C $R_V$} &\colhead{SNe} & \colhead{SN $R_V$}} 
\startdata
 0.85  &      209 & 74.16(3.41) &  2.56  &  0.698 &  -19.13  & -0.23(0.17) &-3.14(0.10)&  0.023  &       37 & 4258 &   $H_{V,I}$ &  3.1  & UBVRI & 2.5 \nl
 0.84  &      209 & 73.97(3.44) &  2.60  &  0.702 &  -19.16  & -0.22(0.17) &-3.15(0.10)&  0.023  &       20 & 4258 &   $H_{V,I}$ &  3.1  & UBVRI & 3.1 \nl
 0.83  &      209 & 75.12(3.43) &  2.56  &  0.702 &  -19.13  & -0.22(0.17) &-3.12(0.10)&  0.010  &       37 & 4258 &   $H_{V,I}$ &  2.5  & UBVRI & 2.5 \nl
 0.82  &      209 & 75.10(3.46) &  2.61  &  0.707 &  -19.15  & -0.21(0.17) &-3.12(0.10)&  0.010  &       20 & 4258 &   $H_{V,I}$ &  2.5  & UBVRI & 3.1 \nl
 0.85  &      209 & 73.27(3.30) &  2.43  &  0.698 &  -19.16  & ------ &-3.17(0.10)&  0.023  &       37 & 4258 &   $H_{V,I}$ &  3.1  & UBVRI & 2.5 \nl
 1.41  &      240 & 77.24(4.05) &  3.30  &  0.698 &  -19.05  & -0.72(0.20) &-3.10(0.13)&  0.023  &       37 & 4258 &   $H_{V,I}$ &  3.1  & UBVRI & 2.5 \nl
 0.86  &      198 & 74.62(3.49) &  2.64  &  0.698 &  -19.12  & -0.31(0.18) &-3.13(0.11)&  0.023  &       37 & 4258 &   $H_{V,I}$ &  3.1  & UBVRI & 2.5 \nl
 0.85  &      209 & 73.19(3.45) &  2.63  &  0.701 &  -19.18  & -0.24(0.17) &-3.14(0.10)&  0.023  &       61 & 4258 &   $H_{V,I}$ &  3.1  & BVRI & 2.5 \nl
 0.81  &      209 & 74.66(3.37) &  2.49  &  0.692 &  -19.09  & -0.22(0.17) &-3.09(0.10)&  0.023  &       28 & 4258 &   $H_{V,I}$ &  2.0  & UBVRI & 2.0 \nl
 0.84  &      209 & 73.51(3.30) &  2.35  & ---- & ---- & -0.23(0.17) &-3.14(0.10)&  0.023  &       42 & 4258 &   $H_{V,I}$ &  3.1  & UBVRI & ---- \nl
 0.82  &      209 & 73.79(3.29) &  2.33  & ---- & ---- & -0.23(0.17) &-3.12(0.10)&  0.023  &       42 & 4258 &   $H_{V,I}$ &  2.5  & UBVRI & ---- \nl
 0.84  &      209 & 74.84(3.37) &  2.40  & ---- & ---- & -0.23(0.17) &-3.15(0.10)&  0.010  &       42 & 4258 &   $H_{V,I}$ &  3.1  & UBVRI & ---- \nl
 0.79  &      209 & 75.64(3.41) &  2.52  &  0.698 &  -19.09  & -0.19(0.17) &-3.01(0.10)&  0.023  &       37 & 4258 &   $H$ &  3.1  & UBVRI & 2.5 \nl
 0.79  &      209 & 74.88(3.31) &  2.39  &  0.698 &  -19.11  & ------ &-3.03(0.10)&  0.023  &       37 & 4258 &   $H$ &  3.1  & UBVRI & 2.5 \nl
 0.83  &      209 & 73.20(3.47) &  2.65  &  0.690 &  -19.13  & -0.19(0.17) &-3.15(0.10)&  0.023  &       26 & 4258 &   $H_{V,I}$ &  3.1  & UBVRI & 3.1 \nl
 0.84  &      209 & 74.05(3.50) &  2.60  &  0.699 &  -19.15  & -0.22(0.17) &-3.15(0.10)&  0.023  &       27 & 4258 &   $H_{V,I}$ &  3.1  & UBVRI & 3.1 \nl
 0.85  &      209 & 74.10(3.38) &  2.53  &  0.692 &  -19.11  & -0.24(0.17) &-3.14(0.10)&  0.023  &       28 & 4258 &   $H_{V,I}$ &  3.1  & UBVRI & 2.0 \nl
 0.85  &      209 & 74.11(3.36) &  2.50  &  0.687 &  -19.08  & -0.25(0.17) &-3.14(0.10)&  0.023  &       29 & 4258 &   $H_{V,I}$ &  3.1  & UBVRI & 1.5 \nl
 0.84  &      219 & 74.91(4.13) &  4.11  &  0.698 &  -19.11  & -0.21(0.17) &-3.20(0.09)&  0.023  &       37 & MW &   $H_{V,I}$ &  3.1  & UBVRI & 2.5 \nl
 0.85  &      219 & 73.68(3.93) &  3.91  &  0.698 &  -19.15  & ------ &-3.22(0.09)&  0.023  &       37 & MW &   $H_{V,I}$ &  3.1  & UBVRI & 2.5 \nl
 0.84  &      219 & 74.68(4.15) &  4.13  &  0.702 &  -19.14  & -0.20(0.17) &-3.21(0.09)&  0.023  &       20 & MW &   $H_{V,I}$ &  3.1  & UBVRI & 3.1 \nl
 0.78  &      219 & 74.56(3.97) &  3.95  &  0.698 &  -19.12  & -0.18(0.17) &-3.06(0.09)&  0.023  &       37 & MW &   $H$ &  2.5  & UBVRI & 2.5 \nl
 1.01  &      262 & 73.27(4.57) &  2.71  &  0.698 &  -19.16  & ------ &-3.17(0.04)&  0.023  &       37 & LMC &   $H_{V,I}$ &  2.5  & UBVRI & 2.5 \nl
\enddata
\end{deluxetable}

In summary, we find that a full propagation of statistical error and
the inclusion of the systematic error gives $H_0 =$ \ho, based on the
cleanest route through NGC 4258, but also consistent with independent
though riskier distance-scale anchors from Milky Way Cepheid parallaxes and the LMC.

\subsection{Error Budget}

As discussed in \S 4, our total error is the sum of the uncertainty
in the three measured terms on the right-hand side of equation (19) and
the systematic error derived from considering alternatives to the
primary analysis.  To illuminate how error propagates along our (and
other) distance ladders, we itemize the contributions in Table 7.

The first term is the distance precision of the anchor, followed by
the mean of its set of Cepheids (i.e., the zeropoint of its \PL
relation).  The next term is the mean of the set of Cepheids in each
SN host and the precision of a single SN, each divided by the number
($n$) of hosts.  For this calculation $n=6$.  Next is the uncertainty
in the SN Ia apparent magnitude vs. $z$ relation; SNe~Ia in the Hubble
flow now provide a Hubble diagram with 240 published SNe~Ia out to $z
\approx 0.1$, yielding an uncertainty of $0.5$\% Hicken et al. (2009).  The
next term arises from the uncertainty in the difference between the
photometric calibration used to observe Cepheids in the anchor and in
the SN hosts in two or more passbands.  These photometric calibration
errors are then amplified by the need to deredden Cepheids with a
reddening law, $R$, of size 2.1 and 0.48 for $VI$ and $VIH$ Cepheid measurements, respectively.  The next two terms arise from the difference in
the mean metallicities and the mean periods of the Cepheids in the
anchor and hosts, and the uncertainty in their respective correlation
with Cepheid luminosity.  The last term contains the uncertainty from
the photometric anomalies of WFPC2, charge transfer efficiency (CTE),
and the ``long versus short effect.'' (Holtzman et al. 1995).

\begin{table}[h]
\begin{small}
\begin{center}
\vspace{0.4cm}
\begin{tabular}{llll}
\multicolumn{4}{c}{Table 7: Error Budget for $H_0$ for Cepheid and SN Ia Distance  Ladders} \\
\hline
\hline
Term & Description &  Previous & Here  \\
\hline
$\sigma_{\rm anchor}$  &   Anchor distance  &  5\%  & 3\% \\
$\sigma_{{\rm anchor}-PL}$  &  Mean of \PL in anchor & 2.5\%  &  1.5\% \\
$\sigma_{{\rm host}-PL}/\sqrt{n}$  &  Mean of \PL values in SN hosts  & 1.5\% & 1.5\% \\
$\sigma_{\rm SN}/\sqrt{n}$  &  Mean of SN Ia calibrators &  2.5\% & 2.5\% \\
$\sigma_{mag-z}$  &  SN Ia $m-z$ relation & 1\% & 0.5\%  \\
$R \sigma_{\lambda,1,2}$  & Cepheid reddening, zeropoints, anchor-to-hosts & 4.5\% & 0.3\% \\
$\sigma_{Z}$ & Cepheid metallicity, anchor-to-hosts  & 3\% & 0.8\%    \\
$\sigma_{\rm PL}$ & \PL slope, $\Delta$ log $P$, anchor-to-hosts & 4\% & 0.5\%  \\
$\sigma_{\rm WFPC2}$ & WFPC2 CTE, long-short & 3\% & 0\% \\
\hline
Total, $\sigma_{H_0}$  &             &     10\%  &  \unc \\
\hline
 \hline
\end{tabular}
\end{center}
\end{small} 
\end{table}

The reduction in total uncertainty in $\sigma_{H_0}$ from 10\% to
5\% is a consequence of a number of improvements along the ladder.
Most come from greater homogeneity in zeropoints, metallicity, and
periods of the samples of Cepheids collected in the anchor and the SN
hosts.  Changing from the optical to the near-IR reduces the reddening
term, $R$, by a factor of 5.  NGC 4258 also provides greater distance
precision than the LMC, and a larger sample of long-period Cepheids.
The recent increase in the sample of SNe Ia at $0.01<z<0.1$ (Hicken et al. 2009) provides a modest improvement.

\section{Dark Energy}

An independent measurement of $H_0$ is a powerful complement to the
measurement of the cosmological term $\Omega_M H_0^2$ derived from the
power spectrum of the CMB.  In the context of a flat Universe, the
fractional uncertainty in the value of an (assumed constant) equation-of-state
parameter ($w$) of dark energy is approximately twice the fractional uncertainty in
$H_0$ ($\sigma_w \approx 2 \sigma_{H_0}$), as long as the fractional
uncertainty in $H_0$ is greater than or equal to that in $\Omega_M
H_0^2$ \citep{hu05}.  A marked improvement in the precision of
$\Omega_M H_0^2$ has been realized in the recent 5-year WMAP analysis
from the localization of the third acoustic peak \citep{komatsu08}.  The
result is a model-insensitive measurement of $\Omega_M H_0^2$ to
better than 5\% precision.  

Using the output of the WMAP 5-year Monte Carlo Markov Chain (MCMC)
from \citet{komatsu08}\footnote {\tt http://lambda.gsfc.nasa.gov/ .}
in a flat, $w$CDM cosmology (i.e., dark energy with constant $w$) 
 yields the degenerate confidence regions
in the $H_0-w$ plane shown in Figure 14.  Combined with our measurement of
$H_0$ we find $w = -1.12 \pm 0.12$, a value
consistent with a cosmological constant ($\Lambda$).  This result is
similar in value and precision to those found from the combination of
baryon acoustic oscillations (BAO) and high-redshift SNe~Ia
\citep{woodvasey07,astier06}.  The important difference from the prior
measurements is that it is independent of the systematic uncertainties
associated with the use of high-redshift SNe~Ia.  Since such
measurements are now dominated by their systematic errors (Wood-Vasey et al. 2007, Kessler et al. 2009, Hicken et al. 2009. Kowalski et al. 2008), independent
measurements are a route to progress.  For comparison, the combination
of the WMAP and BAO data alone gives $w=-1.15 \pm 0.22$ and that
from WMAP and the \citet{freedman01} measurement of $H_0$
yields $w=-1.01 \pm 0.23$.  

The $H_0$+WMAP measurement of $w$ is
quite insensitive to the effect of $w$ on the determination of $a_v$  because the mean
redshift of the Hubble flow sample is only $z=0.04$.  Specifically, the
change in $H_0$ for a change in $w$ of 0.1 (evaluated at $z=0.04$) is only 0.2\%, far less
than the total \unc\ uncertainty in $H_0$ and justifying our use of 
a kinematic expansion to determine $a_v$.  The very mild degeneracy between $a_v$ and $w$ is shown (as a tilt) in Figure 14.

However, fitting a cosmological model with the assumption of a {\it
constant} equation of state (EOS) is itself limiting to the investigation of
dark energy.  It obscures our ability to detect evolution of $w$, an
important test of the presence of a cosmological constant. An
alternative approach is to use a variant of principal-component
analysis \citep{huterer03,huterer05} to extract discrete, decorrelated
estimates of $w(z)$, binned in redshift. This method was used by
\citet{riess07} and \citet{sullivan07} to constrain multiple independent measures of
$w(z)$.  With the improved
constraint on $H_0$, we can use this approach to determine the effect
on the constraints on the components of $w(z)$.  In the following we
employ the implementation of the component analysis from
\citet{sarkar08a} using $N+3$ free parameters in the MCMC
corresponding to $H_0$, $\Omega_m$, $\Omega_K$, and the $N$
independent estimates of $w$.

\begin{table}[!tbh]
\tablenum{8}
\begin{center}
\caption{Decorrelated Estimates of $w$ from Available Datasets (68\% Uncertainty)}
\begin{tabular}{cr|ccc}
\hline
Dataset         & Prior on   & $w_1$ & $w_2$   & $w_3$   \\
 used           &   $H_0$    & z=[0-0.2] & z=[0.2-0.5] & z=[0.5-1.8] \\
\hline \hline
192 SNe + 2 BAO & $72\pm8.0$ &  $-0.976^{+0.142}_{-0.162}$  &  $-0.944^{+0.230}_{-0.235}$   & $-0.471^{+0.327}_{-1.515}$   \\
                & $74\pm 3.5$ &  $-0.940^{+0.102}_{-0.139}$  &  $-0.948^{+0.175}_{-0.160}$   & $-0.692^{+0.301}_{-0.759}$   \\
\hline \hline
\end{tabular}
\end{center}
\end{table}

\subsection{Current Data}

We first examine how the errors on $w(z)$ improve from using the improved constraint on $H_0$. 
For this we use the \citet{davis07} compilation of 192 SNe, 2 BAO
estimates from \citet{percival07}, and the WMAP 5-year constraint
\citep{komatsu08} on the distance to the last-scattering surface
($R_{\rm CMB}$) in the $H_0$-independent form. We also use the WMAP 5-year
constraint on $\Omega_mh^2$ \citep{komatsu08} and allow curvature to
be free. We use the publicly available {\em wzBinned}\footnote{\tt
http://dsarkar.org/code.html .} code and analyze the data using an MCMC
likelihood approach to estimate $w(z)$ in each redshift bin. We take a
total of 3 bins between $z=0$ and $z=1.8$ (see Table 8 for
the redshift ranges) and assume that dark energy at $z>2$ was subdominant by 
fixing $w$ to a constant value of $-1$
from that redshift to the last-scattering surface ($z=1089$). 

We analyze the data using the value of $H_0$ from both {\it HST}
Cepheids \citep{freedman01} and the present work. Our results are
summarized in Table 8. Using the new constraints on the
Hubble constant we get a significant improvement on the 1$\sigma$
errors of the EOS parameters. The improvement in the inverse product of the uncertainties in $w(z)$, widely
referred to as a dark energy "figure of merit", improves by a factor of 3
due to the increased precision in $H_0$, a result of the 
degeneracy between $w$ and $H_0$.  The data remain
consistent with $\Lambda$ within 1$\sigma$ with $[w_1,w_2,w_3]=[-0.940^{+0.102}_{-0.139}, -0.948^{+0.175}_{-0.160},-0.692^{+0.301}_{-0.759}]$ for the ranges $z=[0-0.2],[0.2-0.5],[0.5-1.8]$, respectively.  The data continue to indicate the presence of a dark energy component (i.e., $w<0$) when it was a sub-dominant part of the Universe, in agreement with Riess et al. (2007) (see also Kowalski et al. 2008).

\subsection{Future Surveys}

We now
consider the constraints on $w(z)$ from future surveys in three different scenarios under frequent consideration:

Case (1): An aggressive set of 17 BAO distance measurements. This includes 2 BAO estimates
(as before) at z = 0.2 and z = 0.35, with 6\% and 4.7\% uncertainties,
respectively \citep{percival07}; 5 BAO constraints at
$z=[0.6,0.8,1.0,1.2,3.0]$ from SDSS III and HETDEX with 
respective precisions of $[1.9,1.5,1.0,0.9,0.6]\%$ \citep[scenario V5N5]{seo03}; and
10 BAO estimates from a space mission with precisions of
$[0.36,0.33,0.34,0.33,0.31,0.33,0.32,0.35,0.37,0.37]\%$ from $z =
1.05$ to $1.95$ in steps of $0.05$.

Case (2): An aggressive SN Ia data set of 2300 SNe with 300 SNe uniformly
distributed out to $z = 0.1$, as expected from ground-based low-redshift
samples, and an additional 2000 SNe uniformly distributed in the range
$0.1 < z < 1.7$, as expected from future space mission
\citep{kim04}. We bin the Hubble diagram into 32 redshift
bins (corresponding to a width of the relevant redshift bin of $\Delta
z = 0.05$). The error in the distance modulus for each SN bin is given
by
$\sigma_m = ((\sigma_{\rm int}/N_{\rm bin}^{1/2})^2+\delta
m^2)^{1/2}$, 
where $\sigma_{\rm int} = 0.1$ mag is the intrinsic error for each SN,
$N_{\rm bin}$ is the number of SNe in the redshift bin, and $\delta m$
is the irreducible systematic error.  We take the systematic error to
have the form $\delta m=0.02(0.1/\Delta z)^{1/2} (1.7/z_{\rm max}) (1
+ z)/2.7$, where $z_{\rm max}$ is the redshift of the most distant SNe.
This is equivalent to the form in \citet{linder03}. In generating the
SN catalog, we do not include the effect of gravitational lensing, as
it is expected to be small \citep{sarkar08b} and should not affect our
results much.

Case (3): A combination of the above: 2300 SNe and 17 BAO
estimates.

\begin{table}[!t]
\tablenum{9}
\begin{center}
\caption{$68\%$ Error in the Decorrelated Binned Estimates of $w$ from
Upcoming Surveys}
\begin{tabular}{cr|cccccc}
\hline
Mocks & H$_0$  &$\Delta w_1$&$\Delta w_2$ & $\Delta w_3$ & $\Delta w_4$& $\Delta w_5$ & FoM\\
   used & $74\pm$   & z=[0-0.07]   & z=[0.07-0.15] & z=[0.15-0.30]  & z=[0.3-0.6]   & z=[0.6-1.2] & ($\times 10^4$) \\
\hline \hline
 17 BAO & $8.0$   & 0.549 & 0.462 & 0.323 & 0.202 & 0.158    & 0.038 \\
       & $6.0$       & 0.389 & 0.374 & 0.255 & 0.196 & 0.166 & 0.083 \\
       & $4.0$       & 0.342 & 0.340 & 0.238 & 0.174 & 0.150 & 0.138 \\
       & $3.5$       & 0.331 & 0.329 & 0.224 & 0.163 & 0.143 & 0.176 \\
       & $2.0$       & 0.203 & 0.203 & 0.144 & 0.118 & 0.131 & 1.090 \\
       & $1.0$       & 0.130 & 0.134 & 0.096 & 0.081 & 0.093 & 7.938 \\
\hline \hline
 2300 SNe & $8.0$ & 0.128 & 0.137 & 0.162 & 0.308  & 7.999    & 0.014  \\
         & $6.0$     & 0.127 & 0.132 & 0.156  & 0.294 & 8.799 & 0.015  \\
         & $4.0$     & 0.105 & 0.098 & 0.105  & 0.193 & 1.622 & 0.296  \\
         & $3.5$     & 0.098 & 0.085 & 0.088  & 0.145 & 1.334 & 0.705  \\
         & $2.0$     & 0.091 & 0.070 & 0.064  & 0.083 & 0.291 & 10.16  \\
         & $1.0$     & 0.078 & 0.052 & 0.043  & 0.048 & 0.124 & 96.33  \\
\hline \hline
 2300 SNe & $8.0$ & 0.064 & 0.054  & 0.045  & 0.048   & 0.104    & 129 \\
       + & $6.0$     & 0.063  & 0.051  &0.042  & 0.045   & 0.099 & 166 \\
 17 BAO  & $4.0$     & 0.063  & 0.049 & 0.041  &  0.043  & 0.097 & 189 \\
         & $3.5$     & 0.062  & 0.049 & 0.041  &  0.043  & 0.092 & 203 \\
         & $2.0$     & 0.061  & 0.047 & 0.038  & 0.039   & 0.086 & 274 \\
         & $1.0$     & 0.057  & 0.043 & 0.034  & 0.034   & 0.065 & 543 \\
\hline \hline
\end{tabular}
\end{center}

\label{table:9}
\end{table}

For each of the above-mentioned scenarios, we also use the WMAP 5-year
constraint on $R_{\rm CMB}$ \citep{komatsu08}. Since we are
considering future surveys, we marginalize over $\Omega_m$ prior
obtained from the Planck prior on $\Omega_mh^2$ and different priors
on $H_0$ (see Table 9 for details). As before, we allow the curvature
to be free. In this case, we take a total of 6 bins between $z=0$
and $z=2$ (see Table 9 for the redshift ranges) and fix $w(2<z<1089)=-1$. 
The sixth bin (extending from $z=1.2$ to
$z=2.0$) is suppressed as it is not well constrained. 

Table 9 and Figure 15 summarize our results. A
significant improvement of the $68\%$ error in the decorrelated binned
estimates of $w$ is apparent as we make use of better constraints on
the Hubble constant. 

Further improvement in the measurement
of $H_0$ should allow for the measurement of a fourth independent parameter of the
EOS to an accuracy better than 10\%, even without making use of any
BAO estimate. A combination of next-generation surveys will most
likely be able to measure five independent parameters of the EOS to
better than 10\% accuracy.

An alternative use of a precise measurement of $H_0$ is as an
``end-to-end" test of the best constraints on the cosmological model
from all other data.  As shown in Table 1, the combination of
measurements from WMAP, BAO, and high-redshift SNe~Ia, together with
the assumption of a constant value for $w$, predict $H_0$ to greater
precision than measured here.  This prediction is in
good agreement with our measurement, but belies tension between the
predictions of $H_0$ from BAO and high-redshift SNe~Ia.  Either of
these combined with WMAP results in a 3$\sigma$ difference in their
prediction of $H_0$.  Although our present measurement lies between
these two combinations, it is closer to BAO and inconsistent with WMAP
and high-redshift SNe~Ia at the 2.8$\sigma$ confidence level.
Improvements in all datasets should reveal whether this tension
results from systematic error or is indicative of the need for a
more complex description of dark energy.

\section{Discussion}

Ever more precise measurements of the Hubble constant can contribute
to the determination of the even more elusive nature of dark energy.
The Planck CMB mission is expected to measure $\Omega_M H_0^2$ to 1\%.
A complementary goal would be to reach the same for $H_0$.  We show in
Figure 15 that a measurement of $H_0$ approaching 1\% would be
competitive with ``next generation" measurements of BAO and
high-redshift SNe Ia (Kolb et al. 2006) for constraining the evolution
of $w$, and could supplant either tool should they encounter
insurmountable systematic errors before reaching their goals.
Attempts to explain accelerated expansion without dark energy by
an unexpected failure of the cosmological principle also benefit
from improved measurements of $H_0$.  For example, an approach by \citep{wilt07} in this vein predicts $H_0=62 \pm 2$ which is already inconsistent
with the present measurement at the 3 $\sigma$ confidence level.

How realistic is a measurement of $H_0$ to 1\%? In most respects, the
measurement of $H_0$ to 1\% is no more ambitious than the plans to
push high-redshift SN Ia measurements to their next level of
precision.  Indeed, the dominant sources of systematic uncertainty in
measuring distant SNe Ia do not pertain to $H_0$ as they result from
large redshifts: cross-filter, cross-detector flux calibration,
$K$-corrections, and evolution of SNe~Ia and dust over large changes in redshift \citep{woodvasey07}.

Following Table 7, we consider the two biggest challenges to a 1\%
measurement of $H_0$: the precision of the distance measurement of the
anchor and the size of the calibrator sample of SNe Ia.  The other
terms are near or below 1\% and can be reduced with the collection of more data.

Further improvements in the distance measurement to NGC 4258 requires understanding
and modeling additional complexity in its inner disk, including
eccentricity and the presence of spiral structure (Humphreys et al. 2008, 2009).  We expect progress
with future work, though 1\% or better  would be challenging.
With the present route it may be possible to measure $H_0$ to 2\% or 3\%.

More maser hosts of comparable quality could further reduce the uncertainty in
the anchor through averaging.  The Maser Cosmology Project
\citep[MCP;][]{braatz08} is a large project at NRAO with the goal of
measuring 10 more hosts in the next 5 years (Greenhill et al. 2009).  Of the 112 extragalactic
maser galaxies now known, 30\% show the required high-velocity
features on their limbs and 10\% are disks and are good candidates for
distance measurements.  Two of these, UGC 3789 and NGC 6323, have
already yielded initial distance estimates with 15\% uncertainty and
which, combined with their redshifts, are consistent with the value of
$H_0$ inferred here (Braatz 2008, private comm.; Lo 2008, private
comm.).  Reaching 1\% will require the 10 new MCP maser
hosts to each be measured to the 3\% uncertainty of NGC 4258 (Greenhill et al. 2009),
or some other combination of number of systems and individual precision.
Considering that the majority of maser hosts have been found in the
last 5 years, there is reason for optimism in the future.  
If such a sample of maser hosts is collected, it would then be necessary to correct their recession velocities for peculiar and coherent flows (Hui et al. 2006) to a mean of 1\% or observe their Cepheids to tie their distance scale to the 0.5\% calibration of the Hubble flow from SNe Ia.

Another promising route is offered by GAIA which should collect a few hundred
high-precision parallax measurements for long-period Cepheids in the
Galaxy.  The resulting \PL relations would be more than sufficient
to support a 1\% measurement of $H_0$.  However, the comparison of
bright Galactic Cepheids and faint ones in SN hosts raises the
challenge of measuring fluxes over a range of 20 mag to better than 1\%
precision.  Though formidable, this appears easier than the challenge
facing future high-redshift SN Ia studies because the Cepheid
measurements may all be obtained at the same wavelengths. Accounting
for the difference in crowding between Galactic and extragalactic
Cepheids is also a concern.

The size of the sample of reliable SNe Ia close enough to resolve
Cepheids in their hosts, those within $\sim$ 30 Mpc, presently limits
the determination of their mean fiducial luminosity to 2.5\%.  At least
30 SNe~Ia are needed in this sample and at a rate of $\sim$1 new
object appearing every 3 yr we cannot wait on Nature.  A factor of 2
increase in distance (factor of 8 in volume), and hence 1.5 mag in the
range of resolving Cepheids, is needed.  Ultra-long-period Cepheids
with $80 < P < 180$~d \citep{bird08} and $M_V=-7$~mag are $\sim
2$~mag brighter than the typical, $P = 30$~d Cepheids observed in SN
hosts.  Though these Cepheids appear to obey different \PL relations
than their shorter-period brethren and are rare, their use when
intercompared between galaxies is promising (Bird et al. 2008, Riess et al. 2009).
JWST is expected to routinely resolve Cepheids at $\sim 50$~Mpc and could be
enlisted to help measure $H_0$ to 1\%.

\section{Summary and Conclusions}

(1) We have observed 240 long-period Cepheids in 6 SN Ia hosts and NGC
    4258 using NICMOS in $F160W$.

(2) Unprecedented homogeneity in the periods and metallicity in the
    use of these Cepheids along the distance ladder greatly reduces
    systematic uncertainties

(3) Use of the same telescope, instrument, and filters for all Cepheids
    also reduces the systematic uncertainty related to flux
    calibration.

(4) Our primary analysis gives $H_0 =$ \ho\ .

(5) Alternative analyses using different extinction laws and
    extinction distributions yield consistent results.

\bigskip 
\medskip 

{\bf We are grateful to William Januszewski, William Workman, Neil Reid, Howard Bond, Louis Bergeron, Rodger Doxsey, Craig Wheeler,  Malcolm Hicken, Robert Kirshner, Peter Challis, Elizabeth Humphreys, Lincoln Greenhill, and Ken Sembach for their help in realizing this measurement.  }

Financial support for this work was provided by NASA through programs
GO-9352, GO-9728, GO-10189, GO-10339, GO-10497, and GO-10802 from the Space
Telescope Science Institute, which is operated by AURA, Inc., under
NASA contract NAS 5-26555. A.V.F.'s supernova group at U.C. Berkeley
is also supported by NSF grant AST--0607485 and by the TABASGO
Foundation.

\vfill
\eject

{\bf Figure Captions}

Figure 1: Optical images of SNe Ia near peak (see Figures 2 to 7 for orientations and scales).  These images show the objects used to calibrate
the SN Ia fiducial luminosity.  The images were obtained with CCDs.  The exception is SN 1981B which was observed photoelectrically and with the Texas Griboval electrographic camera (image shown here) which has better sensitivity and linearity that photographic plates.

Figure 2:  {\it HST} ACS $F555W$ image of NGC 3370.  The positions of Cepheids with periods in the range $P>60$ ~d, $30<P<60$ ~d and $10<P<30$ ~d are indicated by red, blue, and green circles, respectively.  A yellow circle indicates the position of the host's SN Ia.  The orientation is indicated by the compass rose whose vectors have lengths of 15$\arcsec$.  The fields of view for the NIC2 follow-up fields in Table 2 are indicated.  

Figure 3.  As Figure 2 for NGC 1309.

Figure 4:  As Figure 2 for NGC 3021.

Figure 5:  As Figure 2 for NGC 4639

Figure 6:  As Figure 2 for NGC 4536, image from {\it HST} WFPC2.

Figure 7:  As Figure 2 for NGC 3982

Figure 8:  Example of scene modeling for the arcsec surrounding each Cepheid in one NIC2 field, NGC3370-GREEN.  For each Cepheid, the stamp on the left shows the region around the Cepheid,  the middle stamp shows the model of the stellar sources, and the right stamp is the residual of the image minus the model.  The position of the Cepheid as determined from the optical data is indicated by the circle.  

Figure 9:  Example of the artificial stars tests in the region around a Cepheid in the field NGC3021-GREEN with $P=82.0$ days.       A thousand artificial stars of the brightness of the Cepheid (as determined from its period)  are randomly added to the image.  The magnitudes of the artificial stars are measured at their known positions (in the same way as the Cepheids).  The difference between the input and measured star magnitudes (i.e., the bias) is shown as a function of the displacement between the injected position and the centroid of the star found nearest this position.  The photometric bias (brighter) increases with the displacement, a direct consequence of blending.   For displacements beyond a pixel, the recovered star is no longer the same as the one injected and the relation between bias and displacement dissipates. Averages and dispersions in bins of the displacement are indicated by the filled dots. For an individual Cepheid, the displacement between the NICMOS and optical position is used to predict and correct for the bias as shown in the vertical dotted line.  The uncertainty is derived from the dispersion of the artificial stars.  

Figure 10:  Example NIC2 fields for the anchor galaxy, NGC 4258 (NIC-POS3, right), and a SN field, NGC3370-GREEN (left).  The Cepheid positions are indicated.  The artificial star tests show that the mean photometric bias in the Cepheid magnitudes due to blending is very similar for these fields (0.14 mag for the anchor and 0.16 mag for the SN host), not surprising from the apparent similarity of their stellar density.  Thus, the {\it difference} in Cepheid magnitudes in these hosts, the quantity used to construct the distance ladder, is quite insensitive to blending.  

Figure 11:  Near-infrared Cepheid period-luminosity relations.  For the 6 SN Ia hosts and the distance-scale anchor, NGC 4258, the Cepheid magnitudes are from the same instrument and filter combination, NIC2 $F160W$.  This uniformity allows for a significant reduction in systematic error when utlizing the difference in these relations along the distance ladder.  The measured metallicity for all the Cepheids is solar-like (12+log [O/H] $\sim$ 8.9).  A single slope has been fit to the relations and is shown as the solid line.  10\% of the objects were outliers from the relations (open diamonds) and are flagged as such for the subsequent analysis.  Filled points with asterisks indicate Cepheids whose periods are shorter than the incompleteness limit identified from their optical detection.  The lower right panel shows the near-IR \PL relation derived from 10 Milky Way Cepheids with precise, individual parallax measurements from Benedict et al. (2007).  

Figure 12: The magnitude-redshift relation of nearby ($z<0.1$) SNe~Ia.
The term $m_V^0$ is the peak apparent magnitude in $V$ corrected for
extinction and to the fiducial luminosity using a light-curve fitter.
The intercept of the linear relation defines the term $a_v = {\rm log}
\ cz - 0.2m_V^0$ used for the determination of $H_0$.  SNe Ia with
redshifts in the range $z>0.01$ or $z>0.0233$ are used in the
analysis.

Figure 13:   Relative distances from Cepheids and SNe Ia.  The x-axis (bottom) shows the peak apparent visual magnitude of each SN Ia (red points) corrected for reddening and to the fiducial brightness (using the luminosity-light curve shape relations), $m_V^0$.  The upper x-axis includes the intercept of the $m_V^0$-log cz relation for SNe Ia, $a_v$ to provide SN Ia distance measures, $m_V^0+5a_v$, which are independent of the light curve shape relations.  The y-axis (right), shows the relative distances between the hosts determined from the Cepheid $VIH$ Wesenheit relations.  The left y-axis shows the same with the addition of the independent geometric distance to NGC 4258 (blue point) based on its circumnuclear masers.  The contribution of the nearby SNe Ia and Cepheid data to $H_0$ can be expressed as a determination of $m_{V,4258}^0$, the theoretical mean of 6 fiducial SNe Ia in NGC 4258.  

Figure 14: Confidence regions in the plane of $H_0$ and the equation of state of dark energy, $w$.  The localization of the 3rd acoustic peak in the WMAP 5 year data (Komatsu et al. 2008) produces a confidence region which is narrow but highly degenerate in this space.  The improved measurement of $H_0$,  \ho, from the SHOES program is complementary to the WMAP constraint resulting in a determination of $w=-1.12 \pm 0.12$ for a constant equation of state.  This result is comparable in precision to determinations of $w$ from baryon acoustic oscillations and high-redshift SNe Ia, but is independent of both.  The inner regions are 68\% confidence and the outer regions are 95\% confidence.  The modest tilt of the SHOES measurement of 0.2\% in $H_0$ for a change in $w$=0.1 results from the mild dependence of $a_v$ on $w$ at the mean $z=0.04$.  The measurement of $H_0$ is made at $j_0=1$ (i.e., $w=-1$).

Figure 15:  Projected constraints on 5 principal components of $w(z)$ as a function of the future precision of $H_0$.  Three future scenarios are considered: an aggressive BAO experiment (black), an aggressive high-z SN Ia experiment (red), or both (blue) along with  a Planck-based prior on $\Omega_mh^2$.  Panels 1 to 5 show the expected constraints in different redshift ranges.  Panel 6 shows a figure of merit, the inverse product of the uncertainties of the 5 components.  As seen, a $\sim$ 1\% measurement of $H_0$ can compensate for either BAO of high-z SNe Ia being limited by systematic errors or can aid their joint use.

\begin{figure}[ht]
\vspace*{140mm}
\figurenum{1}
\includegraphics{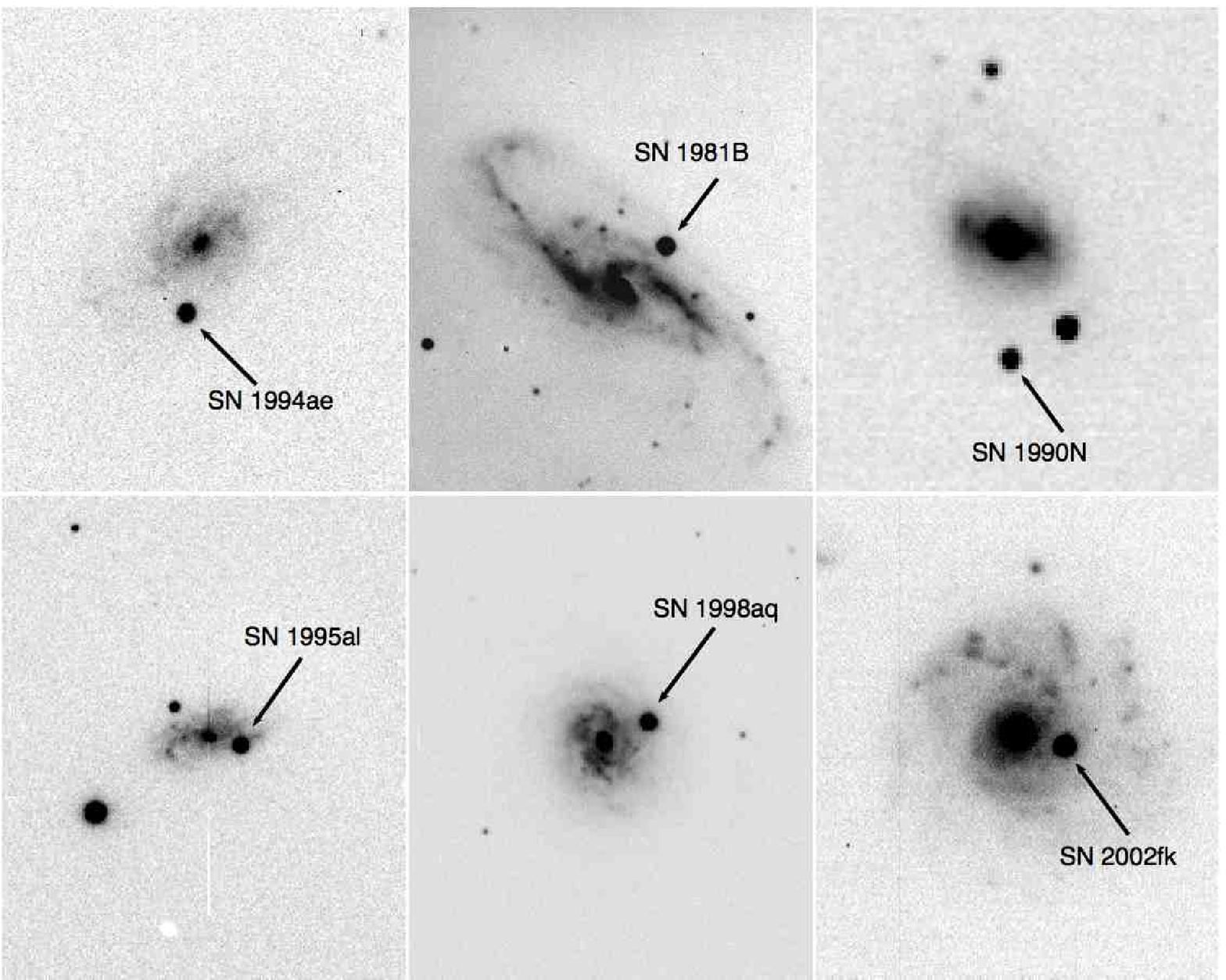}
\caption { }
\end{figure}

\begin{figure}[ht]
\vspace*{140mm}
\figurenum{2}
\includegraphics{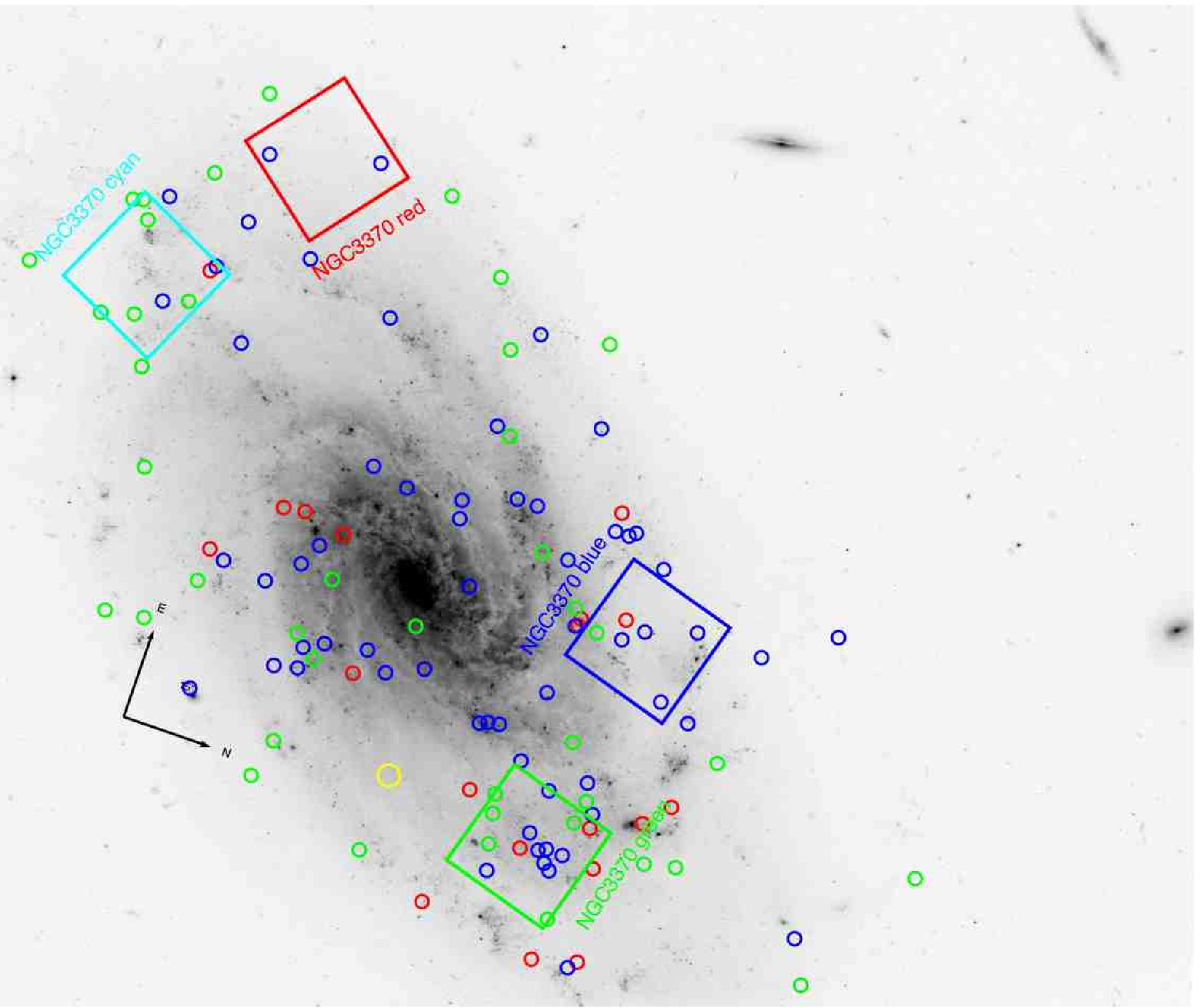}
\caption { }
\end{figure}

\begin{figure}[ht]
\vspace*{140mm}
\figurenum{3}
\includegraphics{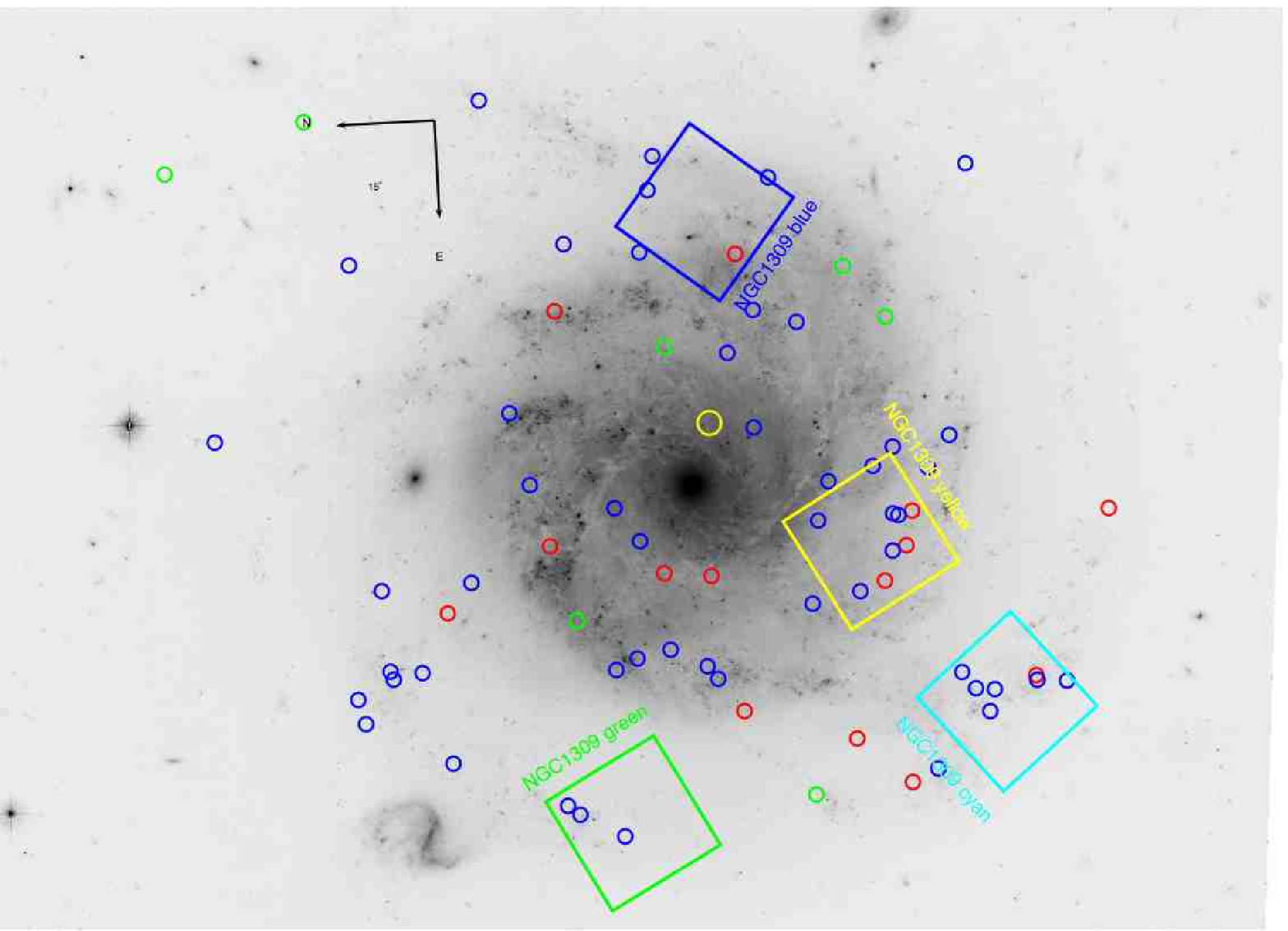}
\caption { }
\end{figure}

\begin{figure}[ht]
\vspace*{140mm}
\figurenum{4}
\includegraphics{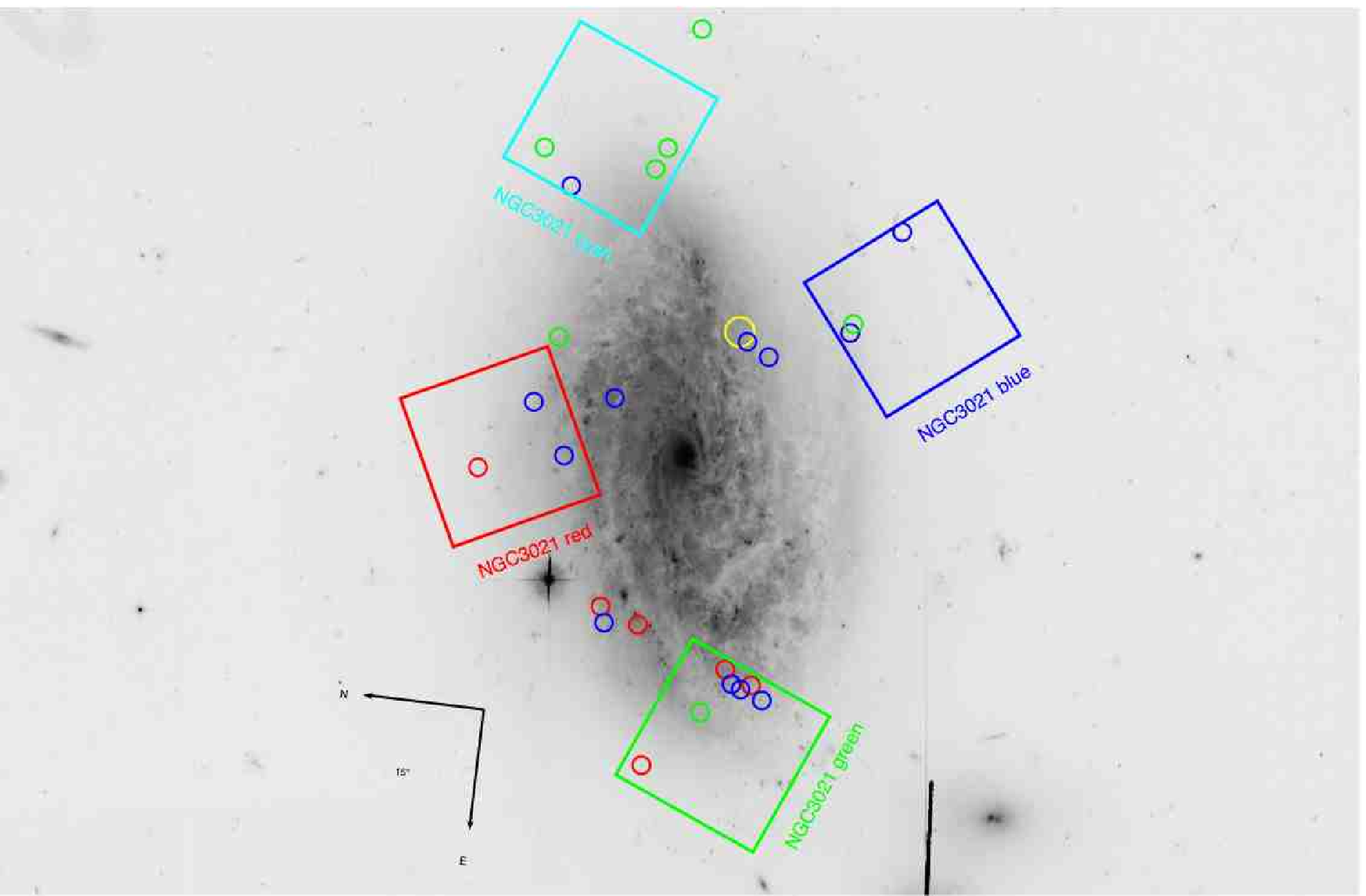}
\caption { }
\end{figure}

\begin{figure}[ht]
\vspace*{140mm}
\figurenum{5}
\includegraphics{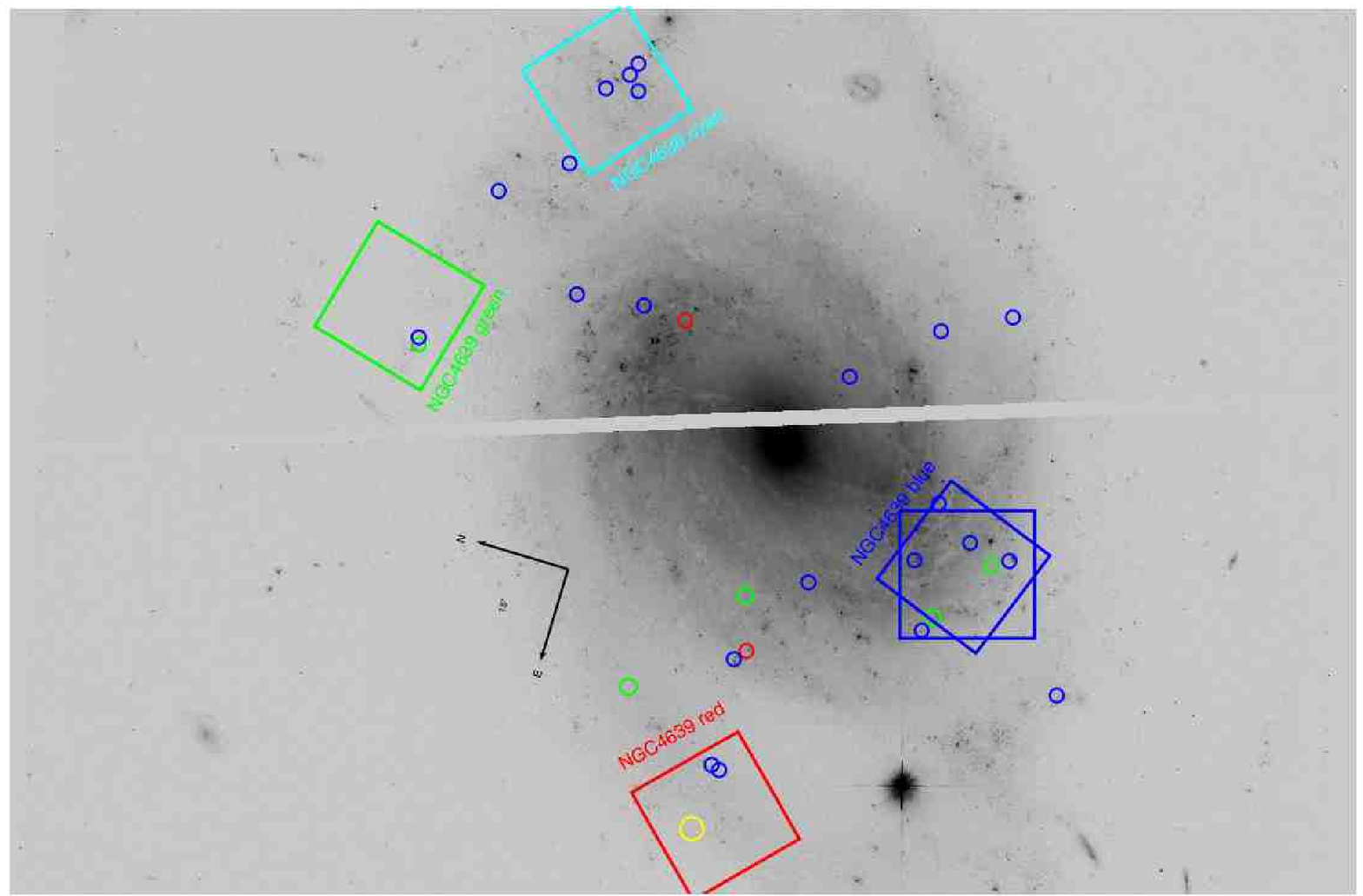}
\caption { }
\end{figure}

\begin{figure}[ht]
\vspace*{140mm}
\figurenum{6}
\includegraphics{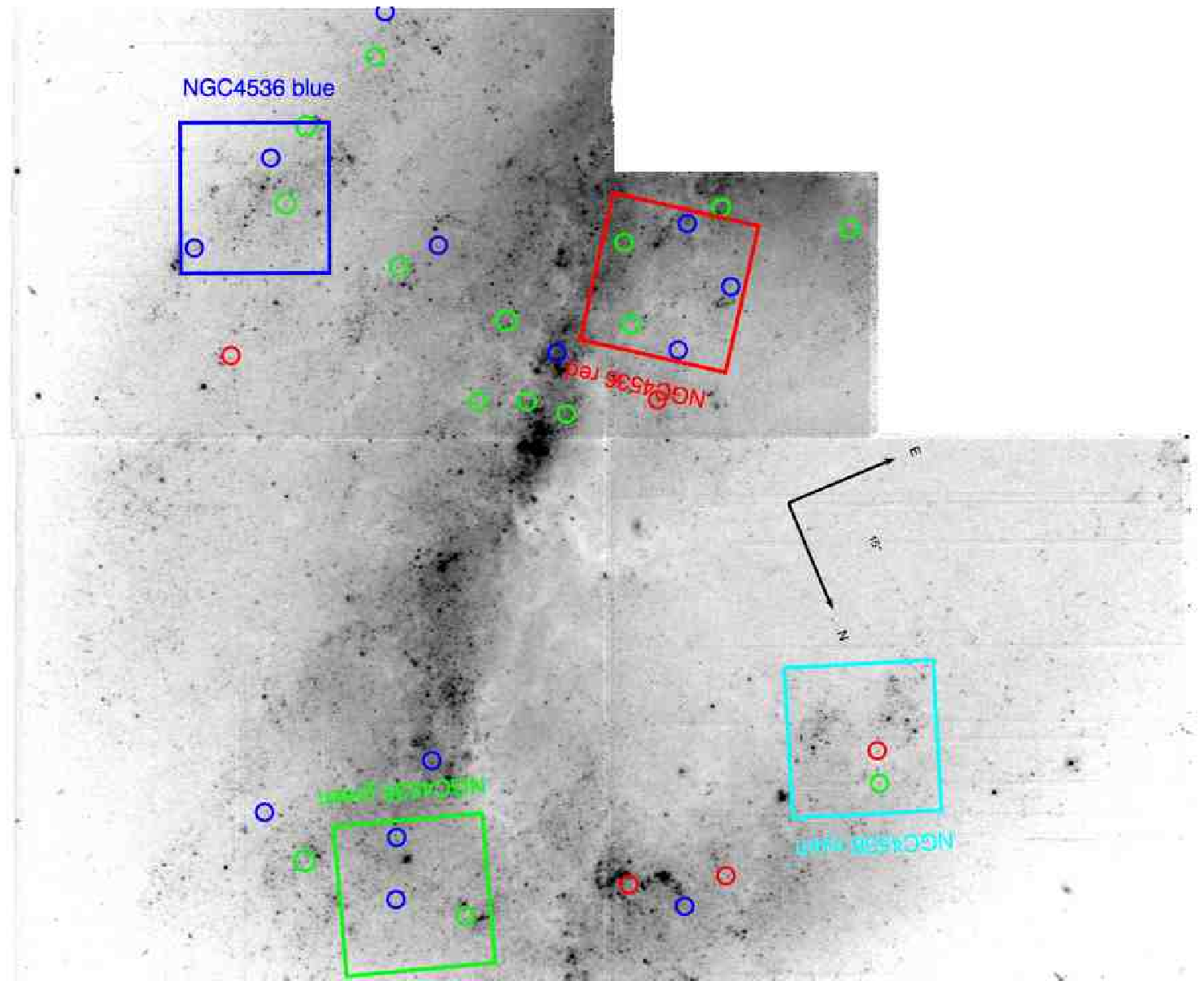}
\caption { }
\end{figure}

\begin{figure}[ht]
\vspace*{140mm}
\figurenum{7}
\includegraphics{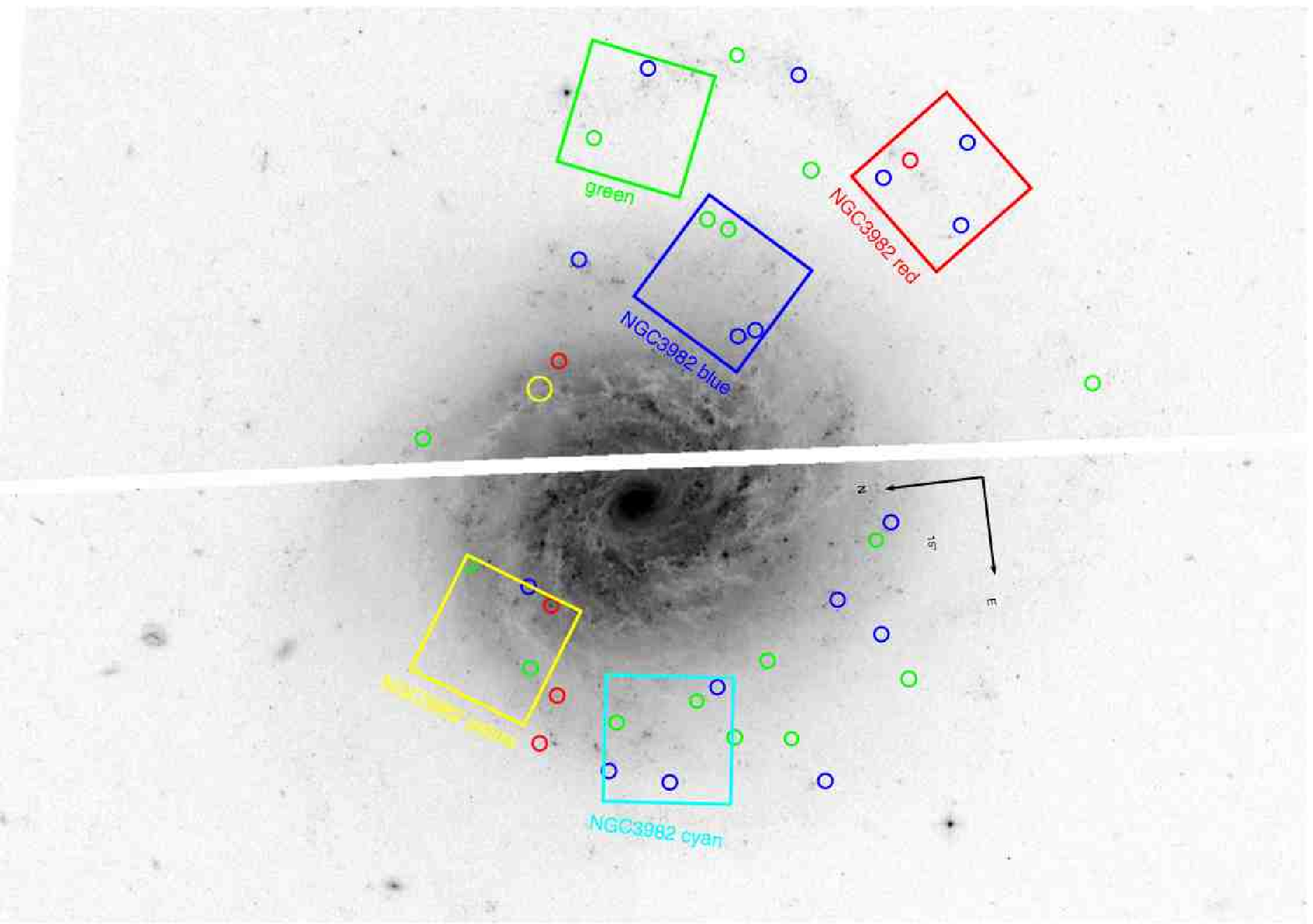}
\caption { }
\end{figure}

\begin{figure}[ht]
\vspace*{140mm}
\figurenum{8}
\includegraphics{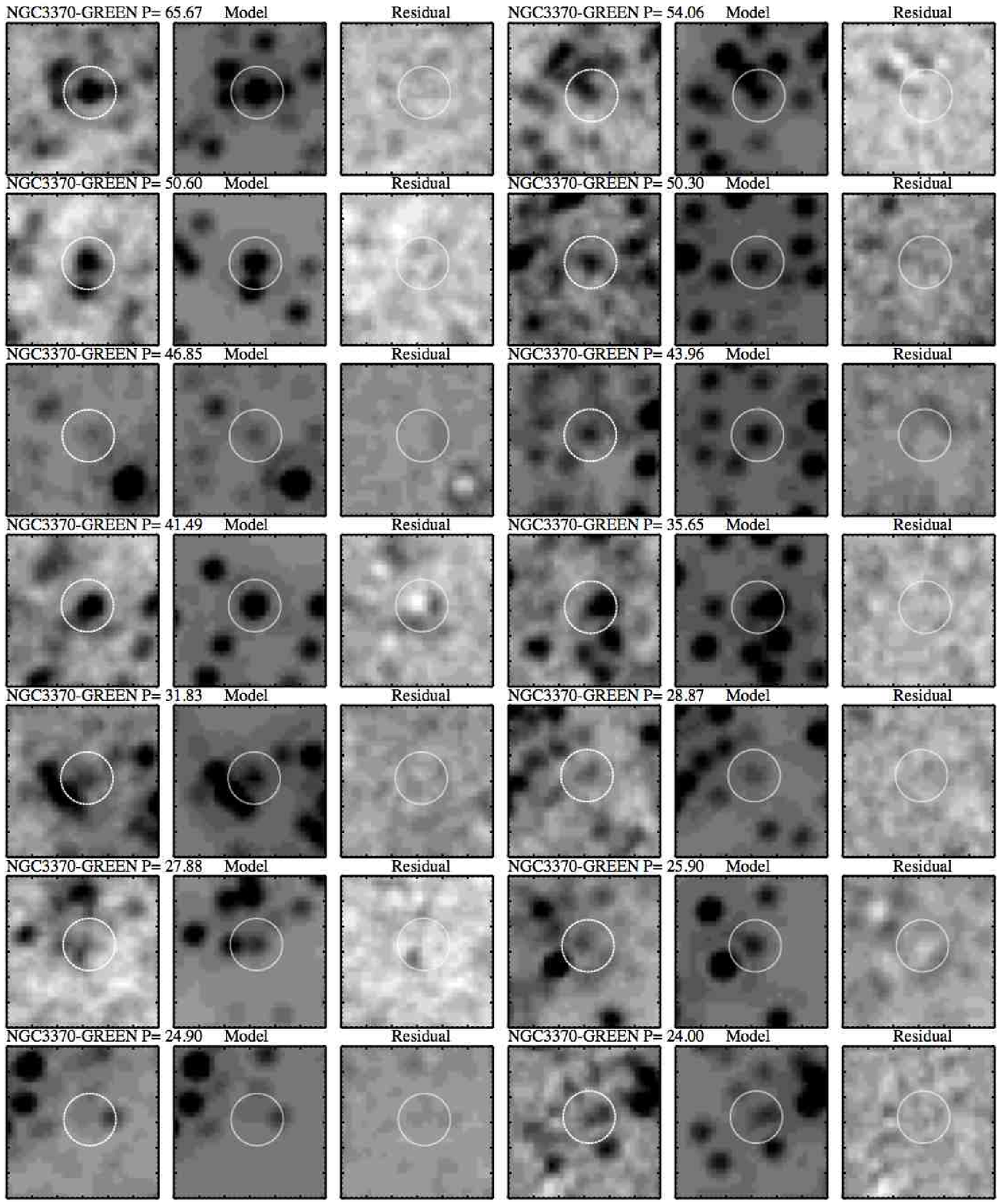}
\caption { }
\end{figure}

\begin{figure}[ht]
\vspace*{140mm}
\figurenum{9}
\includegraphics{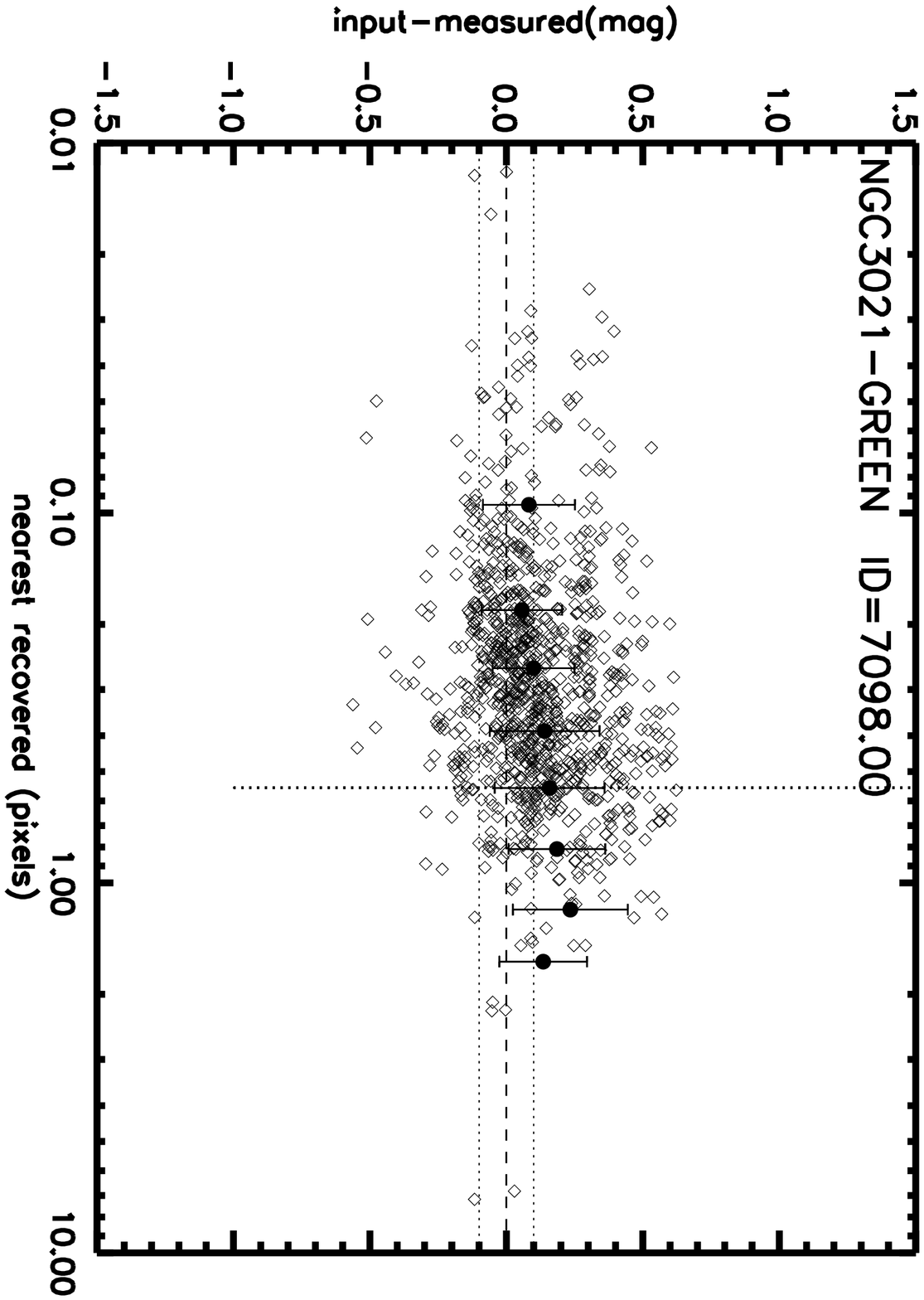}
\caption { }
\end{figure}

\begin{figure}[ht]
\vspace*{140mm}
\figurenum{10}
\includegraphics{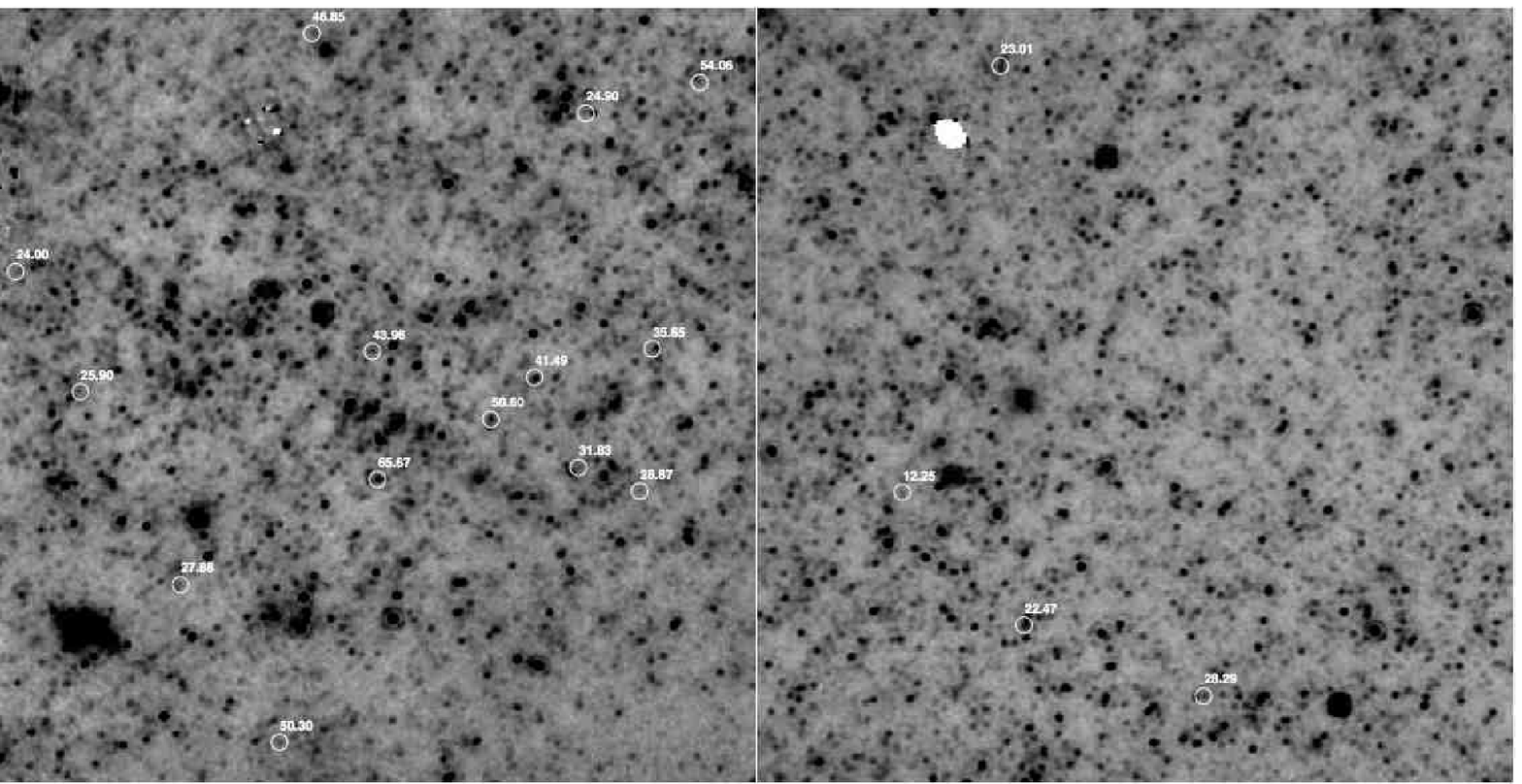}
\caption { }
\end{figure}

\begin{figure}[ht]
\vspace*{140mm}
\figurenum{11}
\includegraphics{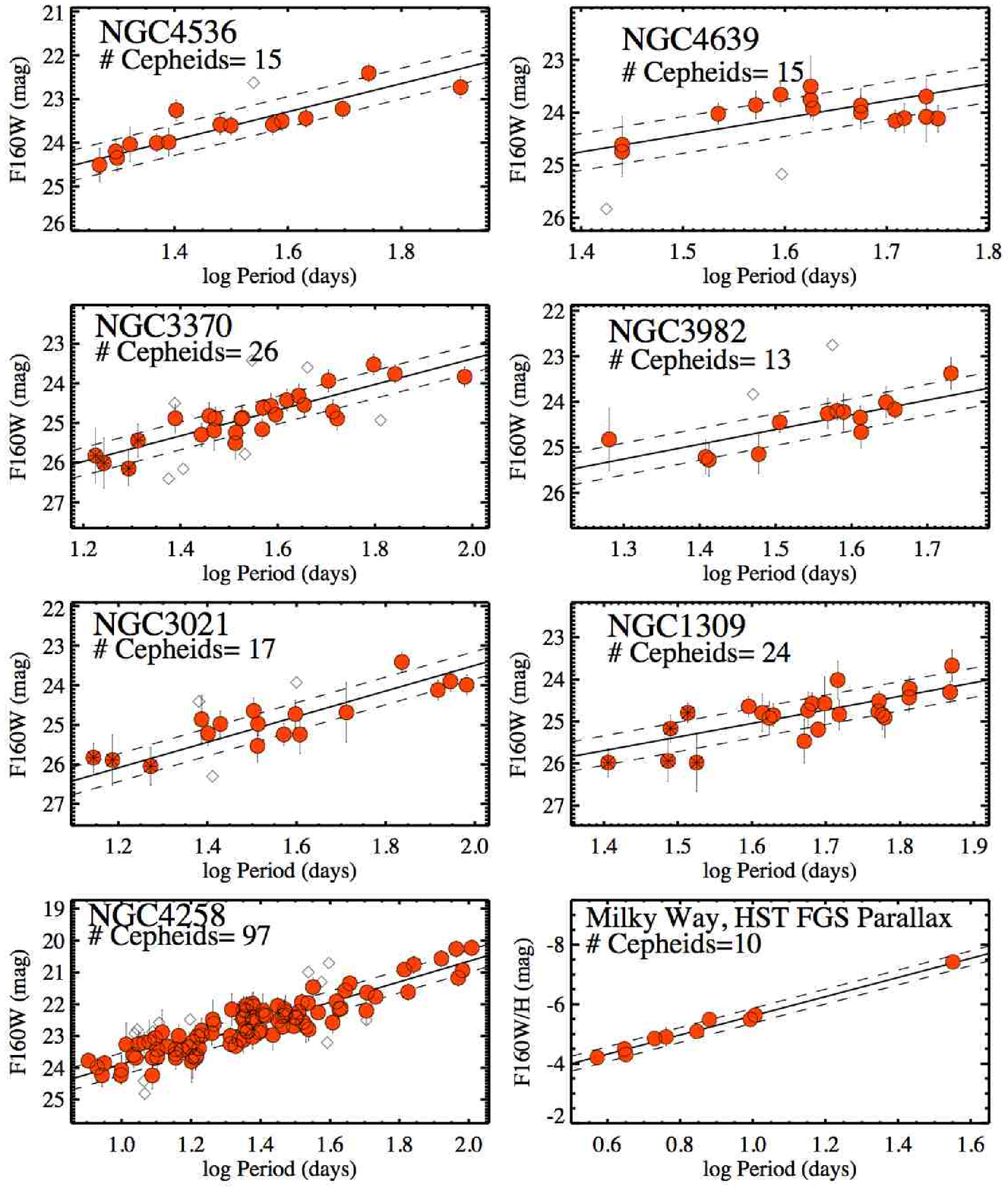}
\caption { }
\end{figure}

\begin{figure}[ht]
\vspace*{140mm}
\figurenum{12}
\includegraphics{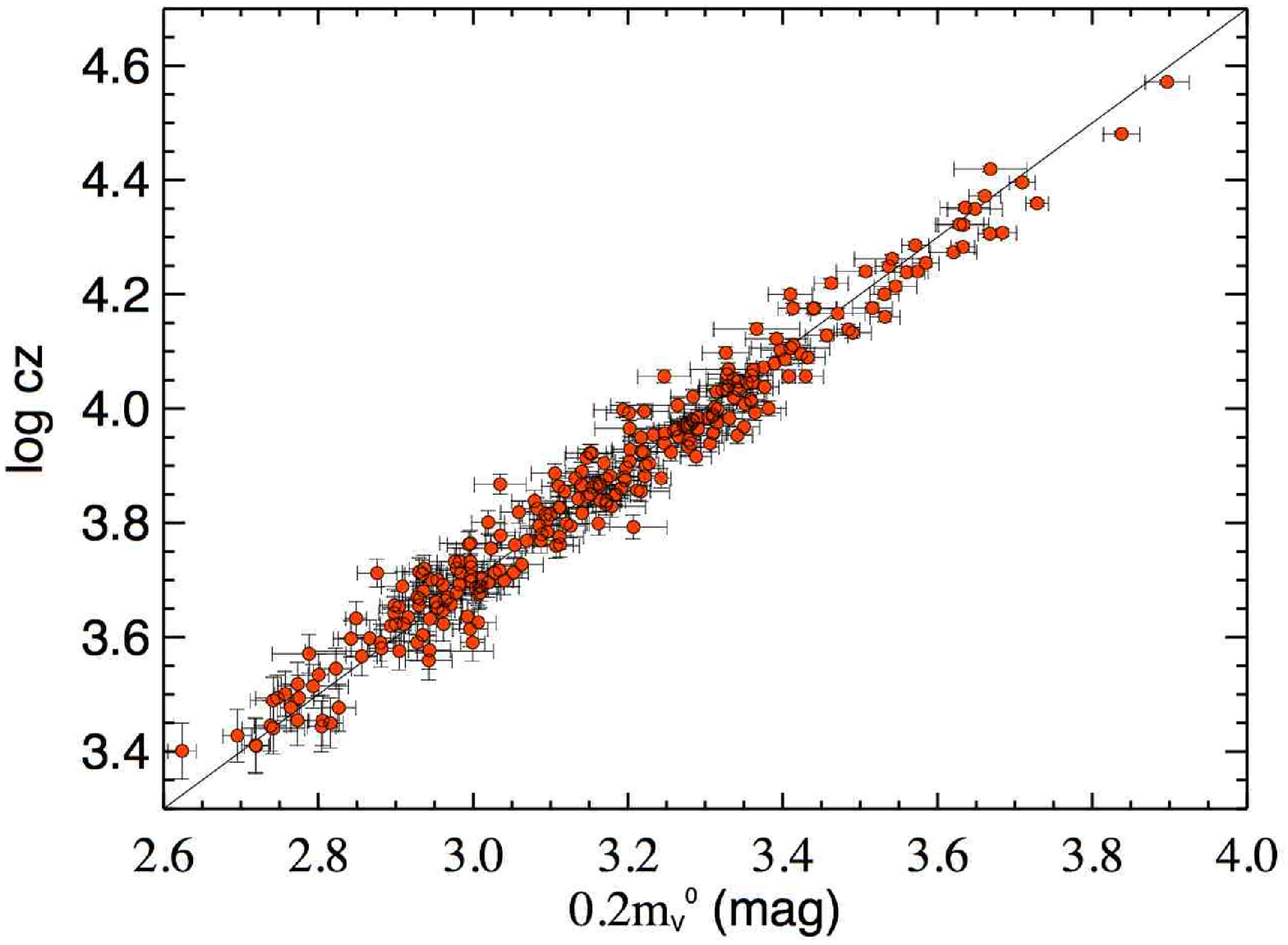}
\caption { }
\end{figure}

\begin{figure}[ht]
\vspace*{140mm}
\figurenum{13}
\includegraphics{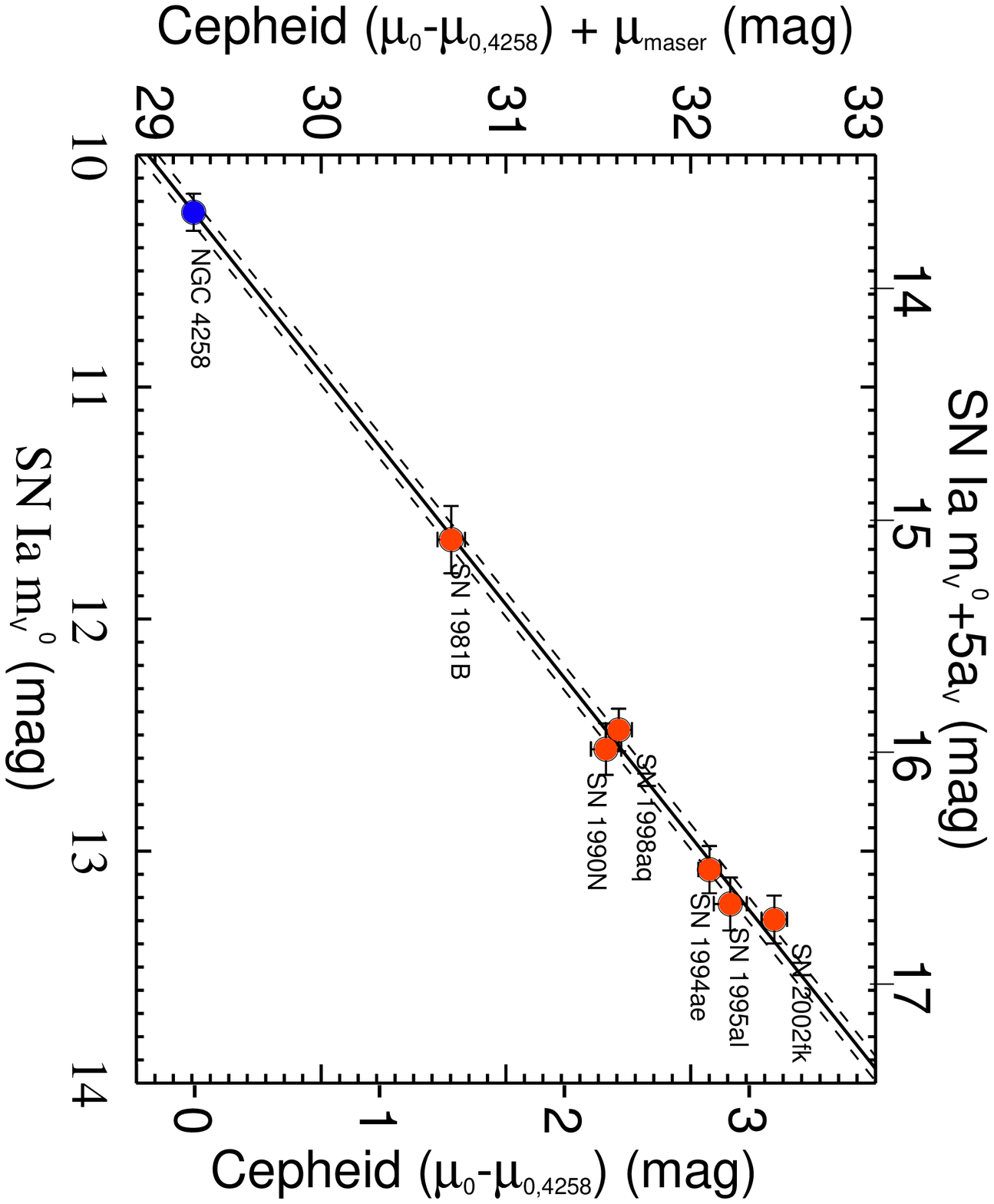}
\caption { }
\end{figure}

\begin{figure}[ht]
\vspace*{140mm}
\figurenum{14}
\includegraphics{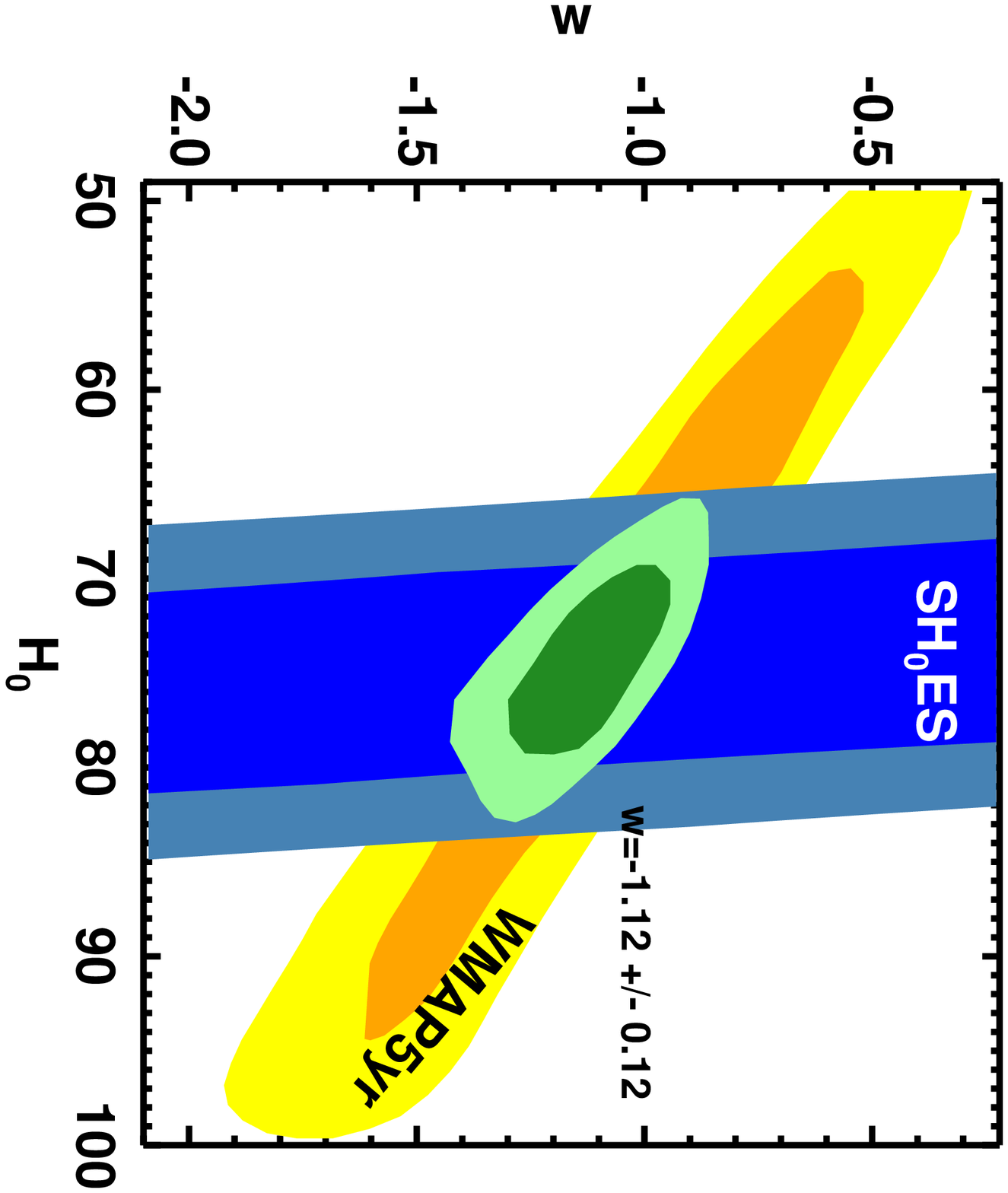}
\caption { }
\end{figure}

\begin{figure}[ht]
\vspace*{140mm}
\figurenum{15}
\includegraphics{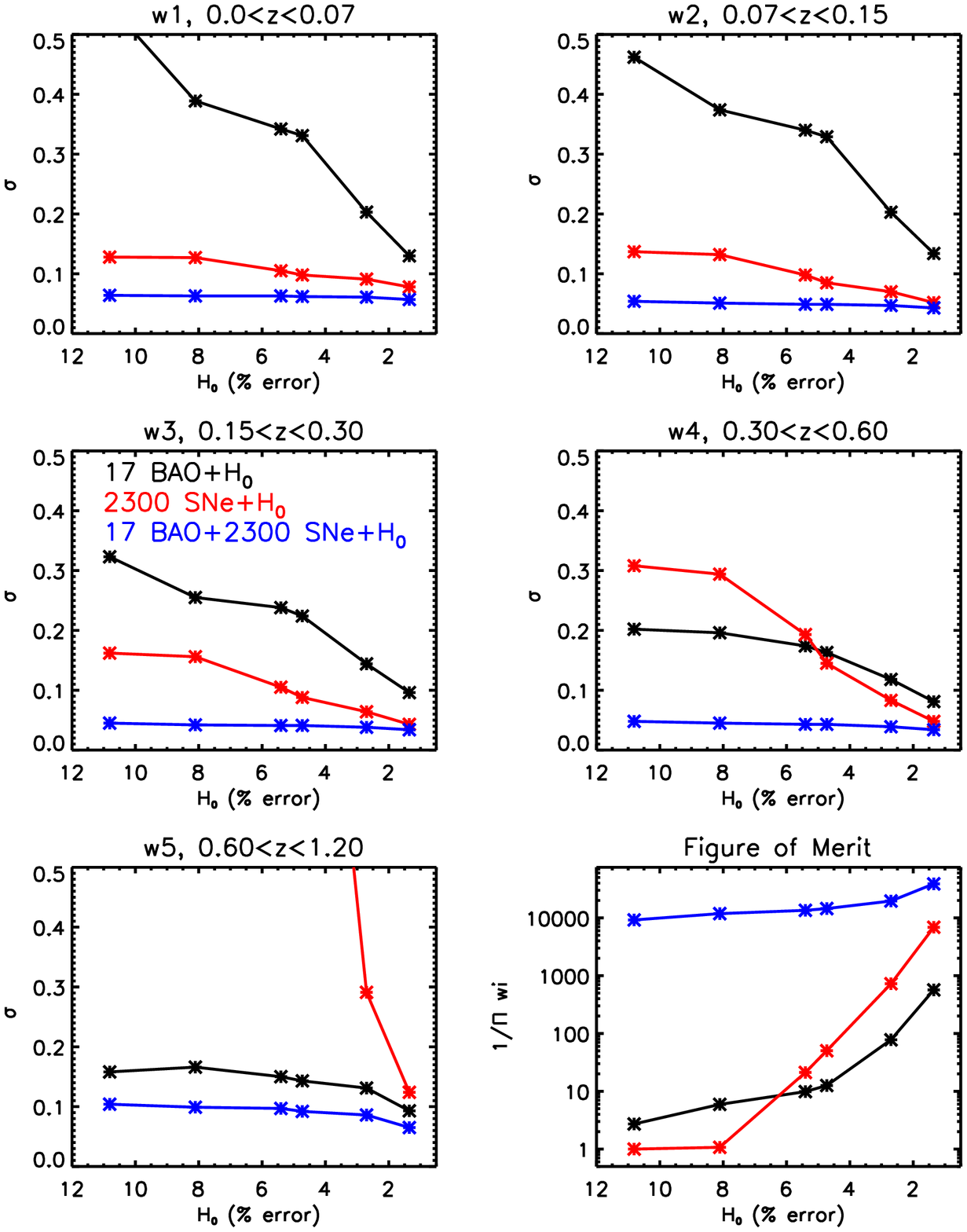}
\caption { }
\end{figure}

\end{document}